\documentclass[
  journal=pasa,
  manuscript=research-paper, 
  year=2021,
  volume=37,
]{cup-journal}

\usepackage{graphicx}
\usepackage{xcolor}
\usepackage{microtype,siunitx}
\usepackage{adjustbox}
\usepackage{amssymb}
\usepackage{amsmath}
\usepackage{hyperref}
\definecolor{dodgerblue}{RGB}{30, 144, 255}
\definecolor{crimson}{RGB}{220, 20, 60}
\definecolor{darkerblue}{RGB}{0, 0, 139}
\definecolor{darkred}{RGB}{150,20,20}
\definecolor{cpink}{RGB}{243, 141, 252}

\hypersetup{colorlinks,citecolor=dodgerblue,linkcolor=blue,urlcolor=magenta}
\usepackage{enumitem, xcolor}
\let\svitem\item

\usepackage{booktabs} 
\usepackage{threeparttable,longtable}  
\usepackage{makecell}

\usepackage[skip=0.5ex]{subcaption}
\usepackage{cprotect}
\usepackage{multirow}

\title{The Rapid ASKAP Continuum Survey (RACS) VIII: total intensity and circular polarization images and catalogues at 887.5\,MHz from RACS-low2}

\author{S.~W.~Duchesne}
\affiliation{CSIRO Space and Astronomy, PO Box 1130, Bentley WA 6102, Australia}
\email[S.~W.~Duchesne]{Stefan.Duchesne@csiro.au}

\author{K.~Rose}
\affiliation{Sydney Institute for Astronomy, School of Physics, The University of Sydney, NSW 2006, Australia}
\alsoaffiliation{CSIRO Space and Astronomy, PO Box 76, Epping, NSW, 1710, Australia}

\author{Tara~Murphy}
\affiliation{Sydney Institute for Astronomy, School of Physics, The University of Sydney, NSW 2006, Australia}

\author{Laura~N. Driessen}
\affiliation{Sydney Institute for Astronomy, School of Physics, The University of Sydney, NSW 2006, Australia}

\author{Joshua Pritchard}
\affiliation{CSIRO Space and Astronomy, PO Box 76, Epping, NSW, 1710, Australia}

\author{V.~A.~Moss}
\affiliation{CSIRO Space and Astronomy, PO Box 76, Epping, NSW, 1710, Australia}
\alsoaffiliation{Sydney Institute for Astronomy, School of Physics, The University of Sydney, NSW 2006, Australia}

\author{T.~J.~Galvin}
\affiliation{CSIRO Space and Astronomy, PO Box 1130, Bentley WA 6102, Australia}

\author{E.~Lenc}
\affiliation{CSIRO Space and Astronomy, PO Box 76, Epping, NSW, 1710, Australia}

\author{A.~W.~Hotan}
\affiliation{CSIRO Space and Astronomy, PO Box 1130, Bentley WA 6102, Australia}

\author{K.~Ross}
\affiliation{Australian SKA Regional Centre (AusSRC), Curtin University, Bentley WA 6102, Australia}
\alsoaffiliation{International Centre for Radio Astronomy Research, Curtin University, Bentley, WA 6102, Australia}

\author{A.~J.~M.~Thomson}
\affiliation{SKA Observatory, SKA-Low Science Operations Centre, 26 Dick Perry Avenue, Kensington WA 6151, Australia}
\alsoaffiliation{CSIRO Space and Astronomy, PO Box 1130, Bentley WA 6102, Australia}

\author{Matthew~T.~Whiting}
\affiliation{CSIRO Space and Astronomy, PO Box 76, Epping, NSW, 1710, Australia}

\doi{10.1017/pasa.20XX.XX}

\received {dd Mmm YYYY}
\revised  {dd Mmm YYYY}
\accepted {dd Mmm YYYY}
\published{22 September 2020}


\definecolor{dodgerblue}{RGB}{30, 144, 255}
\definecolor{rebeccapurple}{RGB}{102, 51, 153}

\def\NsourcesV{221}

\def\SRSCMatches{36}
\def\SRSCNoMatch{25}
\def\StarMeasurements{67}
\def\StarUnique{61}
\def\PulsarMeasurments{105}
\def\PulsarUnique{85}

\def\AGNMeasurements{48}
\def\AGNUnique{43}

\begin{document}

\begin{abstract}
We present a new data release from the Rapid ASKAP Continuum Survey (RACS), a widefield snapshot radio survey conducted with the Australian SKA Pathfinder (ASKAP). This data release contains the second RACS epoch to make use of ASKAP's low-frequency band, centred on 887.5\,MHz with a bandwidth of 288\,MHz, referred to as RACS-low2. This RACS-low2 data release includes both Stokes I and V imaging and catalogue data products covering the whole sky up to a declination of $\approx +48\degr$. RACS-low2 largely follows the observation footprint of the first low-band epoch, though includes additional coverage in the Northern Hemisphere. The observation scheduling and data processing follow previous RACS epochs, making use of autonomous scheduling and holography-derived primary beam models. The Stokes I and V catalogues are derived from images with a median point-spread function (PSF) of $14.2\arcsec \times 11.8\arcsec$, and have median root-mean-square noise properties of $\approx 195$ and $\approx 163$\,\textmu Jy\,PSF$^{-1}$, for the Stokes I and V images, respectively. The consolidated Stokes I catalogue contains 3\,922\,151 sources. We also constructed a catalogue of Stokes V sources by measuring Stokes V images at the Stokes I source positions. For the \NsourcesV\ Stokes V measurements above the estimated leakage and detection thresholds we provide likely identifications, including detections of \StarUnique\ radio stars, \PulsarUnique\ pulsars, \AGNUnique\ AGN (many likely to be residual leakage), and one source that is not associated with a known astronomical object. These data products, including calibrated visibilities, images, source lists, and consolidated catalogues, are made publicly available through the CSIRO ASKAP Science Data Archive (CASDA).

\end{abstract}

\defcitealias{racs1}{Paper~I}
\defcitealias{racs2}{Paper~II}
\defcitealias{Thomson2023}{Paper~III}
\defcitealias{racs-mid}{Paper~IV}
\defcitealias{racs-mid2}{Paper~V}
\defcitealias{racs-high}{Paper~VI}
\defcitealias{Thomson2026}{Paper~VII}

\section{Introduction}
\label{sec:int}

Widefield radio surveys provide an expansive view into the myriad astrophysical phenomena present in the Universe. Among the current suite of radio telescopes with ongoing widefield survey programmes, the Australian SKA Pathfinder \citep{Hotan2021}, the LOw-Frequency ARray \citep{lofar}, the Murchison Widefield Array \citep[MWA;][]{Tingay2013}, the Karl G.\,Jansky Very Large Array \citep[VLA;][]{Perley2011}, and MeerKAT \citep{Jonas2016} are providing data at high angular resolution across multiple wavelength regimes and sensitive  to a range of angular and temporal scales. Altogether, these surveys cover both the Northern \citep[e.g.][]{Shimwell2017,Lacy2020} and Southern Hemispheres \citep[e.g.][]{HurleyWalker2022,Goedhart2024,Mantovanini2025,Hopkins2025}, with surveys covering a significant portion of both. Among these, the now long-running Rapid ASKAP Continuum Survey \citep[RACS;][hereinafter \citetalias{racs1}]{racs1} is making maps of $\approx 88$\% of the sky covering 887.5--1655.5\,MHz at an angular resolution previously not seen over the entire Southern Hemisphere.   

\subsection{The Australian SKA Pathfinder}

{In the Southern Hemisphere, ASKAP is a 36-antenna interferometer operated by CSIRO \footnote{Commonwealth Scientific and Industrial Research Organisation.} on Inyarrimanha Ilgari Bundara, the CSIRO Murchison Radio-astronomy Observatory. The current primary focus of ASKAP is to conduct nine widefield survey programmes, known as Survey Science Projects (SSPs), which cover total intensity, polarization, spectral line, and time-domain science.} As the highest priority, these SSPs occupy the majority of the available telescope time, though ASKAP is also used for Guest Science Projects (GSPs) that do not fit within the existing surveys (up to 300 hr per year), target of opportunity observations, and technical tests for future scientific or functional capabilities with ASKAP. An additional category is the Observatory Projects, which are carried out by ATNF to meet specific goals that are of benefit to the observatory and wider community. 

\subsection{The Rapid ASKAP Continuum Survey}
While the other survey programmes are generally motivated by specific science cases, RACS is an {Observatory Project} intended to construct a global sky model for use in calibration. Multiple epochs of RACS have been observed, covering the three observing bands used by ASKAP, with central observing frequencies ranging from 887.5 to 1655.5\,MHz. Catalogues constructed from these data are already in use for both phase-reference calibration in the ASKAP data processing pipeline, as well as for post-imaging validation and brightness/astrometry corrections for some SSP data products. In addition to the RACS catalogues' usage in the ASKAP operational processing workflow, the catalogues have also seen utility in post-imaging/validation workflows unrelated to the Survey Science teams \citep[e.g.][]{deRuiter2026}, including for non-ASKAP data products \citep[e.g.][]{Frail2024,Rajwade2024,Duchesne2024,Duchesne2025b}.

While the initial goal of RACS was to create a global sky model for ASKAP, the data produced as part of the survey surpassed the sensitivity, angular resolution, and overall sky coverage of previous surveys of the southern sky. The public data products, including visibilities, images, and source catalogues, have seen use in a range of astronomical areas, including studies of radio galaxies \citep[e.g.][]{Bruno2024,Sethi2025,White2025}, high-redshift active galactic nuclei \citep[AGN; e.g.][]{Ighina2025}, variability related to AGN \citep[e.g.][]{Ross2022,Gabanyi2025}, cosmology \citep[e.g.][]{LandStrykowski2025}, tidal disruption events \citep{Anumarlapudi2024}, supernovae \citep[][]{Rose2024}, pulsars \citep[e.g.][]{Lenc2025}, and radio stars \citep[e.g.][]{Driessen2024}. As RACS is observed in full polarization, Stokes V images are also produced for most RACS epochs, providing circular polarization information \citep[with detected emission predominantly from radio stars and pulsars, e.g.][]{Pritchard2021,Rose2023,Kaur2025}. Linear polarisation data is also being produced as part of the ongoing Spectra and Polarisation in Cutouts of Extragalactic sources from RACS project \citep[SPICE-RACS;][]{Thomson2023,Thomson2026}.  

Table~\ref{tab:racs} covers the observed RACS epochs with summary details, including observing dates, frequencies, {basic data product metrics, and references for publications describing the data releases associated with each epoch} \footnote{See also the RACS project website for further general information, including links to public datasets: \url{https://research.csiro.au/racs/}.}. RACS epochs are referred to by the band they are observed in, and for epochs beyond the first a number is typically included to indicate which epoch of that specific band. The initial RACS epoch in the low band is referred to as RACS-low or RACS-low1 to avoid ambiguity\footnote{This may be referred to as RACS-Low or RACS-LOW in the literature, and any form is identical.}.

Processed RACS observations, including calibrated visibilities and images, are made available through the CSIRO ASKAP Science Data Archive\footnote{\url{http://data.csiro.au/domain/casda}.} \citep[CASDA;][]{Huynh2020}.

\section{Data}

\begin{table*}[t]
    \centering
    \begin{threeparttable}
    \caption{\label{tab:racs}{RACS epochs currently observed with properties of the images and catalogues where available}.}
    \begin{adjustbox}{max width=\textwidth}
    \begin{tabular}{c c c c c c c c}\toprule
         Epoch name & Frequency & Date range \tnote{a}  & Polarizations \tnote{b} & Median $\sigma_\text{rms}$ \tnote{c} & Median PSF & $N_\text{sources}$ \tnote{d} &  References \\
         & (MHz) &  & (\textmu Jy\,PSF$^{-1}$) & ($^{\prime\prime} \times ^{\prime\prime})$ & & &\\[0.4em]\midrule
         \multirow{1}{*}{RACS-low} & \multirow{1}{*}{887.5} & 2019-04-21--2020-06-21 & I  & 255 & $ 17.4 \times 11.4$ & - &  \citet[][\citetalias{racs1}]{racs1} \\
         & & & I & 314 & $25.0 \times 25.0$ & 2\,297\,596 &  \citet[][\citetalias{racs2}]{racs2} \\
         & & & Q, U \tnote{e} & 157, 149 & $25.0 \times 25.0$ & 5\,818 &  \citet[][\citetalias{Thomson2023}]{Thomson2023} \tnote{e} \\[0.4em]\midrule
         \multirow{1}{*}{RACS-mid} & \multirow{1}{*}{1367.5} & \multirow{1}{*}{2020-12-20--2022-06-11}  & I, V & 198, 165 & $10.1 \times 8.1$ &  - &  \citet[][\citetalias{racs-mid}]{racs-mid} \\
          & & & I & 182  & $11.2 \times 9.3$ &  3\,105\,668 &  \citet[][\citetalias{racs-mid2}]{racs-mid2} \\[0.4em]\midrule
         \multirow{1}{*}{RACS-high} & \multirow{1}{*}{1655.5} & \multirow{1}{*}{2021-12-24--2022-03-03}  & I, V & 197, 188 & $8.6 \times 6.3$ & - &  \citet[][\citetalias{racs-high}]{racs-high} \\
        &  &  &  I & 195 & $11.8 \times 8.1$ & 2\,677\,509 & \citetalias{racs-high} \\[0.4em]\midrule
         RACS-low2 & 887.5 & 2022-03-18--2022-06-21  & I &  225 & $14.3 \times 11.8$ & - & This work. \\
          &  &   &V & 163 &  $14.3 \times 11.8$ & 221 &  This work. \\
         
          & &  & I & 195 & $15.2 \times 13.0$ & 3\,922\,151 &  This work. \\[0.4em]\midrule
         RACS-low3 & 943.5 & 2023-12-20--2024-04-06  & I, V & 205, -  & $13.4 \times 11.0$ & - &  - \\
          &  &  & I, Q, U & 211, 194, 194 & $14.5 \times 11.0$ & 342\,606 &   \citet[][\citetalias{Thomson2026}]{Thomson2026} \\
          &  &  & I &  $152$ \tnote{f} & $14.1 \times 11.7$ \tnote{f} & $\approx 4.5$\,M \tnote{f} &  Galvin et al. (in prep.) \\[0.4em]\midrule
         RACS-mid2 & 1367.5 & 2024-10-05--2025-08-14  &I, V &  - & - & - &  - \\[0.4em]\bottomrule
\end{tabular}
\end{adjustbox}
\begin{tablenotes}[flushleft]
{\footnotesize \item[a] {Overall date range for the entire epoch, including re-observations usually taken at a later date.} \item[b] {Stokes imaging data products available, however, note all RACS observations are observed in full instrumental polarization which are available in the calibrated visibilities also available on CASDA.} \item[c] {For the relevant Stokes parameters.} \item[d] {From the main consolidated output catalogue of each data release, where available.} \item[e] {Stokes Q, U data are available for a $\approx 1300$-deg$^2$ region in the direction of the Spica Nebula.} \item[f] {Based on work in progress.}}
\end{tablenotes}

\end{threeparttable}
\end{table*}

\subsection{Observations}

\begin{table*}[t]
    \centering
    \begin{threeparttable}
    \caption{\label{tab:holography}Holography observation details and associated science SBIDs for RACS-low2.}
    \begin{tabular}{ c c c c}\toprule
        Holography SBID & Science SBID range \tnote{a} & Date range for science SBIDs \tnote{b} & Number of SBIDs \\\midrule
        38585 & 38307--38528 & 2022-03-18 11:40:00--2022-03-24 22:09:39 & 108 \\
        38709 & 38545--39385 & 2022-03-25 12:20:02--2022-04-14 01:54:49 & 605 \\
        39549 & 39400--40878 & 2022-04-14 10:00:00--2022-05-22 19:23:53 & 356 \\
        41055 & 40989--41829 & 2022-05-26 17:39:29--2022-06-21 13:31:14 & 10 \\\bottomrule
    \end{tabular}
    \begin{tablenotes}[flushleft]
    {\footnotesize \item[a] Inclusive. \item[b] Observation start times.}
    \end{tablenotes}
    \end{threeparttable}
\end{table*}

\begin{figure*}[t]
    \centering
    \includegraphics[width=1\linewidth]{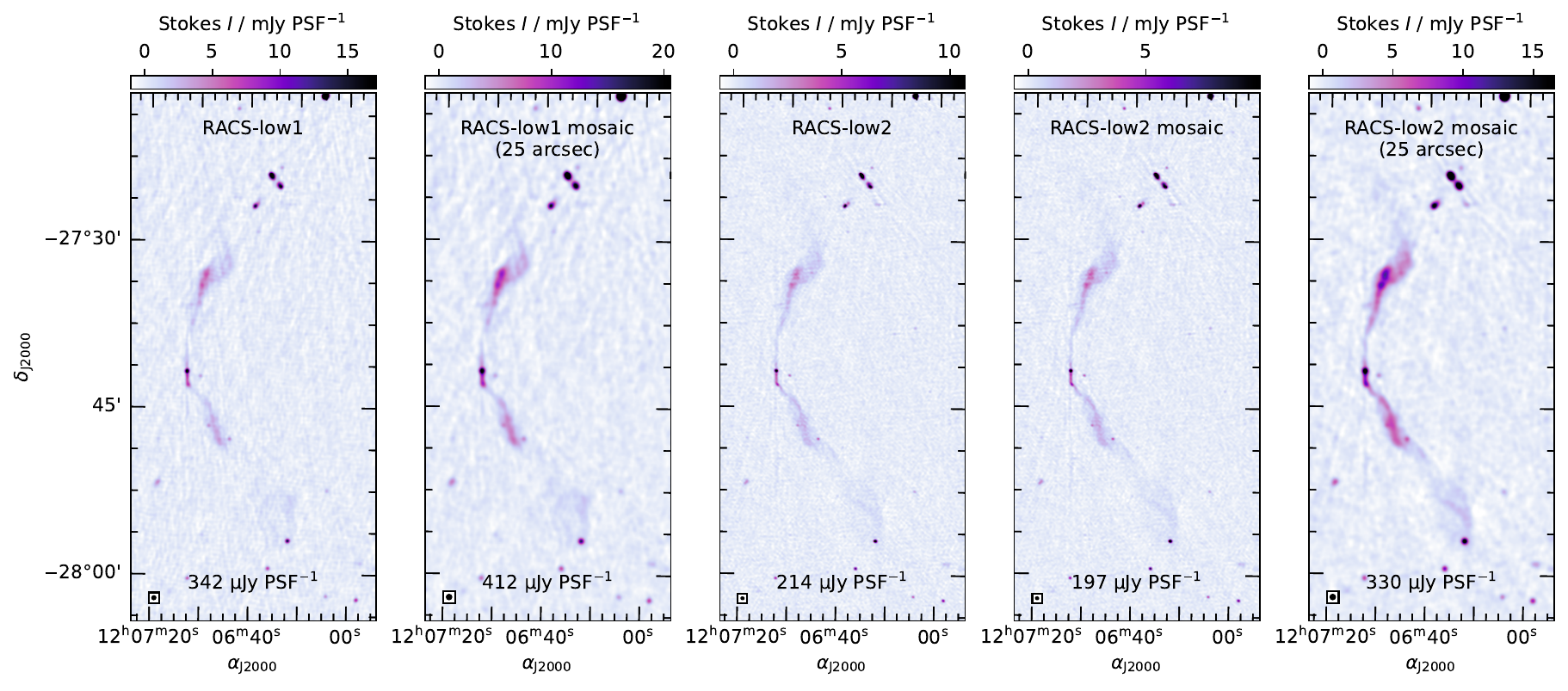}
    \caption{\label{fig:example:low1} Example giant radio galaxy associated with ESO~505$-$G014 \citep{vanVelzen2012,Dabhade2020} viewed in RACS-low1 and RACS-low2 (original images, full-sensitivity mosaics, and at $25\arcsec \times 25\arcsec$ angular resolution). The colour scales are linear between $-3 \sigma_\text{rms}$ and $50\sigma_\text{rms}$ with $\sigma_\text{rms}$ reported at the bottom of each panel. Note that RACS-low1 full-sensitivity mosaics were all produced at $25\arcsec \times 25\arcsec$ angular resolution.}
\end{figure*}

RACS-low2 largely follows the footprint layout of the original RACS-low1, using the regular \texttt{square\_6x6} layout of the Phased Array Feed (PAF) beams with a central frequency of 887.5\,MHz and 288-MHz bandwidth. It was carried out in 2022 based on a specification of 947 pointings.

The RACS-low2 observing strategy matches the later RACS epochs like RACS-mid and RACS-high, executed by the dynamic scheduler SAURON \footnote{Scheduling Autonomously Under Reactive Observational Needs.} as part of the autonomous science operations of ASKAP (Moss et al. in prep). {Each pointing is observed within an individual scheduling block with a unique ID (SBID).} We note that the initial RACS-low survey observations in 2019/2020 preceded the late-2020 shift to this autonomous mode of operations, and provided key input to optimise the transition from semi-automated to autonomous observing. In the case of RACS, we optimise each of the 15-min pointings based on specific constraints in addition to other global constraints and weightings that apply to all surveys being carried out. For RACS, these include restricting observations to $\pm 1$ hour angle off meridian to maintain then best possible angular resolution at any given declination, ensuring the presence of all six outer antennas to also maximise angular resolution, and limiting both solar and lunar proximity to $>$ 10\,deg.

Figure 1 in \citetalias{racs-mid} shows the difference in the PAF beam arrangement for the \texttt{square\_6x6} and \texttt{closepack36} footprints, the latter of which is used for RACS-mid, RACS-high, and RACS-low3. For RACS-low2, pointings were also conducted up to $\delta_\text{J2000} = +43\degr$ following the later epochs to provide survey coverage up to $\delta_\text{J2000} \approx +48\degr$ compared to the $\approx +41\degr$ declination limit of RACS-low1.

RACS-low2 was observed after RACS-high, with data taken between 2022-03-18 and 2022-06-21. Observations were initially processed to check data quality, with poor observations due to miscellaneous system issues flagged to be reobserved. The total number of observations performed was 1079, of which 132 were duplicates of the same field.

As with RACS-high and subsequent RACS epochs, RACS-low2 was observed alongside contemporaneous holography observations of PKS~B0407$-$658 to allow us to model the PAF beam responses over the course of the observations. The PAF beams are measured after every major beam weight update, and for RACS-low2 there were four separate holography-derived primary beam models used. Table~\ref{tab:holography} reports the SBID range for each holography observation and derived beam model. Note that unlike RACS-high, widefield leakage models derived from those holography observations are available for all of the holography observations, though for the work here we did not make use of the widefield leakage models.

\subsection{Processing}


The RACS-low2 data were processed using the ASKAP pipeline (version 2.9.11) \footnote{
\url{https://www.atnf.csiro.au/computing/software/askapsoft/sdp/docs/current/pipelines/index.html}.} built around the \texttt{ASKAPsoft} software packages \citep{Guzman2019}. Processing took place on the Pawsey Supercomputing Centre cluster \textit{Setonix} \footnote{\url{https://doi.org/10.48569/18sb-8s43}.}. We obtained 7.4M CPU hours as part of the Pawsey Partner Scheme (PPS) under project \textit{pawsey1014} (PI Duchesne). Some additional reprocessing (described in Section~\ref{sec:patching}) made use of a separate PPS project, \textit{ja3} (PI Whiting), allocated for extra post-processing and related work for ASKAP surveys. Data processing for RACS-low2 is largely identical to {previous RACS epochs}, including initial bandpass and complex gain calibration and on-axis leakage correction based on observations of PKS~B1934$-$638 (assuming it is unpolarized), two rounds of phase-only self-calibration, and final imaging with the \texttt{ASKAPsoft} \texttt{cimager} \footnote{\url{https://www.atnf.csiro.au/computing/software/askapsoft/sdp/docs/current/calim/cimager.html}.} for each PAF beam dataset. 

{We make both Stokes I and Stokes V images, making use of multi-term, multi-scale deconvolution using the \texttt{Basis\-function\-MFS} CLEAN deconvolution algorithm \footnote{\url{https://www.atnf.csiro.au/computing/software/askapsoft/sdp/docs/current/calim/solver.html}.}. Two Taylor terms are used to represent the apparent sky brightness distribution. Widefield gridding is performed with the $W$-projection algorithm \citep{Cornwell2008}, and image weighting uses Wiener preconditioning \citep{Rau2010} with a robust parameter of 0.0 (similar to `Briggs' robust 0.0, \citealt{Briggs95}) providing a good trade-off between sensitivity and angular resolution. Stokes I images go through up to ten major cycles of CLEAN deconvolution, down to a final depth of 0.5 mJy. Stokes V images only use up to three major iterations due to the comparatively empty circularly-polarized sky.} 

A subset of the data also go through a bespoke off-axis source subtraction process prior to self-calibration, where bright sources (and observations they affect) are manually selected. This process is not part of the official ASKAP pipeline, though the \texttt{ASKAPsoft} off-axis source subtraction implementation has recently been enabled in the ASKAP operation pipeline. The bespoke workflow \footnote{\url{https://gitlab.com/Sunmish/potato}.} is described in \citetalias{racs-mid} and \citetalias{racs-high}, and involves imaging a small region around the offending source and either directly subtracting the CLEAN component model or first deriving complex gain amplitudes and phases in the direction of the source and then subtracting the model. This process is similar to peeling \citep[see][]{Noordam2004,Smirnov2011b}, though gain solutions are not directly applied to the data {and only operates on XX and YY instrumental polarizations}. This process uses \texttt{WSClean} \citep{wsclean1,wsclean2} for imaging the regions around sources and CASA \footnote{Common Astronomy Software Applications: \url{https://casa.nrao.edu/}.} \citep{CASA2022} for deriving gain solutions. The list of subtracted sources is included in Appendix~\ref{app:subtracted}, though we note sources are only subtracted from a PAF beam dataset if they are not within 1.2\,deg of the beam centre. This choice of radius resulted in over-subtraction in some images---see Section~\ref{sec:patching} for further details.

After imaging, the individual PAF beam images are then mosaicked using the \texttt{ASKAPsoft} linear mosaicking package \texttt{linmos}, which applies the primary beam models and also provides Stokes I leakage removal for the Stokes V images based on the leakage models derived from holography. {We use a primary beam attenuation cutoff of 12\% during linear mosaicking so the mosaic images do not extend beyond this cutoff.} Individual PAF beam images are convolved to a common resolution {for that specific set of PAF beams} prior to mosaicking, and mosaics are weighted with a combination of primary beam attenuation and the total number of gridded visibilities. An output mosaic weight map is also produced.

Unlike RACS-mid, we do not apply any post-image brightness scale corrections. See Section~\ref{sec:brightness} for an assessment of the resultant brightness scale for the RACS-low2 data.

\section{Catalogues}
\subsection{Stokes I mosaics}\label{sec:mosaics}

\begin{figure*}[t]
    \centering
    \includegraphics[width=1\linewidth]{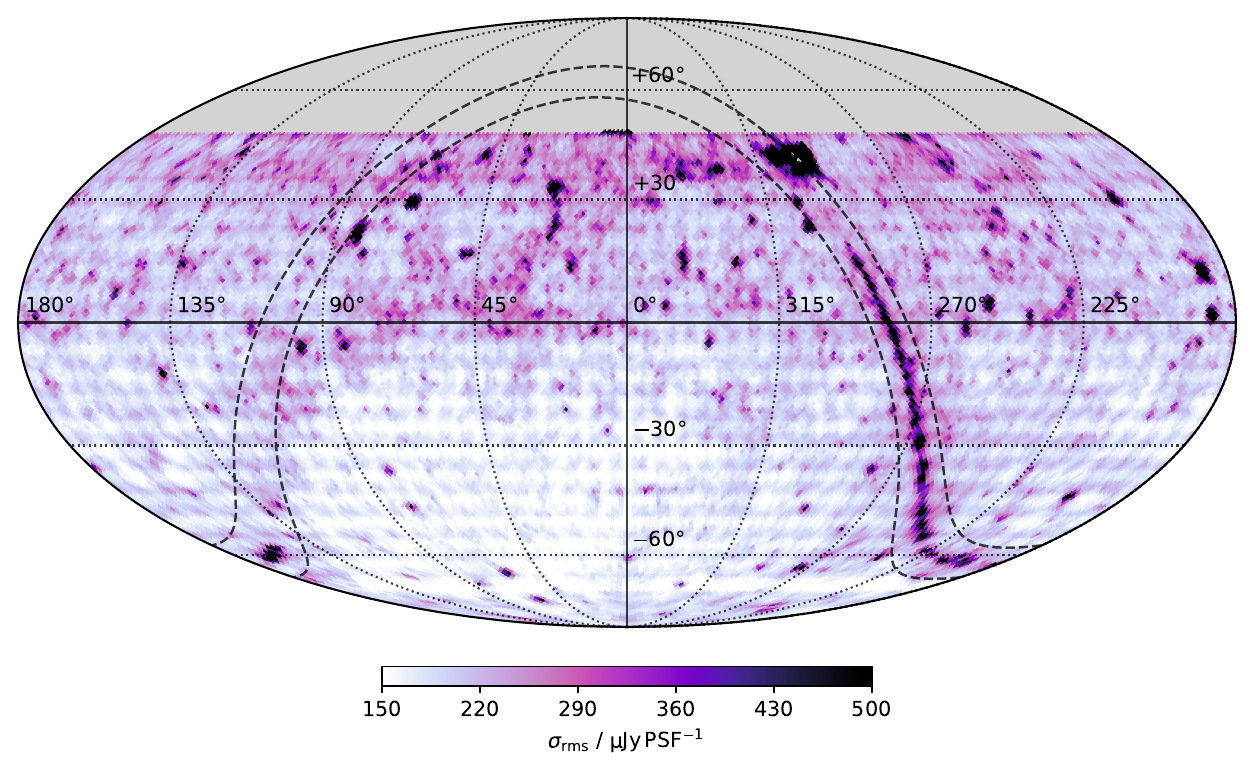}
    \caption{\label{fig:rms} The HEALPix-binned rms noise across the full-sensitivity mosaics as reported in the catalogue. {We use $N_\text{side}=64$ for HEALPix binning, corresponding to an area of $\approx 55 \times 55$\,arcmin$^2$.}  The dashed black lines are drawn at Galactic latitudes of $|b| \pm 5\degr$.}
\end{figure*}

To reduce sensitivity roll-off at the edge of the individual images due to the primary beam attenuation, we mosaic neighbouring observations following \citetalias{racs2}. Mosaicking and subsequent source-finding made use of the internal CSIRO supercomputer, \texttt{petrichor}. As the spacing between fields is equivalent to the spacing of the PAF beams, there is significant overlap between each individual image allowing us to create a `full-sensitivity' image for each field as a linear mosaic. These mosaics make use of the weight maps generated earlier so that the mosaicking is functionally a continuation of the mosaicking of individual PAF beams. In this case, we make use {of} \texttt{SWarp} \footnote{\url{https://www.astromatic.net/software/swarp/}.} \citep{swarp} for regridding and co-adding the images. For each full-sensitivity mosaic, we also convolve all individual images to a common resolution for that mosaic, following \citetalias{racs-mid2}, using \texttt{beamcon\_2D} from RACS-tools \footnote{\url{https://github.com/AlecThomson/RACS-tools}.}. This allows us to retain an angular resolution that is close to the original resolution of the images while ensuring flux density measurements will be accurate in the full-sensitivity images. 

In Figure~\ref{fig:example:low1} we show example cutouts of a giant radio galaxy \citep[associated ESO~505$-$G014;][]{vanVelzen2012,Dabhade2020} to highlight the differences in the available RACS-low1 and RACS-low2 images. The left and centre-left panels show the original RACS-low1 images (before and after mosaicking and convolution to $25\arcsec \times 25\arcsec$ angular resolution) and the remaining three images show cutouts from the same field from RACS-low2 (original image, full-sensitivity mosaic image, and the full-sensitivity mosaic image convolved to $25\arcsec \times 25\arcsec$ angular resolution). This particular field saw a dramatic improvement to the angular resolution and root-mean-square (rms) noise in the RACS-low2 observation, going from $18.3\arcsec \times 17.3\arcsec$ (with an rms noise of $\sigma_\text{rms} = 342$\,\textmu Jy\,PSF$^{-1}$) in the RACS-low1 observation to $13.1\arcsec \times 11.1\arcsec$ ($\sigma_\text{rms} = 214$\,\textmu Jy\,PSF$^{-1}$) in the equivalent RACS-low2 field. Part of the improvement in rms noise follows from the improvement in angular resolution, though we also see reduction in the rms noise for the full-sensitivity mosaic $25\arcsec$ images as well, highlighting the general data quality improvement with this second RACS-low epoch.  

\subsection{Stokes I catalogue}\label{sec:catalogue:stokesi}

\subsubsection{Initial source finding}
With the full-sensitivity mosaics we use \texttt{PyBDSF} \footnote{\url{https://pybdsf.readthedocs.io/en/latest/.}} \citep{pybdsf} for source-finding. The process is identical to that described in \citetalias{racs-high} for RACS-high: we use a $5\sigma_\text{rms}$ threshold for initial source detection, and $2\sigma_\text{rms}$ threshold for extending the pixel boundaries of the source detection for fitting and measurement. Each source is decomposed into a collection of 2-D Gaussian components, and since we typically find \texttt{PyBDSF} is generous in how many components are fit to what should be a point source, we further regroup multiple components into a single component when the ratio of integrated to peak flux density is less than the median for a particular image \citepalias[see section~2.4 in][]{racs-mid2}. Source lists from each full sensitivity mosaic are then merged into a contiguous catalogue, with duplicate entries in overlap regions removed following \citetalias[][]{racs-mid2}. Details of the columns included in the Stokes I catalogue are included in Appendix~\ref{app:columns}.

\subsubsection{Catalogue patching}\label{sec:patching}

After creating the catalogue we identified two issues affecting some of the brightest sources in the sky.

In initial assessments of the catalogue, we found that some of the peeled sources had a reduced flux density. We used the same 1.2 deg radius originally used for RACS-mid and RACS-high to determine when an off-axis source should subtracted, and in some cases the source was subtracted when within the imaged field-of-view even after primary beam clipping. After mosaicking, this results in a reduction in source brightness by up to 10\%. To correct for this in the catalogue, we re-image the PAF beam closest to each of the peeled sources and then apply a primary beam correction before re-running \texttt{PyBDSF} on the single PAF beam images. We then patch entries in the catalogue within the apertures originally used to define the subtraction regions. Note that images that appear on the CASDA retain the underestimated flux densities for these bright sources (see Appendix~\ref{app:subtracted} for a full list of subtracted sources that may be affected).  

Due to our rms thresholding, some large extended sources with fairly uniform surface brightness distributions were missed in the source-finding process. This includes the Galactic sources Taurus~A, M~42, Orion~B, Kes~62, Sgr~A*, RCW~38, and the radio galaxy PKS~J0133$-$3629. To correct this, we re-run \texttt{PyBDSF} on cutouts around these sources with modified rms thresholding to ensure the source itself does not hinder the local rms noise calculation. The exact thresholding varies by source, but generally is an increase in the box size used for calculating the local rms noise. The new Gaussian components and sources are then added to the catalogue by replacing any existing components/sources within an aperture slightly larger than the extent of the source, as above. For Taurus~A, M~42, and Orion~B, we do this on the single-beam images described above as they had been originally partially subtracted. Similarly, Pictor~A is was also catalogued as two separate sources (i.e.\ the two lobes) so we repeat the same process for Pictor~A here to ensure it is a single source with multiple components. We note as well that PKS~J0133$-$3629 remains split into two sources (as individual lobes). 

After patching entries in the catalogue we report 3\,922\,151 unique sources (with 5\,324\,477 Gaussian components).

\subsection{Initial Stokes V catalogue}\label{sec:catalogue:stokesv}

Circular polarization affords a unique opportunity to search for sources such as pulsars and radio stars due to the overall lower density of circularly polarized sources in the radio sky \citep[e.g.][]{Gaensler1998}. As many sources of strong circularly polarized emission tend to be time-varying \citep[e.g.,][including long-period radio transients; \citealt{Lee2025}]{Pritchard2021, Rose2023}, performing a similar mosaicking process with the Stokes V images would potentially result in averaging a time-dependent signal between observations that might be separated by up to a few months. 

To generate a catalogue of Stokes V sources, we instead use the individual Stokes I and V images prior to constructing the full-sensitivity mosaics described in Section~\ref{sec:mosaics}. For individual SBIDs, we first run \texttt{PyBDSF} on the Stokes I images and make measurements on the corresponding Stokes V images using the positions from Stokes I source lists. This ensures that each Stokes I and V measurement are taken from the exact same observation, and that each Stokes I measurement has a corresponding Stokes V measurement. {We use the whole image for source-finding, which has been constructed with a primary beam cutoff of 12\% during mosaicking.} The Stokes V measurements are made as an absolute peak Stokes V flux density within a $3\times3$ pixel box at the Stokes I source position reported by \texttt{PyBDSF}. We take a local measurement of the Stokes V rms noise as the uncertainty on the Stokes V flux density measurement. The resulting Stokes V source lists are then concatenated, retaining the relevant Stokes I and Stokes V information for each entry in the resultant table. For each measurement, we also record the direction cosines, $l$ and $m$, relative to the image centre to provide information about where on a particular image the measurement was made. We note that the construction of the Stokes I portion of this catalogue is identical to the `time-domain' catalogue produced for RACS-mid \citepalias{racs-mid2} and no de-duplication is done so sources may have multiple entries and measurements corresponding to measurements in adjacent or overlapping observations.

Note that we do not patch this catalogue with fixed measurements for bright sources as described in Section~\ref{sec:patching}. While the off-axis source-subtraction does not operate on XY and YX correlations, the underestimated Stokes I flux (formed from XX and YY correlations) may boost the fractional polarization reported for those sources. For this reason, they are removed from the raw Stokes V catalogue. Similarly, the extended sources that were originally missed are not added.

In total, we report 5\,727\,958 Stokes I measurements in this concatenated catalogue which includes multiple measurements of sources {in} regions that overlap between images. From this, we obtain 32\,098 Stokes V measurements with absolute peak flux density above $5\sigma_\text{V,rms}$. These Stokes V measurements includes sources of Stokes I leakage into Stokes V. In Section~\ref{sec:leakage} we will assess the leakage and appropriate thresholds to help remove sources of leakage.

\begin{figure}
    \centering
    \includegraphics[width=1\linewidth]{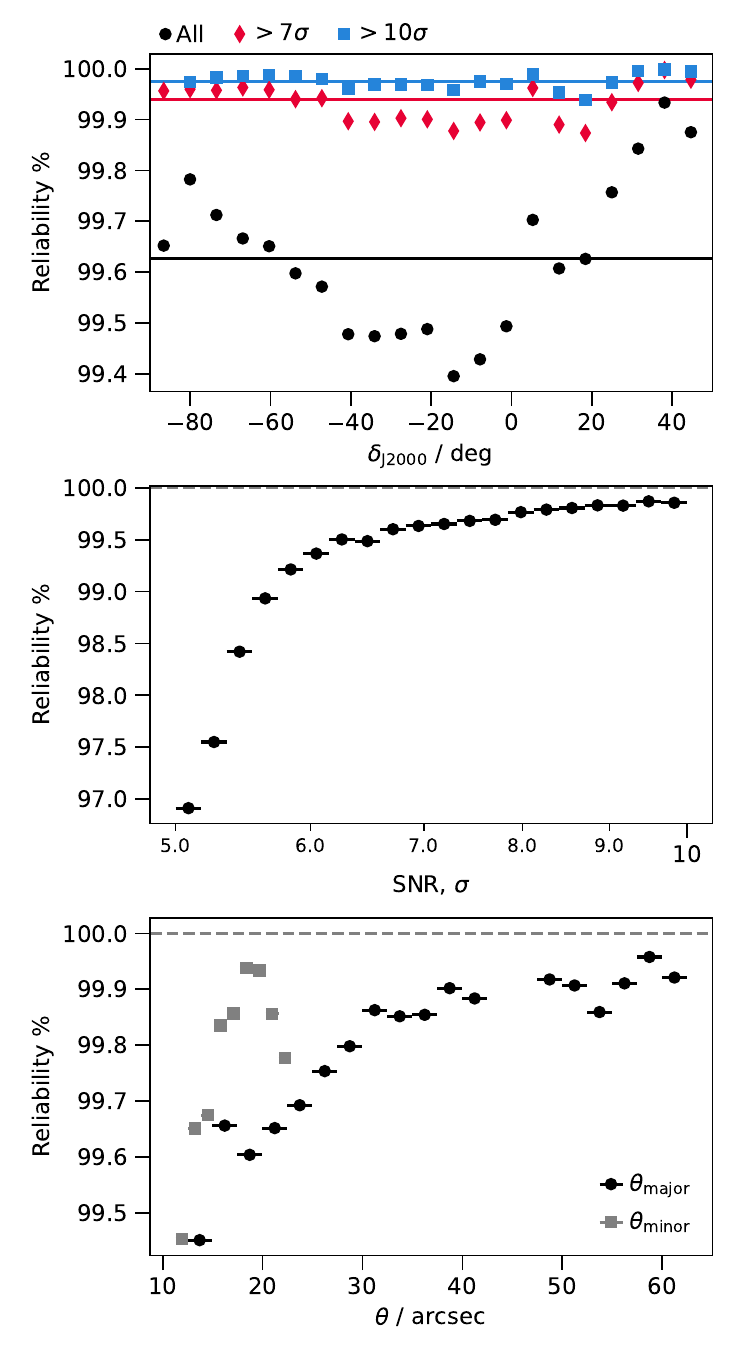}
    \caption{\label{fig:reliability} Estimates of the reliability of the catalogue, binned as a function of declination (\emph{top}), SNR (\emph{middle}), and angular resolution (\emph{bottom}). In the top panel, we also show the declination-dependent reliability at three separate SNR limits, with medians for each limit shown as a horizontal line. In the centre and bottom panels, a dashed horizontal line is drawn at 100\% reliability.}
\end{figure}

\section{Stokes I data assessment}

\subsection{Noise and reliability}\label{sec:reliability}

As a measure of image and catalogue fidelity, we report the general rms noise properties and reliability of our source finding process. Figure~\ref{fig:rms} shows the position-dependent rms noise across the survey as reported in the catalogue {with $N_\text{side} = 64$ HEALPix binning (corresponding to bins of $\approx 55 \times 55$\,arcmin$^2$)}. This highlights regions of higher noise like the Galactic Plane, around Cygnus~A and other bright sources, and generally above $\delta_\text{J2000} \approx +0\degr$, common with other RACS data products. Due to the overlap in SBIDs where the sky tiling changes to cover the South Celestial Pole {(SCP)}, the rms noise is also lower around $\delta_\text{J2000} \approx -70\degr$, also a feature of the common tiling pattern used across RACS epochs. The median rms noise across the whole survey is $195_{-32}^{+48}$\,\textmu Jy\,PSF$^{-1}$ {(when measured at the locations of sources in the catalogue)}. This is an overall improvement from RACS-low1 which reports \ $314_{-72}^{+123}$\,\textmu Jy\,PSF$^{-1}$ when obtained from catalogue rms noise values, though note a fixed resolution of $25\arcsec \times 25\arcsec$ is used in that case. {For comparison, median noise across the individual SBID images prior to the additional mosaicking of neighbouring images is $\approx 225$\,\textmu Jy\,PSF$^{-1}$.}

As with other RACS data releases \citepalias{racs2,racs-mid2,racs-high}, we define an estimate of the catalogue reliability based on the number of negative components found with identical source-finding parameters below $-5\sigma_\text{rms}$. We assume that noise is Gaussian and that for all negative components found an equal number of positive components would be found and should be considered artefacts/false detections \citep[e.g.][]{Intema2017}. Repeating source-finding after multiplying the images by $-1$ yields an overall 15\,834 negative Stokes I sources with $|S_\text{peak}| \geq 5\sigma_\text{rms}$ after merging source lists. For comparison, we take the source catalogue prior to patching (Section~\ref{sec:patching}, with 3\,922\,159 sources) and report a total reliability of 99.60\%. Figure~\ref{fig:reliability} shows how this reliability metric varies as a function of declination (\emph{top panel}), SNR (\emph{middle panel}), and angular resolution (\emph{bottom panel}). As with other RACS catalogue at varied angular resolutions, we find the reliability varies as a function of declination, decreasing in regions of finer angular resolution (and as, as expected, a function of SNR). We note that overall reliability increases to 99.94\% for sources above $7\sigma_\text{rms}$.

\subsection{Angular resolution and unresolved sources}\label{sec:resolved}

\begin{figure}[t]
    \centering
    \includegraphics[width=1\linewidth]{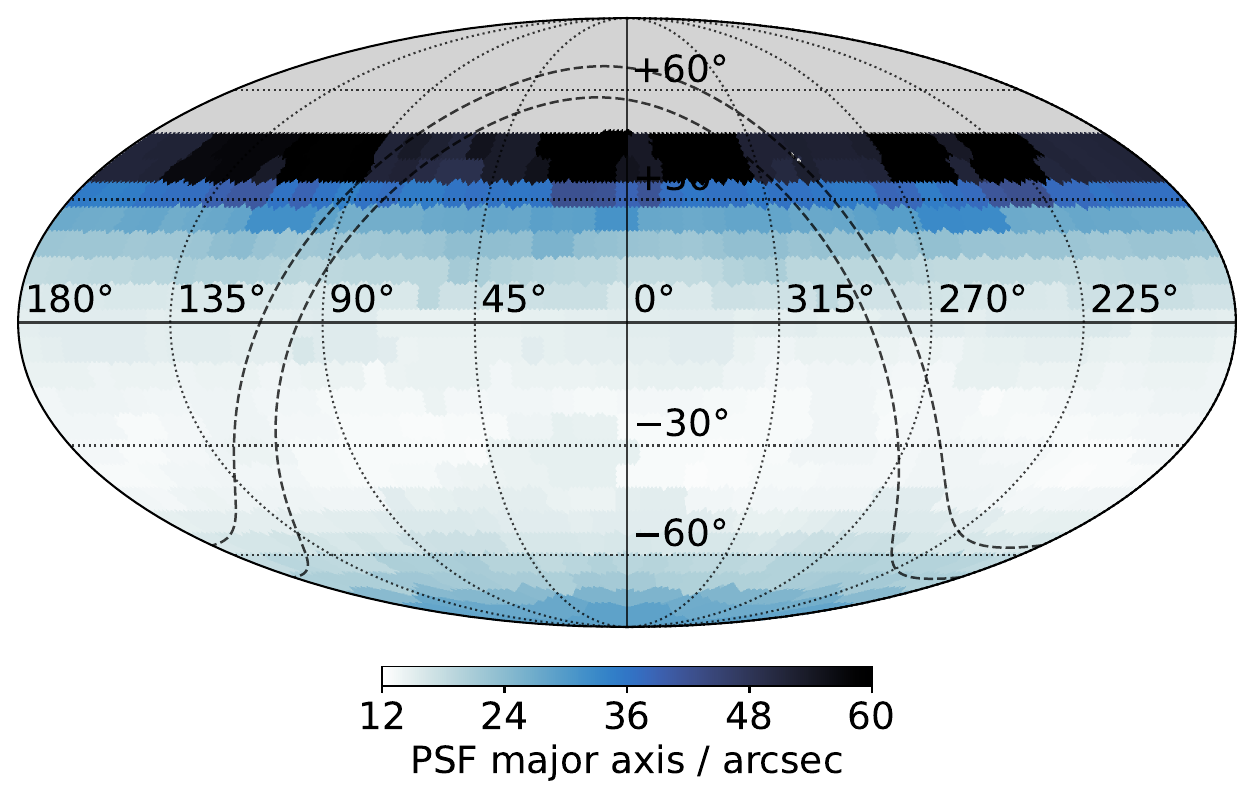}\\
    \includegraphics[width=1\linewidth]{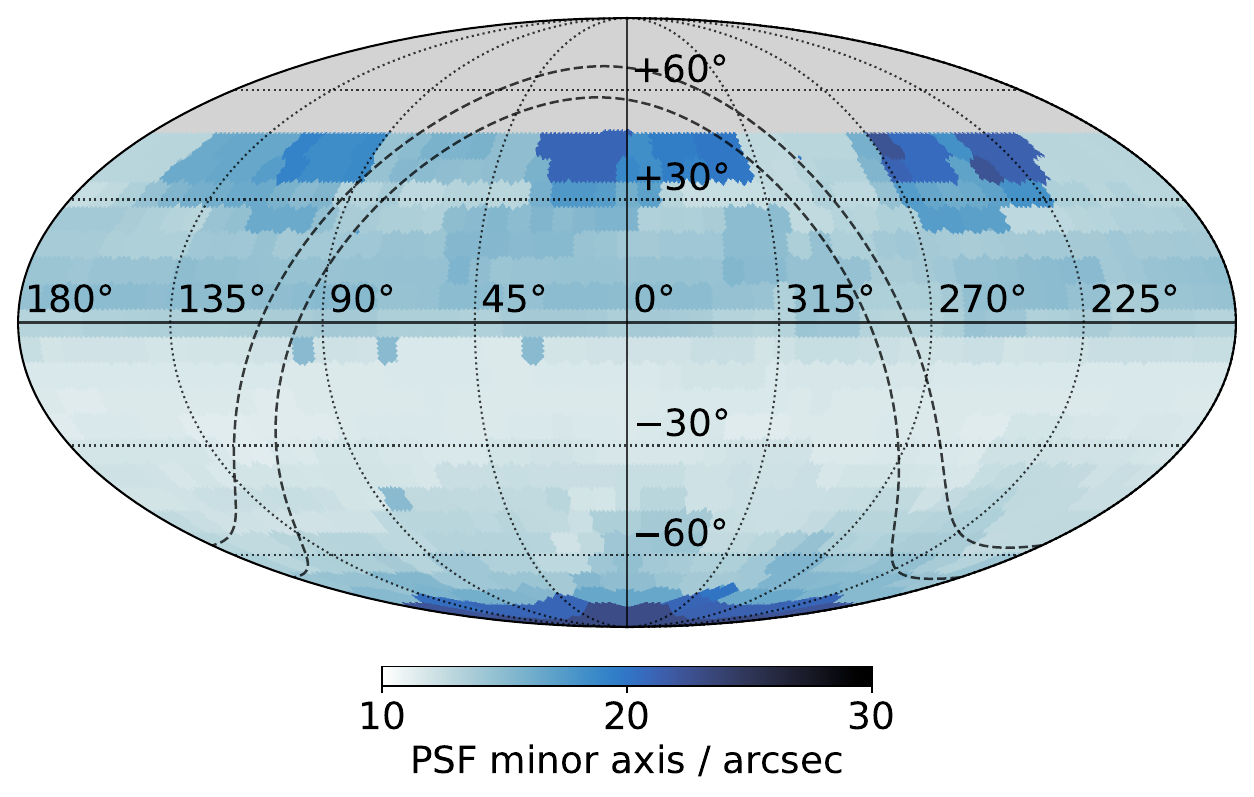}
    \caption{\label{fig:psf} The The HEALPix-binned FWHM of the PSF major (\emph{top}) and minor (\emph{bottom}) axes as a function of position across the sky, as reported at catalogue positions. {The binning is the same as in Figure~\ref{fig:rms}.}}
\end{figure}

\begin{figure}[t]
\includegraphics[width=1\linewidth]{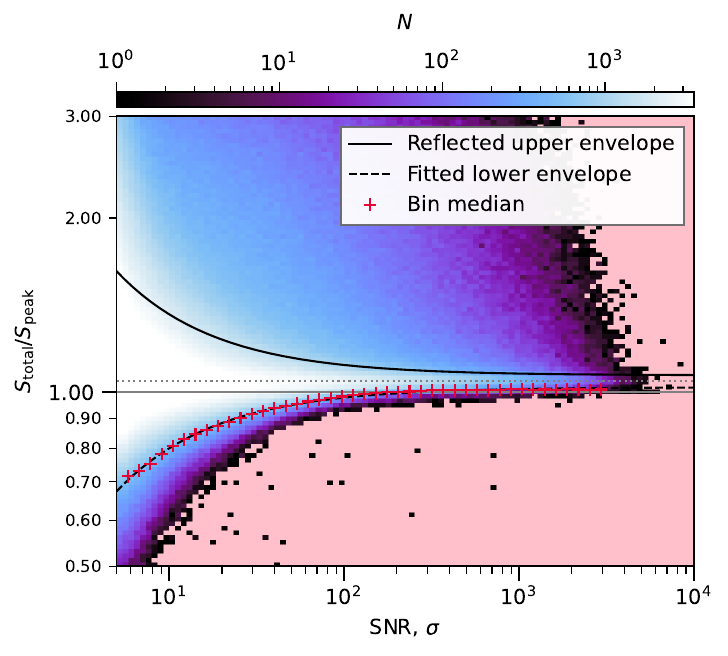}
\caption{\label{fig:resolved} $S_\text{total}/S_\text{peak}$ distribution for single component sources as a function of SNR to define sources that are unresolved. We show the lower fitted envelope and the reflected upper envelope used to define an unresolved source, along with the binned medians used for fitting. The horizontal solid line is drawn at $S_\text{total}/S_\text{peak} = 1$, and the horizontal dotted line is drawn at the sample median.}
\end{figure}

As the catalogue is constructed by combining source lists from images with their own angular resolution, it has a variable angular resolution, in contrast to the fixed angular resolution of the RACS-low1 catalogue. This variation is largely a function of declination, and Figure~\ref{fig:psf} shows how the angular resolution varies as a function of position. The median PSF is $15.2\arcsec \times 13.0\arcsec$, and the major and minor axes range from $12.5$\arcsec--$69.2$\arcsec\ and $11.3$\arcsec--$22.9$\arcsec, respectively. {For comparison, the original SBID images prior to additional mosaicking have a median PSF of $14.3\arcsec \times 11.8\arcsec$ with a range in major and minor axes of $11.8$\arcsec--$50.8$\arcsec\ and $10.6$\arcsec--$21.0$\arcsec, respectively, so in general there is only a small decrease in angular resolution when making the full-sensitivity images.}

For comparisons against other surveys in later sections, it is important to consider whether all sources or just unresolved sources should be compared. For the RACS-low2 catalogues, we follow the method described in \citetalias{racs2} where we assume that the ratio of total flux density to peak flux density, $S_\text{total} / S_\text{peak}$ should be approximately 1 for an unresolved source, but will deviate away from 1 for faint sources (due to scatter in the measurements of the Gaussian component parameters) and will have a bulk shift above 1 for all sources due to blurring from astrometric offsets between images when mosaicking. 

Figure~\ref{fig:resolved} shows the distribution of $S_\text{total} / S_\text{peak}$ as a function of SNR for single component sources, highlighting the spread at low SNR and the overall marginal increase above 1. Following \citetalias{racs2} we create bins in SNR and take the lower 95-th percentile of the logarithmic distribution for each bin to consider as a lower envelope fit with $S_\text{total} / S_\text{peak} = A\sigma^{B} + \text{med}(S_\text{total} / S_\text{peak})$. The reflected upper enveloped (shown on Figure~\ref{fig:resolved}) is used to define an unresolved source, and includes an additional offset to account for blurring that is not accounted for with the lower envelope. Sources below that line are considered unresolved. Overall we find $\approx 66$\% of sources can be considered unresolved. In the FITS tables we use \texttt{Flag = 0} to define unresolved sources and \texttt{Flag = 1} for resolved sources.

\begin{figure}[t]
    \centering
    \includegraphics[width=1\linewidth]{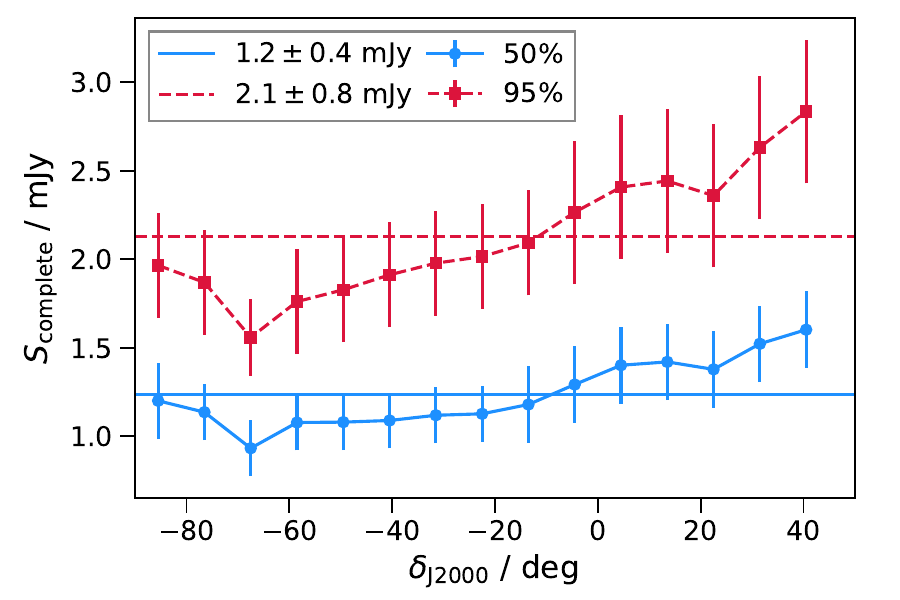}
    \caption{\label{fig:compl:dec} {Estimated RACS-low2 catalogue 50\% (blue circles) and 95\% (red squares) completeness as a function of declination in equally-spaced 9-deg bins between $-90 \leq \delta_\text{J2000} \leq +45\degr$. The horizontal lines indicate the median flux density completeness limits.}} 
\end{figure}

\begin{figure}[t]
    \centering
    \includegraphics[width=1\linewidth]{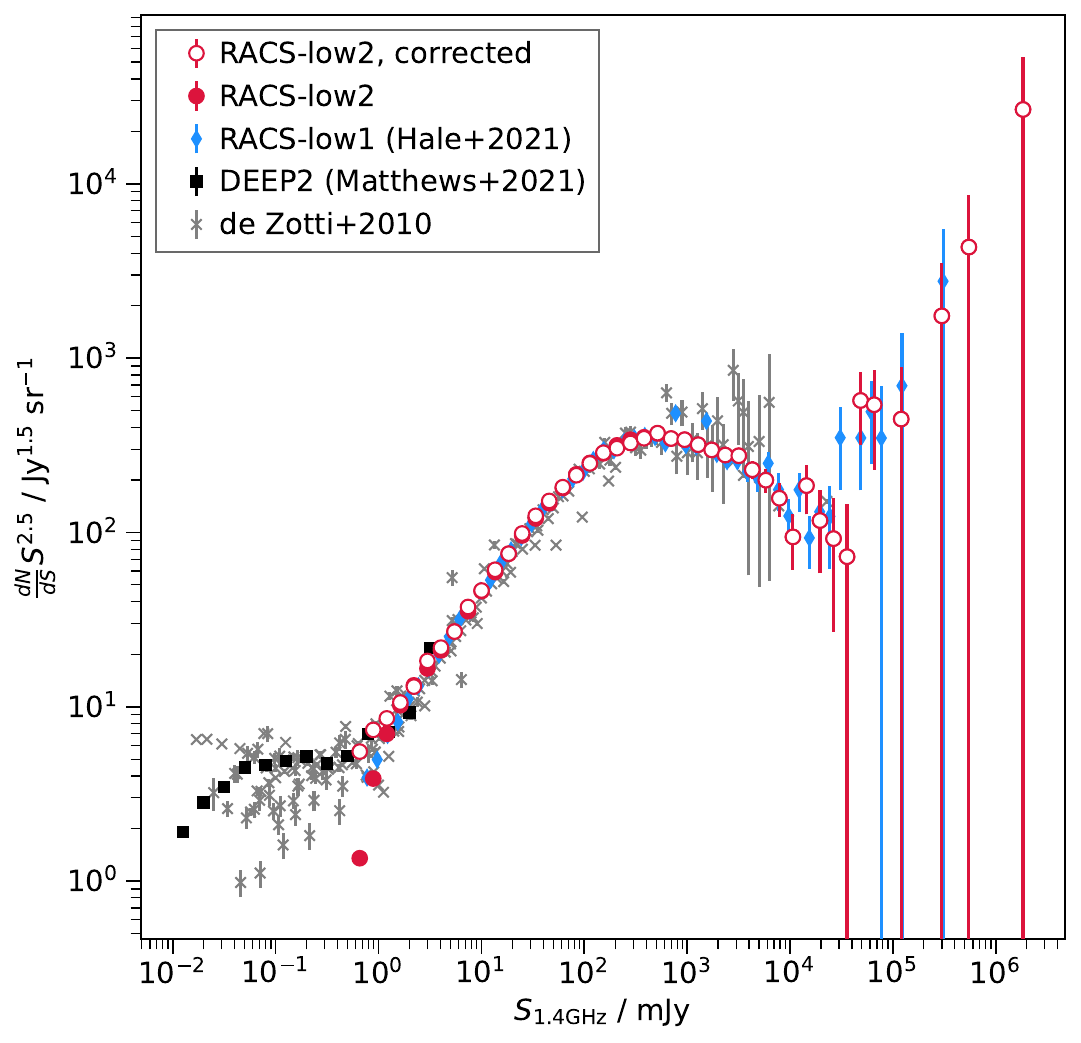}
    \caption{\label{fig:counts} {The RACS-low2 Euclidean normalised source counts, before (red filled circles) and after correction for source-finding incompleteness (red empty circles). We also show the equivalent corrected source counts from RACS-low1 \citepalias[blue diamonds][]{racs2}, the MeerKAT DEEP2 images \citep[black squares][]{Matthews2021}, and the source counts compilation from \citet[grey crosses][]{deZotti2010}, highlighting the RACS-low2 catalogue follows the expected source-count distribution when incompleteness is modelled and corrected for. Note that RACS-low1 and RACS-low2 source counts are scaled to 1.4 GHz for comparison with other work assuming $\alpha=-0.83$.}}
\end{figure}

\subsection{Source-finding completeness}\label{sec:completeness}

{To check how complete the catalogue is at different flux density limits, we follow a process is similar to previous RACS epochs \citepalias{racs2,racs-mid2}. This takes simulated models of the sky and injects sources onto residual images after removing the 2-D Gaussian components modelled by \texttt{PyBSDF} during source-finding. As a change from RACS-low1 and RACS-mid, we make use of a model generated by the Tiered Radio Extragalactic Continuum Simulation \citep[T-RECS;][]{Bonaldi2019}, which includes star-forming galaxies and AGN. We use the `medium' tier, 25-deg$^2$ catalogue with a 1.4-GHz flux density limit of 10\,nJy.}

{After source-finding is done for each full-sensitivity image, we obtain an image of the residuals after subtraction of the 2-D Gaussian components. To inject sources onto the residual image, we randomise positions of each simulated source within a $5\degr \times 5\degr$ region within the image boundaries, and take sources in range of $3\sigma_\text{rms} \leq S_\text{simulated} < 2\times\text{max}(S_\text{image})$, where $S_\text{image}$ is the maximum brightness in the original image and $S_\text{simulated}$ is the peak brightness of the simulated source, assuming it is a point source. Simulated flux densities are scaled to 887.5\,MHz from the 780-MHz simulated value assuming a spectral index of $\alpha=-0.83$. Sources in the T-RECS catalogue are reported with a projected angular size in one dimension and we assume both AGN and star-forming galaxies will have some varied ellipticity, $e$. We take a random value of $e$ from a uniform distribution in the range $1 \leq e \leq 4$ to model the sources as 2-D Gaussians. These are convolved to the angular resolution of the image and injected onto the residual images. \texttt{PyBDSF} is then re-run to generate a simulated source list. This process is run five times for each image, with the source positions and $5\degr \times 5\degr$ box placement randomised for each iteration to provide a range of samples across each full-sensitivity image. Each output source list is then matched to the input simulated catalogue independently and finally all source-lists and matched source-lists are concatenated.}

{We estimate the flux densities above which the RACS-low2 catalogue is complete to 50\% and 95\% as the ratio of sources in the input T-RECS simulated source lists to the output \texttt{PyBDSF} source lists after matching to the respective input lists. Figure~\ref{fig:compl:dec} shows the flux density limits for 50\% and 95\% completeness in 9-deg declination bins, covering $-90\degr \leq \delta_\text{J2000} \leq +45\degr$. Completeness generally decreases as declination increases, likely corresponding to a similar increase in rms noise as a function of declination (Figure~\ref{fig:rms}). The lower flux density limit around $ \delta_\text{J2000} \approx -70\degr$ also corresponds to the higher sensitivity in the region where the observation tiling changes to cover the SCP. We estimate the mean 50\% and 95\% flux density limits as $1.2\pm0.4$ and $2.1\pm0.8$\,mJy, respectively (shown as horizontal lines on Figure~\ref{fig:compl:dec}).}

\subsection{Source counts}\label{sec:counts}

{To check the accuracy of our source-finding completeness assessment, we produce the Euclidean normalised source counts \citep[e.g.][]{Heywood2013} for the RACS-low2 catalogue. For each declination band described in the previous section, sources are counted in 51 logarithmically-spaced flux density bins in the range $0.82$\,mJy (corresponding to approximately five times the 16-th percentile of the rms noise distribution) to $\approx 3\,100$\,Jy (ensuring we include the brightest source in our survey, Cygnus~A). The raw counts per declination band are corrected for incompleteness using the T-RECS simulated models, and are then summed and normalised by the total area of 32\,362\,deg$^2$ between $-90\degr \leq \delta_\text{J2000} \leq +45\degr$, excluding the Galactic latitudes of $|b| < 5\degr$.}     

{Figure~\ref{fig:counts} shows the Euclidean normalised source counts for RACS-low2, both before and after correction of source-finding incompleteness, and scaled to 1.4 GHz assuming $\alpha=-0.83$. We also show the equivalent corrected source counts from RACS-low1 \citepalias{racs2}, the faint source counts from the MeerKAT DEEP2 image \citep{Matthews2021}, and the source counts from the compilation of \citet{deZotti2010}. At the high flux density end of the distribution, we see significant scatter and uncertainties due to very low numbers of extragalactic sources this bright, with most flux density bins above $\approx 100$\,Jy only featuring a single source at most, consistent with other widefield surveys including RACS-low1. A decrease in counts at $\approx 50$\,Jy is present in the RACS-low2 catalogue compared to RACS-low1, which may be the result of a small number of bright extended sources being decomposed into multiple Gaussian components by \texttt{PyBDSF} that were not patched as in the case of PKS~J0133$-$3629 (see Section~\ref{sec:patching} and Section~\ref{sec:brightness}). The comparatively low angular resolution of the RACS-low1 catalogue is better suited to modelling extended sources with simple Gaussian components. At low flux densities the RACS-low2 corrected counts agree well with \citet{Matthews2021} and \citet{deZotti2010}, though tend to deviate from RACS-low1.}

\subsection{Efficacy of the primary beam models}\label{sec:tile_brightness}


\begin{figure*}[p]
    \centering

    \begin{subfigure}[b]{0.33\linewidth}
    \includegraphics[width=1\linewidth]{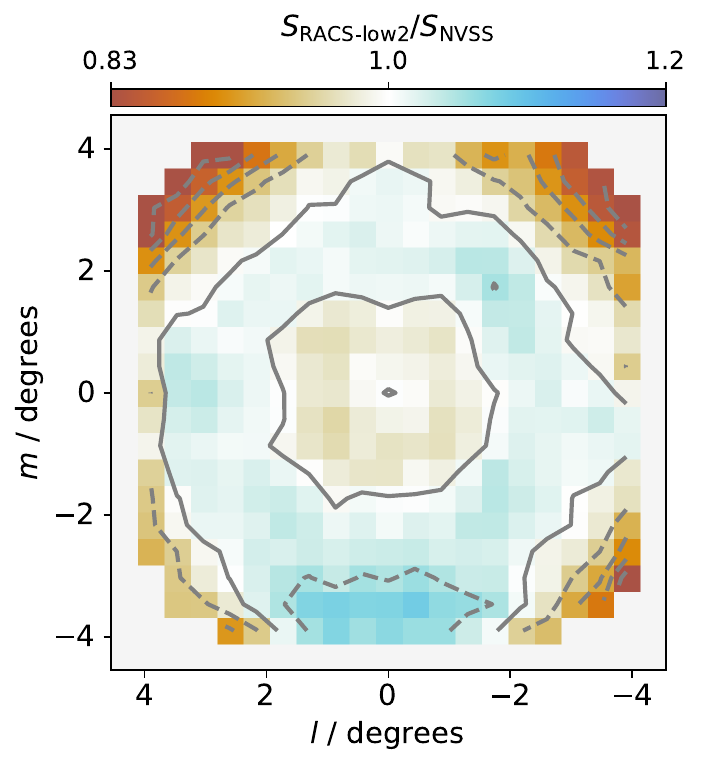}
    \caption{\label{fig:tile_scale:nvss} NVSS comparison.}
    \end{subfigure}%
    \begin{subfigure}[b]{0.33\linewidth}
    \includegraphics[width=1\linewidth]{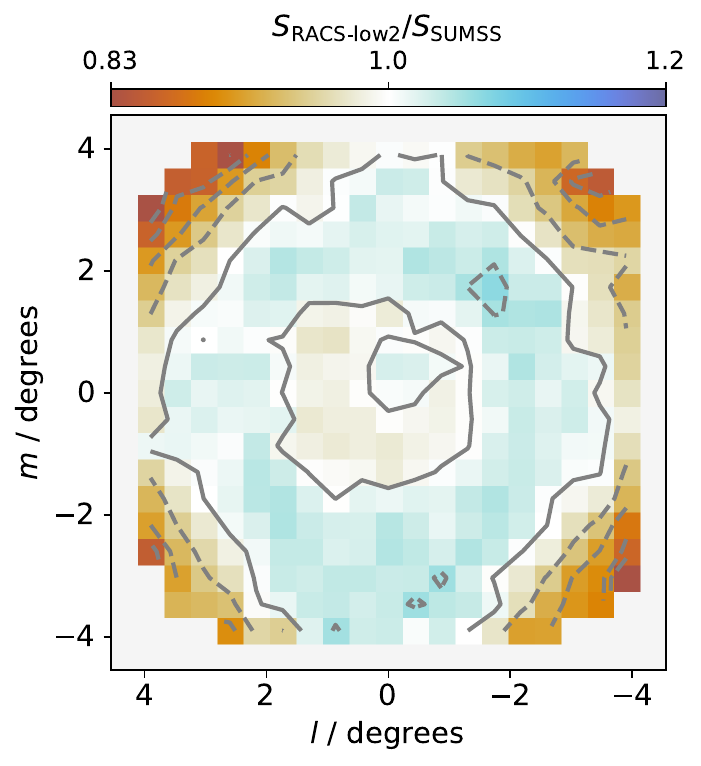}
    \caption{\label{fig:tile_scale:sumss} SUMSS comparison.}
    \end{subfigure}%
    \begin{subfigure}[b]{0.33\linewidth}
    \includegraphics[width=1\linewidth]{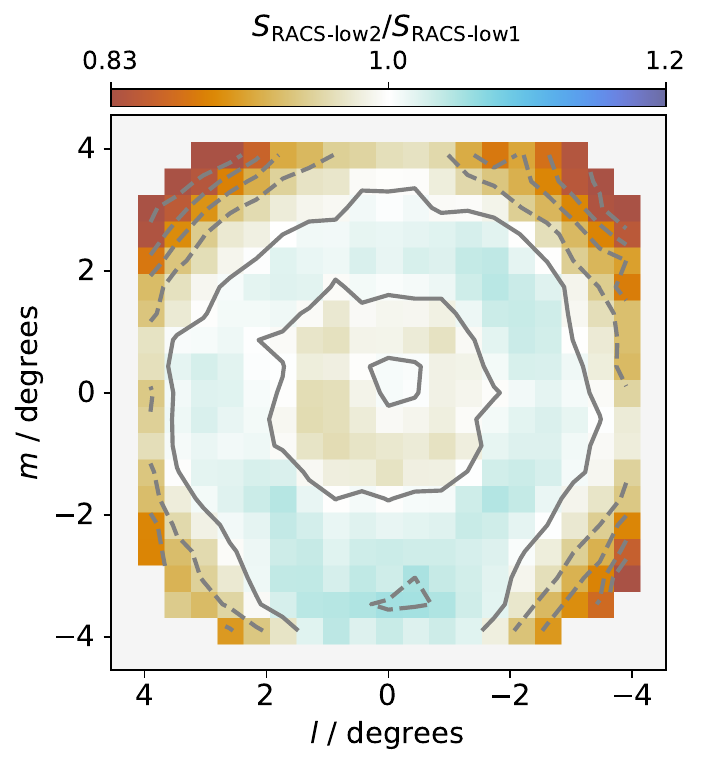}
    \caption{\label{fig:tile_scale:low} RACS-low1 comparison.}
    \end{subfigure}\\%
     \begin{subfigure}[b]{0.33\linewidth}
    \includegraphics[width=1\linewidth]{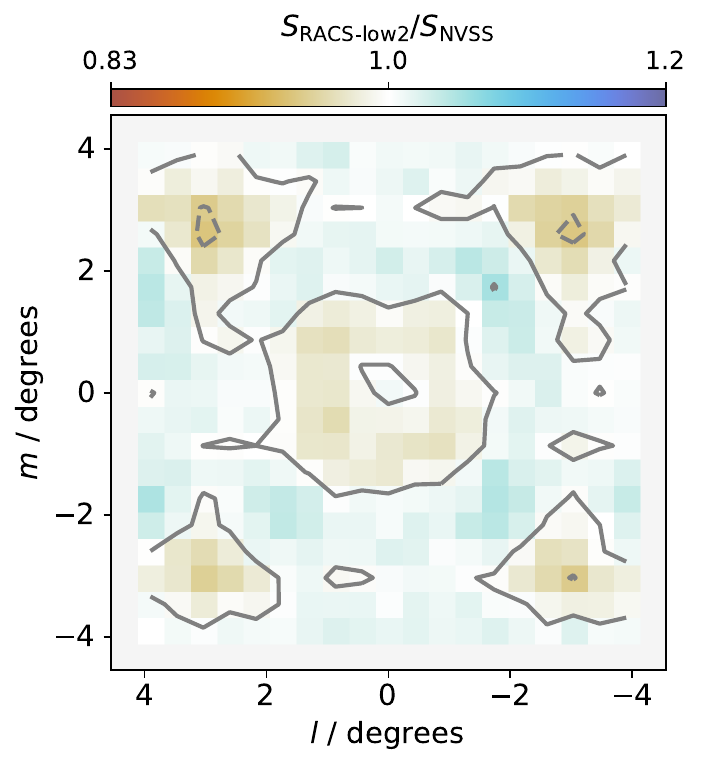}
    \caption{\label{fig:mosaic_scale:nvss} NVSS mosaic comparison.}
    \end{subfigure}%
    \begin{subfigure}[b]{0.33\linewidth})
    \includegraphics[width=1\linewidth]{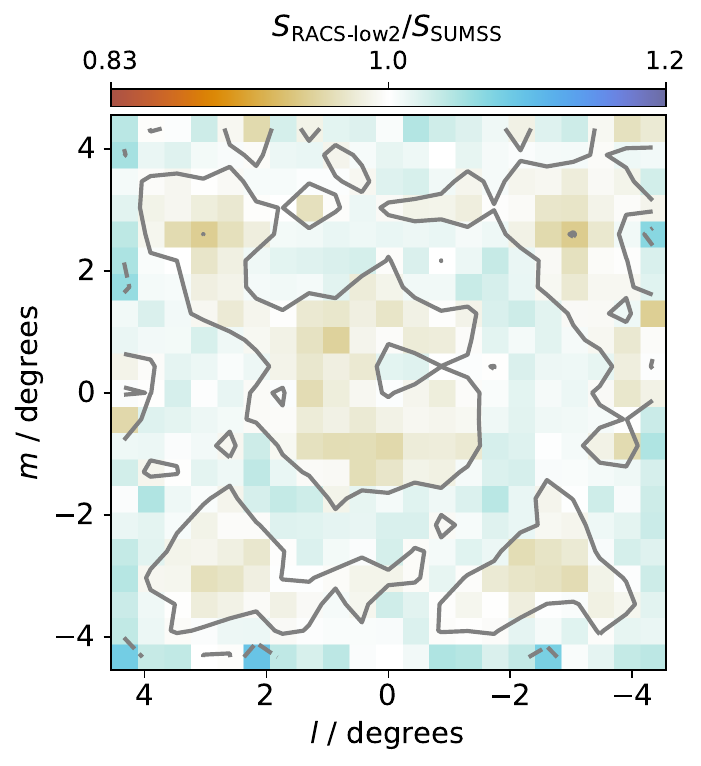}
    \caption{\label{fig:mosaic_scale:sumss} SUMSS mosaic comparison.}
    \end{subfigure}%
    \begin{subfigure}[b]{0.33\linewidth}
    \includegraphics[width=1\linewidth]{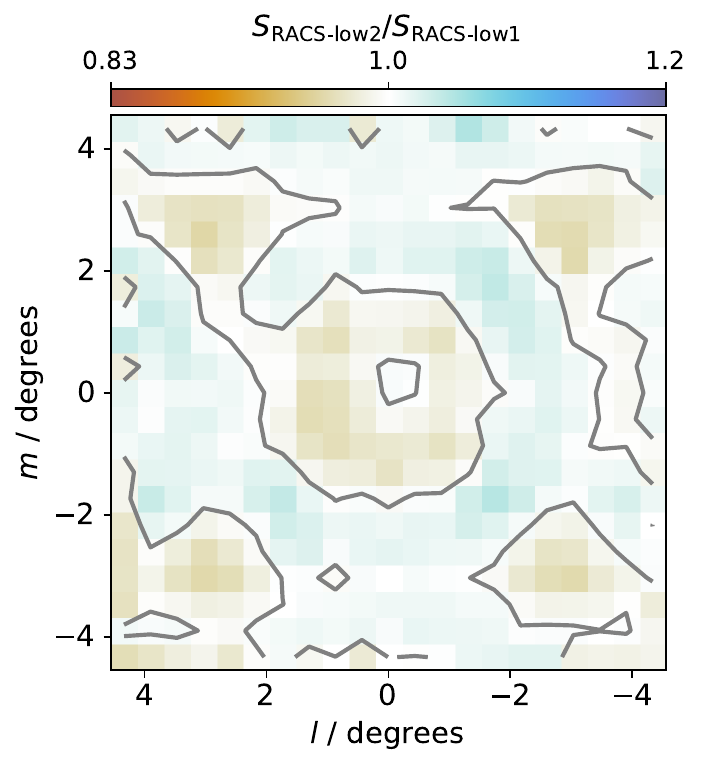}
    \caption{\label{fig:mosaic_scale:low} RACS-low mosaic comparison.}
    \end{subfigure}%
    \caption{\label{fig:tile_scale} The binned flux density ratios between RACS-low2 and NVSS (\ref{fig:tile_scale:nvss} and \ref{fig:mosaic_scale:nvss}), SUMSS (\ref{fig:tile_scale:sumss} and \ref{fig:mosaic_scale:sumss}), and the original RACS-low (\ref{fig:tile_scale:low} and \ref{fig:mosaic_scale:low}) as a function of position on the individual images (top row) and full-sensitivity mosaics (bottom row). The ratios are normalised to the median of each comparison. The bin sizes are $15.8 \times 15.8$\,arcmin$^{2}$. The solid contour is drawn at 1, and dashed contours are drawn at $[\pm5\%, \pm10\%, \pm 15\%, \pm 20\%]$.}
\end{figure*}

\begin{figure*}[p]
    \centering
    \begin{subfigure}[b]{0.33\linewidth}
    \includegraphics[width=1\linewidth]{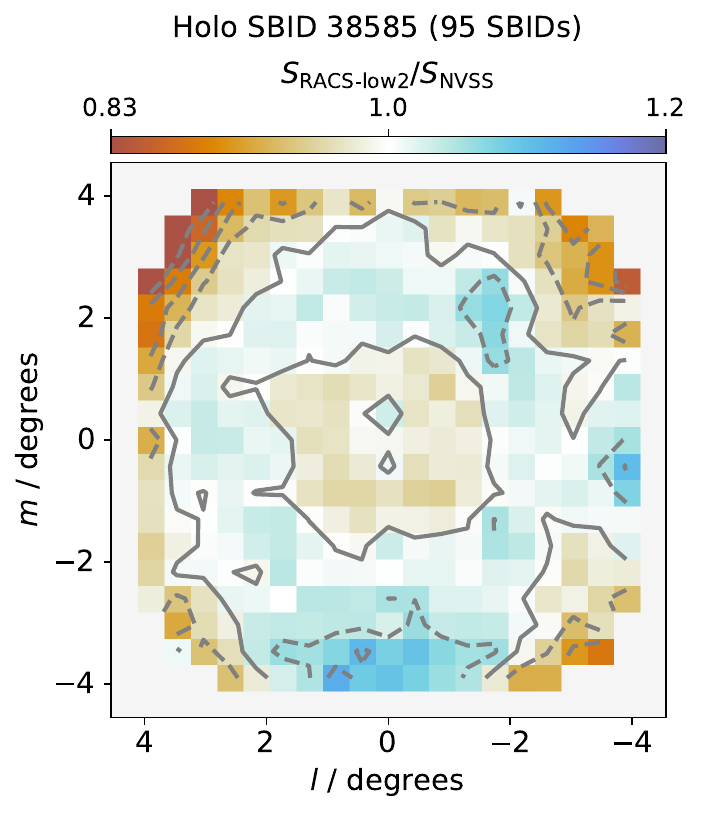}
    \caption{\label{fig:app:scale:low2:38585} Weights ID 38240.}
    \end{subfigure}%
    \begin{subfigure}[b]{0.33\linewidth}
    \includegraphics[width=1\linewidth]{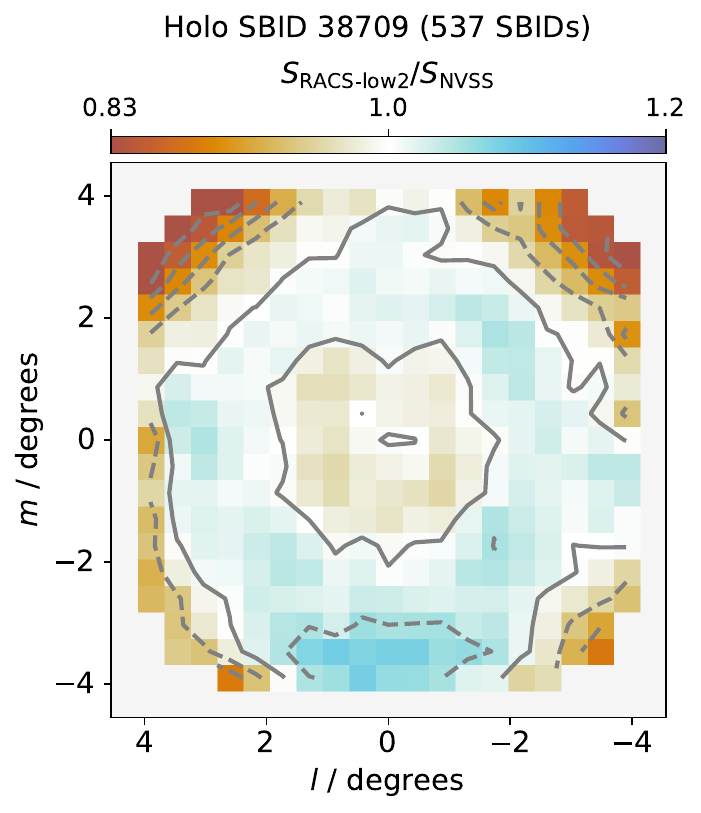}
    \caption{\label{fig:app:scale:low2:38709} Weights ID 38532.}
    \end{subfigure}%
    \begin{subfigure}[b]{0.33\linewidth}
    \includegraphics[width=1\linewidth]{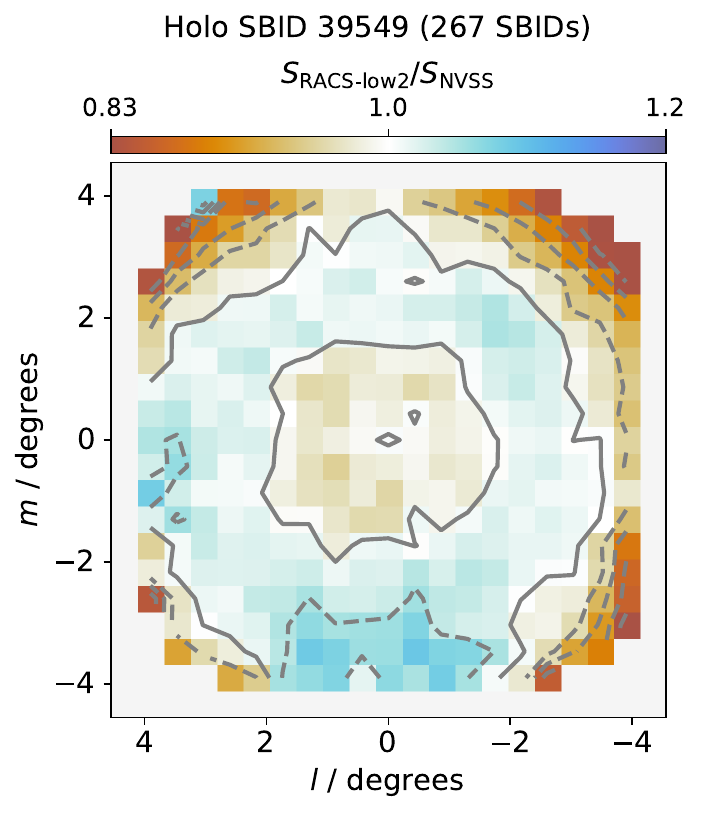}
    \caption{\label{fig:app:scale:low2:39549} Weights ID 39390.}
    \end{subfigure}\\%
    \begin{subfigure}[b]{0.33\linewidth}
    \includegraphics[width=1\linewidth]{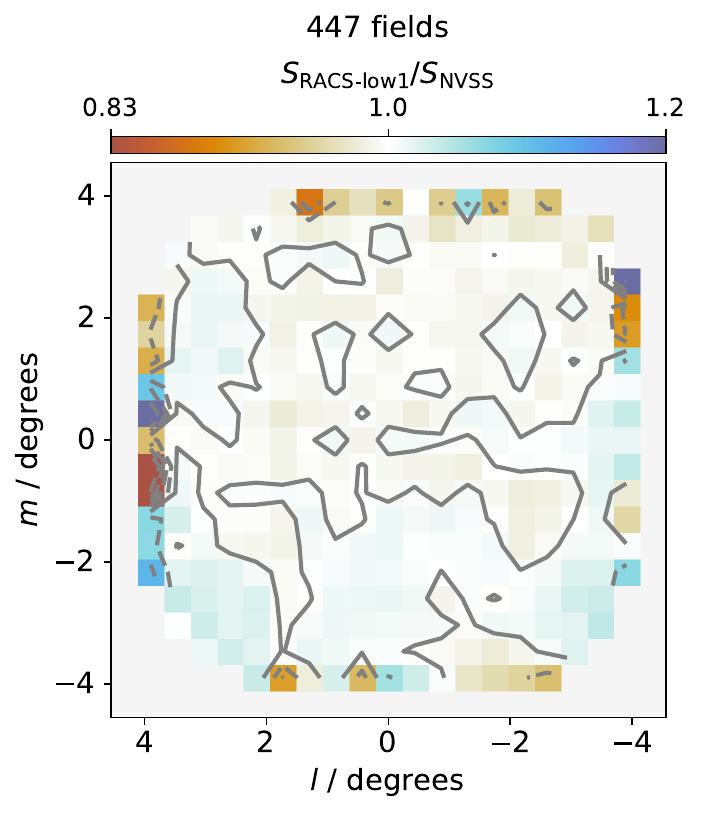}
    \caption{\label{fig:app:scale:low1:8247} Weights ID 8247.}
    \end{subfigure}%
    \begin{subfigure}[b]{0.33\linewidth}
    \includegraphics[width=1\linewidth]{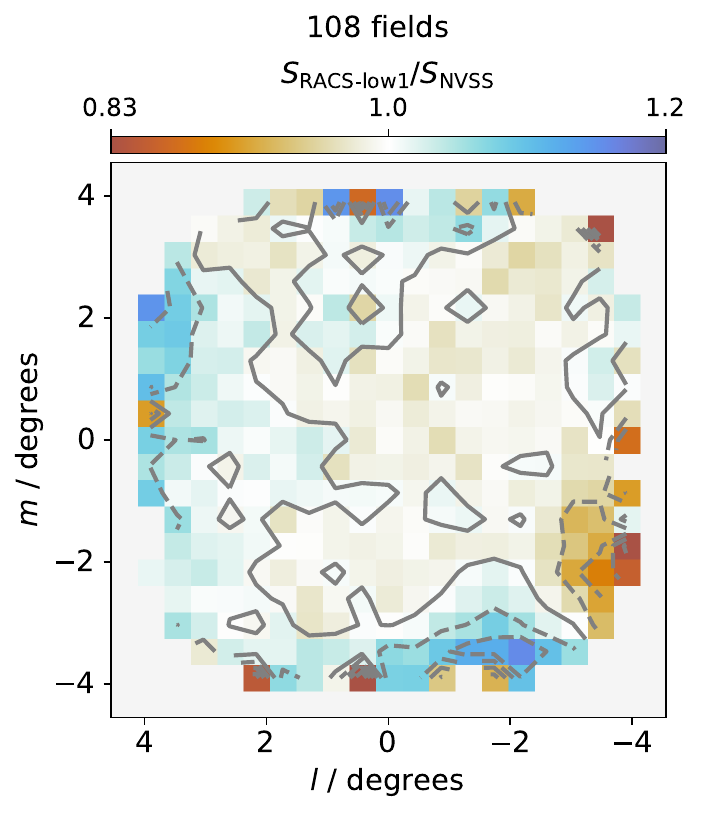}
    \caption{\label{fig:app:scale:low1:11371} Weights ID 11371.}
    \end{subfigure}%
    \begin{subfigure}[b]{0.33\linewidth}
    \includegraphics[width=1\linewidth]{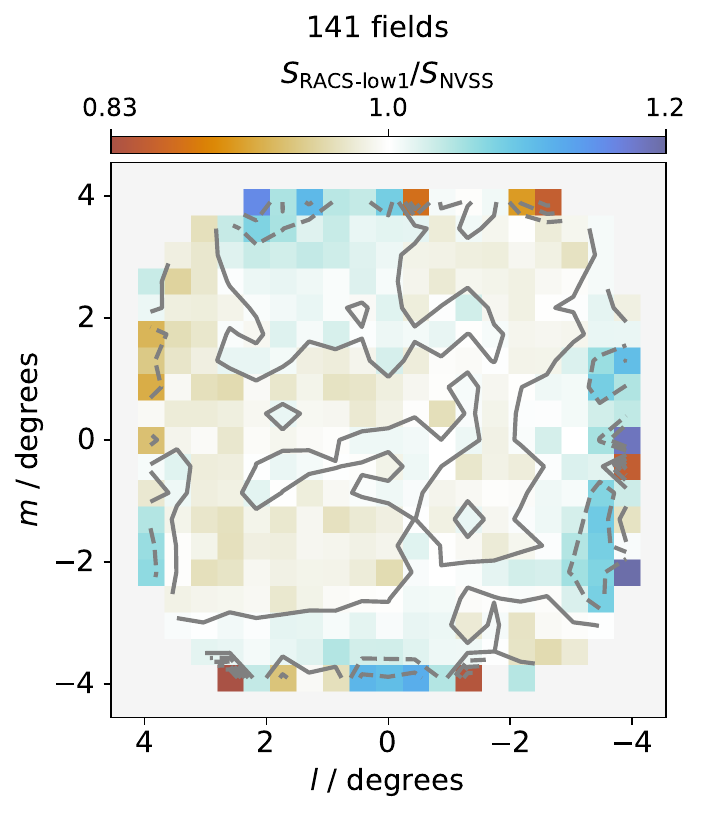}
    \caption{\label{fig:app:scale:low1:13624} Weights ID 13624.}
    \end{subfigure}%
    \caption{\label{fig:app:scale:low2} \emph{Top row.} The binned flux density ratios between RACS-low2 and NVSS for three of the four common beam-former weights periods during RACS-low2 (each with their own primary beam models) as a function of position on the individual images. The number of individual observations that go into each comparison are reported at the top of each panel, and the specific holography observation is also reported. \emph{Bottom row.} The binned flux density ratios between RACS-low1 and NVSS for three of the five common beam-former weights periods during RACS-low1, as a function of position on the individual images. The number of individual observations that go into each comparison are reported at the top of each panel. Note because RACS-low1 SBIDs contained multiple pointings we have referred to the individual pointings as `fields'. Other details are as in Figure~\ref{fig:tile_scale}. }
\end{figure*}

As RACS-low2 is the first RACS epoch to make use of holography-derived models of the \texttt{square6x6} PAF footprint at 887.5\,MHz, we explored the resulting brightness scale in our images, as a function of position on the images. \citetalias{racs-mid} and \citetalias{racs-high} explored how the mid- and high-band \texttt{closepack36} PAF footprint beam models affect the resultant brightness scale, finding that the image edges suffered from discrepancies of up to $\approx 20$\% that was suggested to be a residual telescope pointing error, though the definite cause is not yet known.

We initially look at only the individual SBID images (i.e.\,prior to mosaicking neighbouring observations) to assess how effective the primary beam correction is for the individual observations. As part of the ASKAP pipeline run, we generate a source list of each individual SBID image with \texttt{Selavy} \citep{selavy} \footnote{We note that the RACS-low2 \texttt{Selavy} source lists that are publicly available on CASDA contain some duplicate components ($\lesssim 100$ of the total) in the lower left corner of each image. These duplicated entries have identical component names, and are likely a result of mismatched \texttt{Selavy} parallelisation parameters for the CPU parallelisation configuration used specifically for RACS-low2 processing. No other issues were identified with these lists, and we have removed these prior to cross-matching.} which is cross-matched to the Sydney University Molonglo Sky Survey \citep[SUMSS;][]{bls99,mmb+03}, the NRAO \footnote{National Radio Astronomy Observatory.} VLA Sky Survey  \citep[NVSS;][]{ccg+98}, and the original RACS-low1 catalogue \citepalias[][specifically after making the equivalent full-sensitivity images]{racs2}. {We only consider sources that are isolated (no neighbours within 90\arcsec) and unresolved ($0.8 \leq S_\text{total} / S_\text{peak} \leq 1.2$) \footnote{Note that the selection of unresolved sources described in the previous section is specifically relevant for the catalogue constructed from full-sensitivity mosaics, which introduces additional blurring to sources in overlap regions, and so we do not use the same model for the individual SBID image source lists.}, resulting in an average of $\approx 2\,500$ sources per image available for cross-matching.} We scale the flux densities of each external survey to 887.5 MHz assuming a fixed spectral index $\alpha = -0.83$ \citep[the median spectral index between SUMSS and NVSS reported by][]{mmb+03}, though note for the purpose of assessing the brightness scale across the image we normalise the ratios to the overall median.

With the cross-matched {source lists, we remove sources with RACS-low2 SNR $< 10$, and} assign direction cosine, $l$ and $m$, to the source positions corresponding to the directions from the image centres. We then bin the ratios in $l$ and $m$, taking a median for each bin. Figure~\ref{fig:tile_scale} shows the binned medians as a function of $l$ and $m$ for the NVSS, SUMSS, and RACS-low1 comparisons. We see a residual brightness scale error across the images in all three comparisons. There is a ring of $\approx 4$--$5$\% elevated brightness at $\approx 2$--$3\degr$ from the image centre, ending in up to $\lesssim 20$\% reduced brightness at the image corners. 

As noted, similar effects were seen with RACS-mid and RACS-high for the \texttt{closepack36} PAF footprint, though the change in footprint has altered the pattern of the residual error. The comparison with NVSS (for images $\delta_\text{J2000} \gtrsim -40\degr$) in particular highlights the same {north to south} bias in the error that was observed in \citetalias{racs-mid}, though the magnitude is smaller here at $\approx 5$\% difference in the median between $m > +2.5\degr$ and $m<-2.5\degr$ parts of the image. The SUMSS comparison (for images $\delta_\text{J2000} \lesssim -30\degr$) shows a more symmetric error ($\approx 3$\% difference) The comparison with RACS-low1 also features the N-S effect with a reduced magnitude ($\approx 4$\% difference) compared to the comparison with NVSS, highlighting a potential declination-/elevation-dependent residual primary beam error.

We split the RACS-low2 \texttt{Selavy} source list and NVSS cross-matches and {grouped} by which holography-based model was used for primary beam correction. Figure~\ref{fig:app:scale:low2} shows the equivalent of {Figure~\ref{fig:tile_scale:nvss}} for three of the four holography models, with the fourth not shown due to having only a small number of SBIDs and therefore few sources for cross-matching. We see the same ring-shaped pattern occur in all cases, though with a slight lateral shift which is to be expected as beam-former weight updates may shift the beam positions. 
 
In the bottom row of Figure~\ref{fig:app:scale:low2} we also show the equivalent common beam-former weights periods for RACS-low1, using the same cross-matching outlined in Section~\ref{sec:tile_brightness} with the \texttt{Selavy} source lists available on CASDA. For RACS-low1, a post-mosaic primary beam correction was applied using a primary beam model based on holographic observations of Virgo~A (see Section 3.3.2 and Appendix A in \citetalias{racs1}). A single correction is applied for all observations, rather than individual models for each common weights period. We see marginal decrease in the residual errors, and the pronounced ring feature from RACS-low2 is not present.  
 The Virgo A holography observation was recorded with different beamformer weights to the entirety of RACS-low1, and beam models derived from Virgo~A for RACS-low1 extend out to the first sidelobe at 887.5\,MHz. Conversely, RACS-low2 primary beam models are only evaluated out to some fraction of the primary beam main lobe, where the extent of the modelled main lobe for each beam varies based on coma distortion and the general beam shape variation for a given set of beamformer weights. This can be up to $\approx 21$\% for some corner beams (e.g.\ beam 31 from holography SBID 38709), whereas our mosaicking procedure clips images when attenuation is $<12$\%, so the actual coverage of the primary beam models is less than ideal.
 
We also cross-matched the \texttt{PyBDSF}-derived source lists for each of the full-sensitivity mosaic to explore how the edge effects change through the mosaicking process. We repeat the same cross-matching {(from an average of $\approx 3\,300$ unresolved and isolated sources per image) and subsequent} binning and show the result with comparison to NVSS, SUMSS, and RACS-low1 in Figure~\ref{fig:mosaic_scale:nvss}--\subref{fig:mosaic_scale:low}. We see a reduction in edge effects as individual images overlap with the same spacing as the PAF beam images, so the edge contribution is significantly down-weighted in the full-sensitivity mosaics. As the tile centres are largely unchanged, we still see some residual errors, with maximum difference of up to $\approx 5$\%. 

\subsection{Time-dependence of the brightness scale}\label{sec:time_brightness}

\begin{figure*}[t]
\centering
\includegraphics[width=1\linewidth]{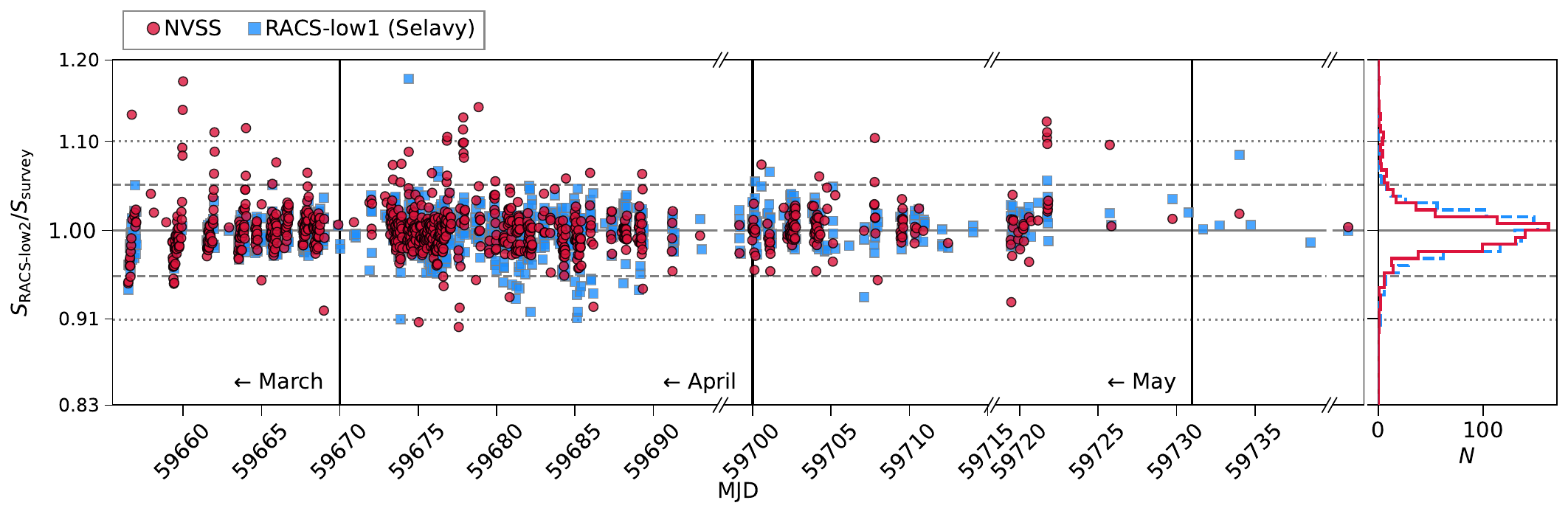}
\caption{\label{fig:flux:time} Median flux density ratios against NVSS {(red circles)} and RACS-low1 {(blue squares)} for each SBID as a function of time, highlighting the time-dependent variation of the brightness scale common to all RACS epochs. The solid horizontal line is drawn at a ratio of 1, and the dashed and dotted horizontal lines are drawn at $\pm 5$\% and $\pm 10$\%, respectively. {Vertical black lines mark the end of the labelled month.}}
\end{figure*}

In all released RACS epochs so far we have observed a time-dependent fluctuation of the brightness scale \citepalias{racs1,racs-mid,racs-high}. This fluctuation has been investigated down to only the 15-min interval as an SBID-dependent fluctuation. Figure~\ref{fig:flux:time} shows the median flux density ratios for the NVSS and RACS-low1 $\text{SNR}>10$ cross-matches for each SBID described in the previous section. In general, we see the familiar scatter in the brightness scale as a function of time with an overall standard deviation of $\approx 4$\% for both surveys. While there is some periodicity to the fluctuations, the magnitude of this periodic signal is reduced compared to RACS-high and RACS-mid and is similar to RACS-low1. We do not rescale the RACS-low2 data due to the lack of significant periodic signature, and attempts at rescaling with the same method as RACS-high with a sinusoidal model provides no significant reduction in the variance of the residual flux density ratios (as with RACS-low1; \citetalias{racs1}, see their section 3.3.3).

We suspect the magnitude of these fluctuations is driven by temperature variations of the PAFs, which is more significant in the Australian Summer months. RACS-mid, RACS-high, and RACS-low3 (Galvin et al., in prep.) had the bulk of their observations conducted during Summer months, whereas RACS-low2 was observed predominantly over the comparatively cooler Autumn. In addition, RACS-low1 data was observed over Winter, and while the time-dependent brightness scale fluctuation is still present the magnitude is also smaller in that case. {Despite these fluctuations, the source lists (and images) from individual RACS-low1 and RACS-low2 observations are still suitable for studying time-variability if taking into account the appropriate brightness scale uncertainty for each epoch.}

\subsection{Absolute brightness scale}\label{sec:brightness}

\begin{figure}[t]
    \centering
    \includegraphics[width=1\linewidth]{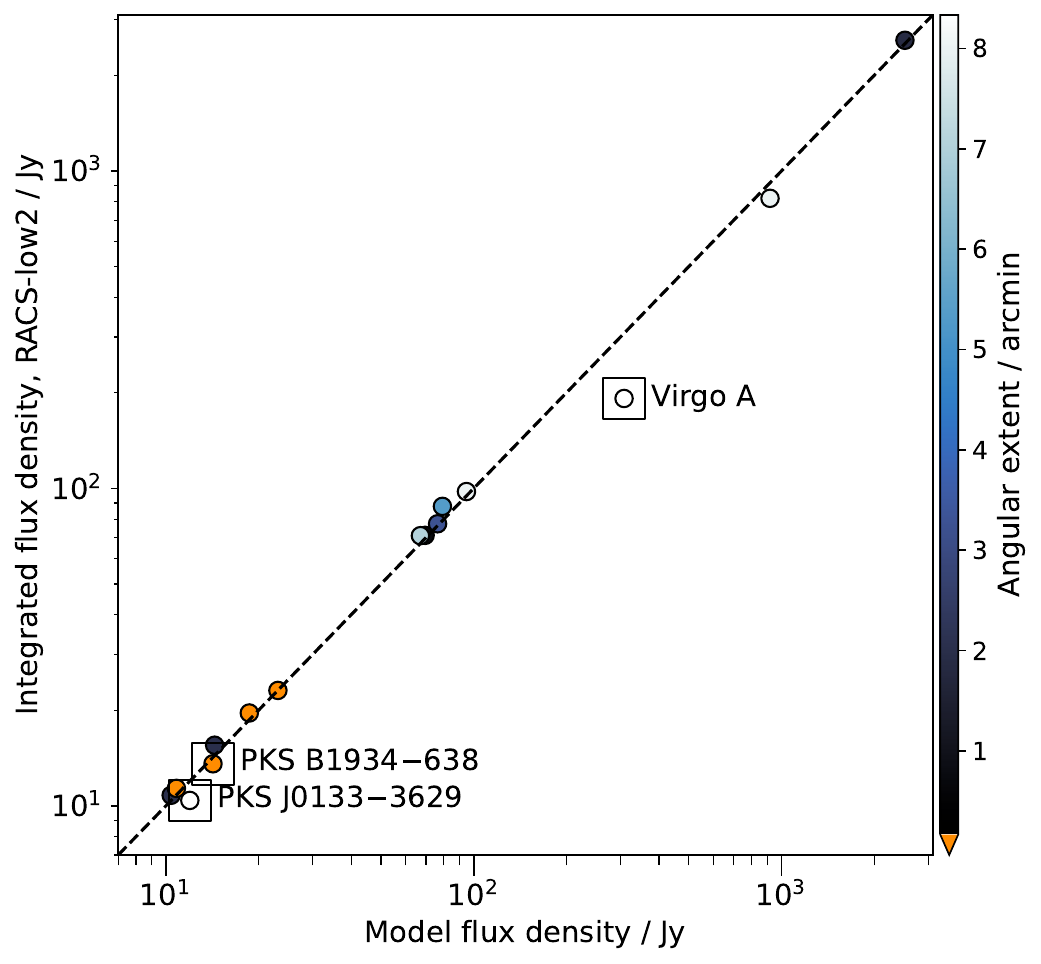}
    \caption{\label{fig:flux:abs} RACS-low2 measured flux densities for bright calibrator sources as a function of their model values from \citet[][for PKS~B1934$-$638, highlighted with a box]{Reynolds1994} and \citet[][for all others]{Perley2017}. The sources are coloured by their angular size, with orange colouring indicating a source is unresolved in the RACS-low2 image. The diagonal dashed line indicates measured and model flux densities are equal. Note that measurement uncertainties are drawn but are smaller than the circular markers. A brightness scale uncertainty is not included.}
\end{figure}

As a last assessment of the brightness scale for Stokes I, we also compare 887.5-MHz flux density measurements of bright calibrator sources from \citet{Perley2017} and PKS~B1934$-$638 \citep{Reynolds1994}. For sources in the survey region, we compare the catalogued flux density against the model flux density evaluated at 887.5\,MHz if they are cross-matched within their source size. Fornax~A and PKS~J0133$-$3629 are detected as multiple sources. In the case of Fornax~A, we simply consider its full extent beyond the detectability of the RACS snapshot observations, but for PKS~J0133$-$3629 we combine the two sources it {resolves} into for the purpose of this comparison as each lobe is well-detected.  The measured flux densities are shown on Figure~\ref{fig:flux:abs} against their model values at 887.5\,MHz, coloured by source size. We see good general agreement, with a standard deviation of 4\% for the full sample. As expected due the $(u,v)$ cut, larger sources tend to have their total flux density underestimated {(e.g. PKS~J0133$-$3629 and Virgo~A, highlighted on the plot) or not catalogued as a single source as in the case of Fornax~A (not present on the plot)}. We note that the median flux density ratio between the measured and model values is $1.03$. We suggest our overall brightness scale is similar to RACS-low1, with a similar bulk uncertainty of 7\% in the full sensitivity images, though note this will increase if measuring sources at the edges of the pre-mosaic single-SBID images. 

As noted in Section~\ref{sec:patching}, some of these sources will have their flux densities slightly underestimated in the images (but not in the catalogue). A full list of sources and the patched flux densities versus their pre-patched values are reported in Appendix~\ref{app:subtracted}, and care should be taken if measuring them from the images directly.

\subsection{Astrometry}\label{sec:astrometry}

\begin{figure}[t]
    \centering
    \includegraphics[width=1\linewidth]{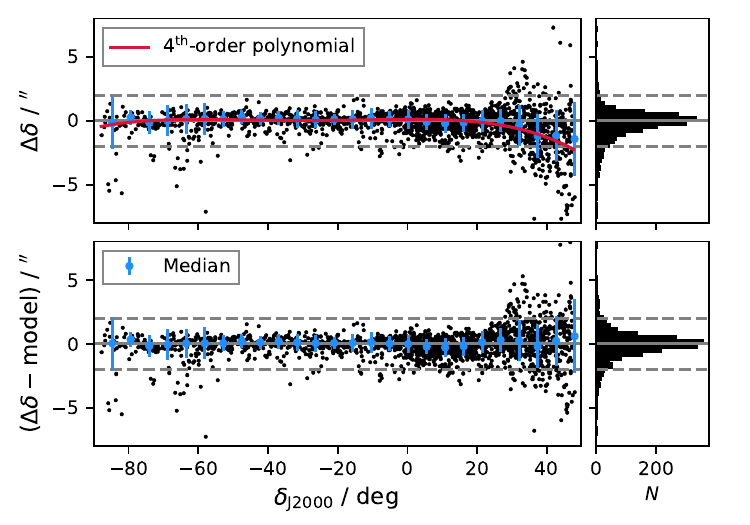}
    \caption{Declination offsets with respect to the RFC, as a function of declination. The blue markers are medians within 5-deg bins, and the solid red line in the top panel is the polynomial fit. The horizontal solid gray line is drawn at an offset of 0, with the dashed gray lines indicating the pixel size. The bottom panel shows the residuals after applying the polynomial model.\label{fig:astrometry:dec}}
\end{figure}

\begin{figure*}[t]
    \centering

    \begin{subfigure}[b]{0.33\linewidth}
        \includegraphics[width=1\linewidth]{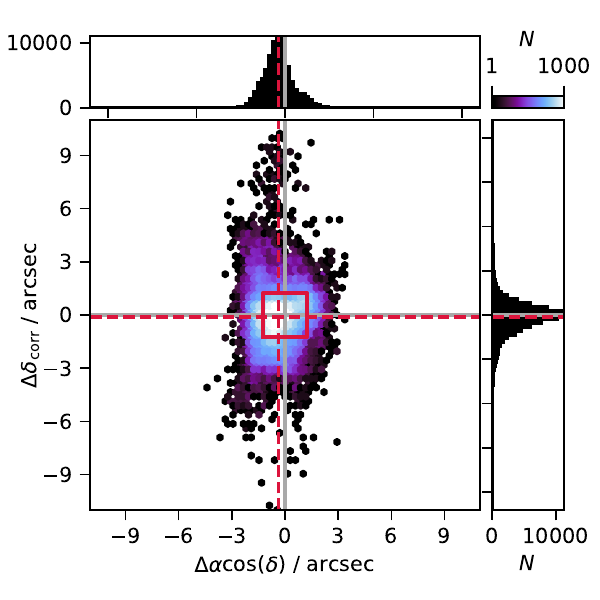}
        \caption{\label{fig:astrometry:hex:vlass} RACS-low2 $-$ VLASS-QL.}
    \end{subfigure}%
    \begin{subfigure}[b]{0.33\linewidth}
        \includegraphics[width=1\linewidth]{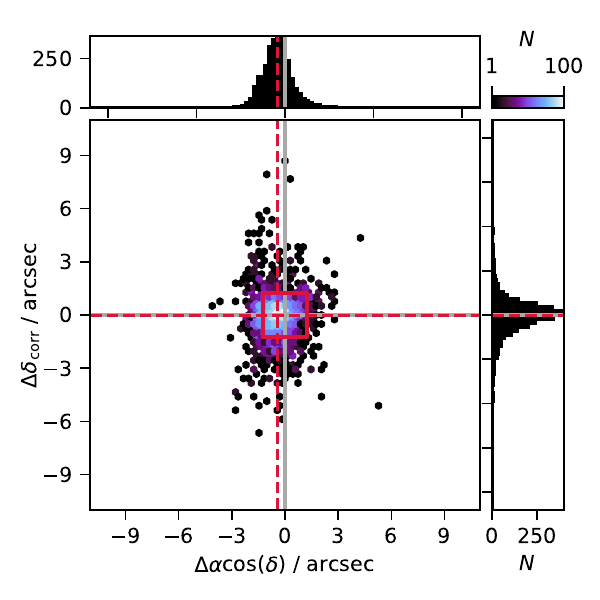}
        \caption{\label{fig:astrometry:hex:rfc} RACS-low2 $-$ RFC.}
    \end{subfigure}%
    \begin{subfigure}[b]{0.33\linewidth}
        \includegraphics[width=1\linewidth]{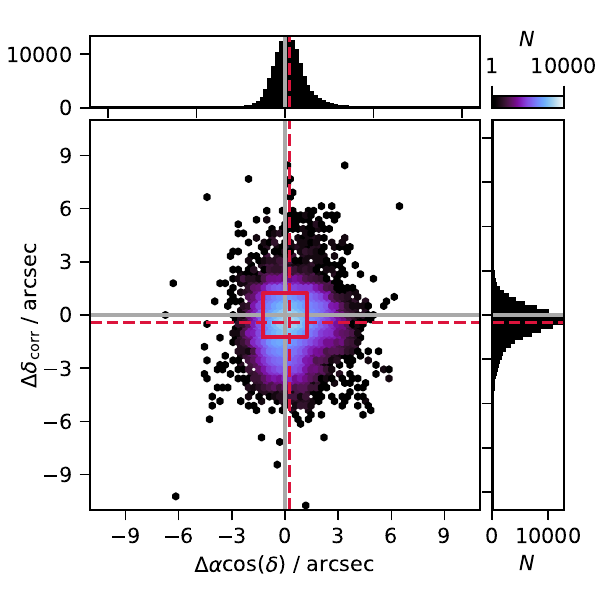}
        \caption{\label{fig:astrometry:hex:low1} RACS-low2 $-$ RACS-low1.}
    \end{subfigure}
    \caption{\label{fig:astrometry:hex} Astrometric offsets between the RACS-low2 catalogue and the VLASS-QL catalogue (\ref{fig:astrometry:hex:vlass}), the RFC (\ref{fig:astrometry:hex:rfc}), and RACS-low1 (\ref{fig:astrometry:hex:low1}), after applying the declination offset correction. The solid gray vertical and horizontal lines are drawn at an offset of zero, and the red box indicates the pixel size of {$2.5\arcsec \times 2.5\arcsec$}. The dashed red vertical and horizontal lines are drawn at the sample means.}
\end{figure*}

RACS data products tend to have astrometric precision of order 1--2\arcsec\ for most of the sky, though this becomes worse for low-elevation pointings, reducing to $\gtrsim 2\arcsec$ (typically high declination, $\delta_\text{J2000} \gtrsim +20\degr$; see e.g.\,\citetalias{racs-mid2}, \citetalias{racs-high}, and \citealt{Jaini2025}). For RACS-mid and RACS-high, a declination-dependent correction to the offsets in declination was applied to the catalogues (but not the images). While there is ongoing work to improve the astrometric precision of RACS (and ASKAP data in general) through post-imaging corrections on single image catalogues or individual images (\citealt{Jaini2025}, Galvin et al. in prep.), RACS-low2 made use of the standard observing and processing so does not benefit from these efforts.

We initially look at the the declination-dependent declination offsets with respect to the Radio Fundamental Catalogue \citep[RFC;][]{rfc} \footnote{{Where we have previously used the ICRF3 for declintion-dependent astrometry corrections \citepalias{racs-mid2,racs-high}, we now use the RFC simply due to the larger sample size and better coverage in the Southern Hemisphere. The Very Long Baseline Interferometry data of both produces similar $\approx$\,mas astrometric accuracy for our purposes.}}. We cross-match {the 2\,121\,918 unresolved (Section~\ref{sec:resolved}) and isolated sources}, and show the declination offsets in Figure~\ref{fig:astrometry:dec} {after removing sources with SNR $< 10$}. We see an overall increase in scatter at high declination ($\gtrsim +20\degr$) as well as an additional bulk offset, similar to what was seen in RACS-mid and RACS-high. As with RACS-mid and RACS-high, we fit polynomial model to the offsets following previous work, largely to provide a correction for high declinations. The model is obtained by trialling polynomial orders up to five, with the fitted model chosen from the Akaike Information Criterion \citep[AIC;][]{Akaike1974}. This results in a 4-th order polynomial model (as with \citetalias{racs-high}) with fitted parameters 
\begin{multline}
    \Delta\delta = +(0.15 \pm 0.04) - (29 \pm 21)\times 10^{-3} \delta \\ - (3.1 \pm 0.4) \times 10^{-4} \delta^2 - (8.3 \pm 1.2) \times 10^{-6} \delta^3 \\ -(6.5\pm 1.4)\times 10^{-8} \delta^4 \, \text{arcsec} ,
\end{multline}
The resulting model, and residuals after application, are shown on Figure~\ref{fig:astrometry:dec}. This does not reduce overall scatter inherent to RACS, but can be used to aid in cross-matching to high-declination data.

\begin{table}[t!]
    \centering
    \caption{\label{tab:astrometry} Astrometric offsets of sources cross-matched between the RACS-low2 catalogue and external catalogues, with offsets defined as $\Delta\delta = \delta_\text{RACS-low2}-\delta_\text{survey}$ and $\Delta\alpha\cos\delta = (\alpha_\text{RACS-low2} - \alpha_\text{survey})\cos\delta$ for declination and R.~A. offsets, respectively.}
    \begin{tabular}{cc c c c}\toprule
         Survey & $N_\text{sources}$ & $\Delta\alpha\cos\delta$ & $\Delta\delta$ & $\Delta\delta_\text{corr}$ \\
          & & (\arcsec) & (\arcsec) & (\arcsec) \\[0.5em]\midrule
          \multicolumn{5}{c}{Full sky} \\[0.5em]\midrule
         RFC & 2\,749 & $-0.42 \pm 0.80$ & $-0.25 \pm 1.17$ & $-0.05 \pm 1.12$ \\
         VLASS-QL & 78\,257 & $-0.36 \pm 0.80$ & $-0.31 \pm 1.10$ & $ -0.12 \pm 1.05$ \\
         RACS-low & 104\,046 & $+0.23 \pm 0.82$ & $-0.42 \pm 1.00$ & $-0.44 \pm 1.00$ \\[0.5em]\midrule
         \multicolumn{5}{c}{$\delta_\text{J2000} > +0\degr$} \\[0.5em]\midrule
         RFC & 1\,428 & $ -0.37 \pm 0.89$ & $-0.42 \pm 1.40$ & $ 0.01 \pm 1.34$ \\
         VLASS-QL & 43\,543 & $-0.21 \pm 0.88$ & $ -0.46 \pm 1.28$ & $ -0.06 \pm 1.22$ \\
         RACS-low & 36\,161 & $+0.40 \pm 0.97$ & $-0.60 \pm 1.07$ & $-0.56 \pm 1.07$ \\[0.5em]\bottomrule
    \end{tabular}
    
\end{table}

To further check RACS-low2 astrometric accuracy, we cross-match to the VLA Sky Survey \citep[VLASS;][]{Lacy2020} quick-look catalogue version 2 \citep[VLASS-QL;][after correction of a known astrometry issue]{Gordon2020}, as well as RACS-low1. Figure~\ref{fig:astrometry:hex} shows the cross-match results for isolated and unresolved sources $>100\sigma_\text{rms}$ for VLASS-QL, the RFC, and RACS-low1. We use the `corrected' declination for the purpose of cross-matching sources within 10\arcsec\ and show the corrected declination offsets {in Figure~\ref{fig:astrometry:hex}}. In general we see the typical scatter we expect from RACS, with larger scatter in declination due to the increased PSF size in the N-S direction. Table~\ref{tab:astrometry} summarises the weighted mean offsets for each survey{, both before and after applying declination offset corrections}. Overall, we suggest that RACS-low2 retains the 1--2\arcsec\ astrometric uncertainty of previous RACS epochs, with larger offsets still expected at high declination.

\section{Stokes V data assessment}

\begin{figure*}[t]
    \centering
    \includegraphics[width=1\linewidth]{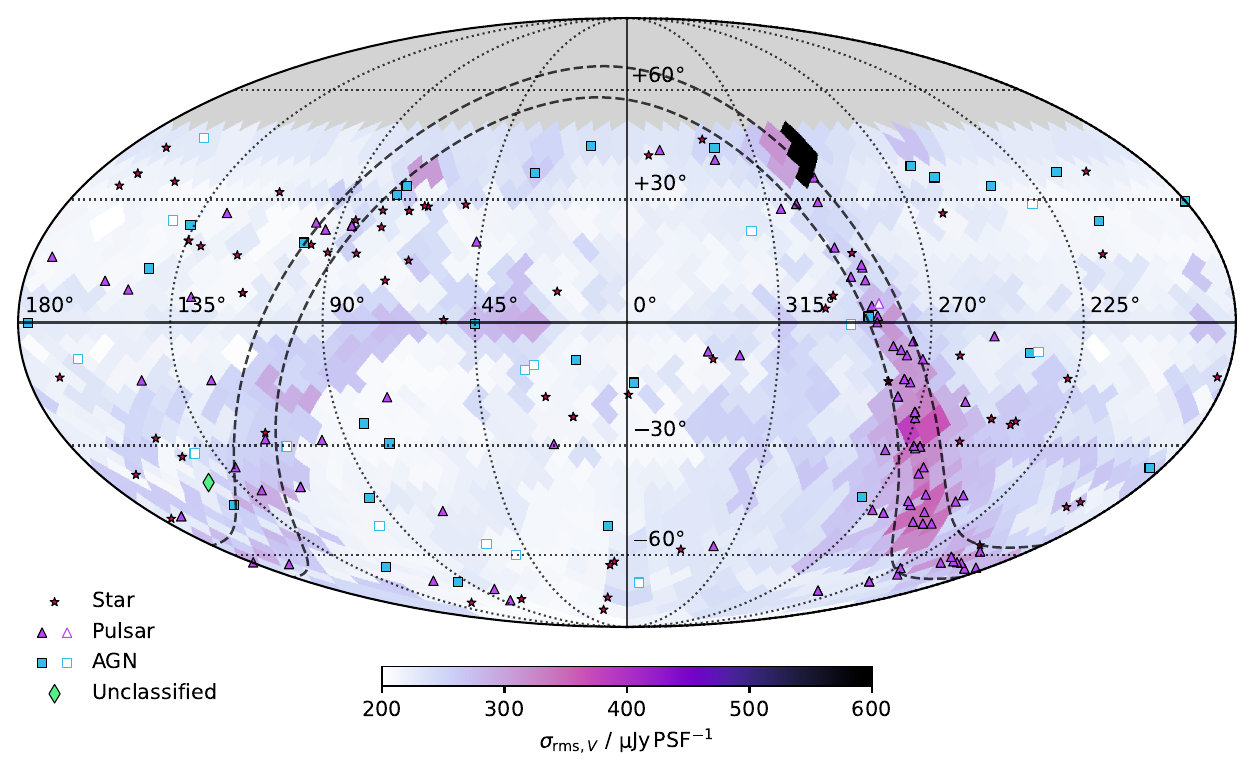}
    \caption{\label{fig:circ:rms} The HEALPix-binned rms noise across the Stokes V images reported as a {mean} for each field. {HEALPix-binning with $N_\text{side}=16$ is done after interpolating the single rms values per field to an oversampled grid to allow plotting. The HEALPix bins are $\approx 13.4$\,deg$^2$, which is smaller than size of the individual fields.} Note that without mosaicking of neighbouring images, the rms noise in the overlap region between tiles is generally higher than the Stokes I equivalent in Figure~\ref{fig:rms}, but this increase in noise towards image edges is not visible with this binning. The markers indicate sources with circularly polarized emission discussed in Section~\ref{sec:circ:sources}, with open markers representing probable source associations.}
\end{figure*}

In general, the rms noise in the Stokes V images is lower than the Stokes I counterparts in part due to the relatively lower source density in the Stokes V image plane as compared to Stokes I. In Figure~\ref{fig:circ:rms} we show the position-dependent Stokes V rms noise in HEALPix binning across the survey. Note that as we do not mosaic neighbouring Stokes V images, the rms noise is high at the image edges due to the correction of primary beam attenuation. Otherwise, the median rms noise is {$163_{-31}^{+134}$}\,\textmu Jy\,PSF$^{-1}$ with only minimal variation across the sky except towards the Galactic Centre and Cygnus~A {(which is not removed from the Stokes V data during the off-axis source removal process that only operates on the XX and YY correlations)}, where leakage and calibration errors around the bright radio sources are significant.

We note that the \texttt{ASKAPSoft} imager produces Stokes V images that follow the IAU polarization convention where right-hand circularly polarized emission is positive, and left-hand circularly polarized emission is negative.

\subsection{Leakage}\label{sec:leakage}

\begin{figure*}[t]
    \centering
    \begin{subfigure}[b]{0.33\linewidth}
        \includegraphics[width=1\linewidth]{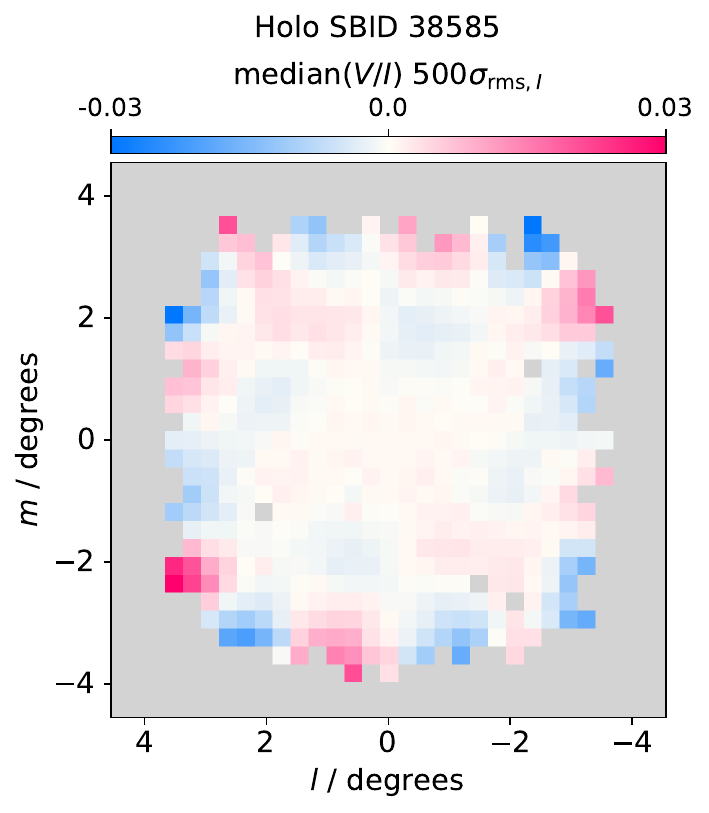}
    \end{subfigure}%
    \begin{subfigure}[b]{0.33\linewidth}
        \includegraphics[width=1\linewidth]{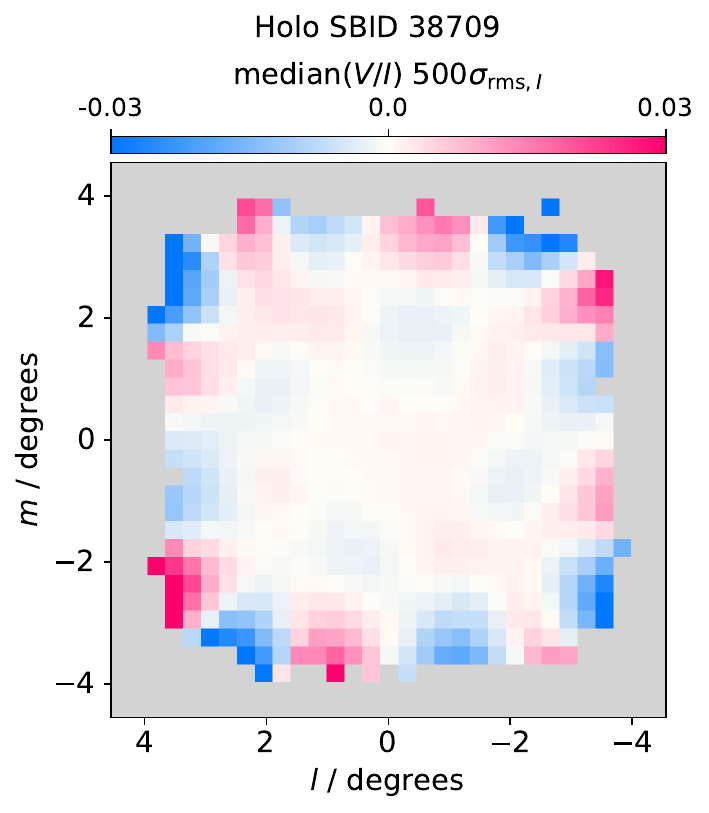}
    \end{subfigure}%
    \begin{subfigure}[b]{0.33\linewidth}
        \includegraphics[width=1\linewidth]{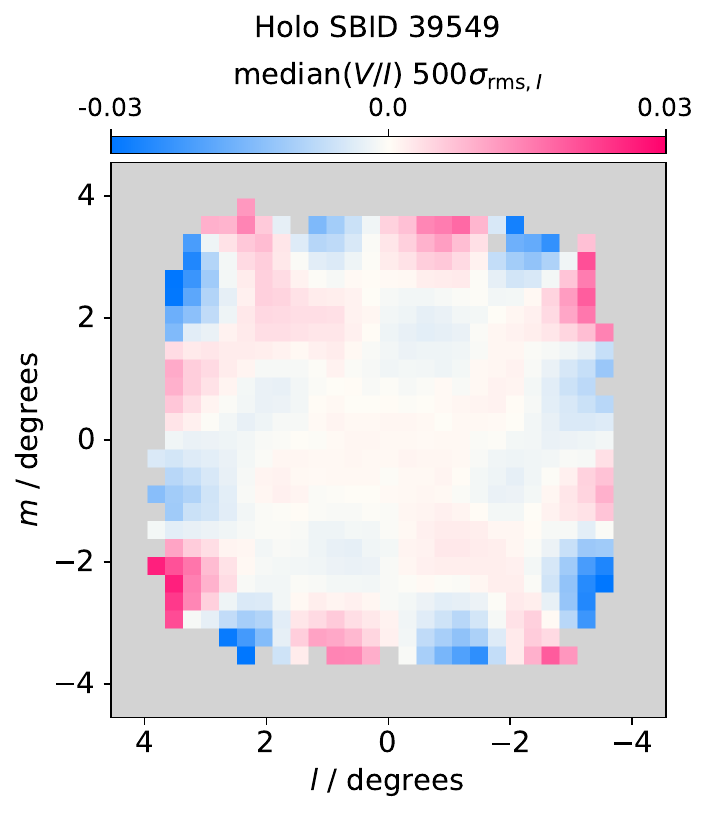}
    \end{subfigure}%
    \caption{\label{fig:circ:tile} Binned median residual $V/I$ as a function of position across the PAF footprint for Stokes I sources above $500\sigma_\text{rms}$, highlighting the residual widefield leakage pattern. Each panel shows a separate holography grouping (see Table~\ref{tab:holography}). We do not show holography SBID 41055 due to the low number of observations associated with it. }
\end{figure*}

\begin{figure}[t]
    \centering
    \includegraphics[width=1\linewidth]{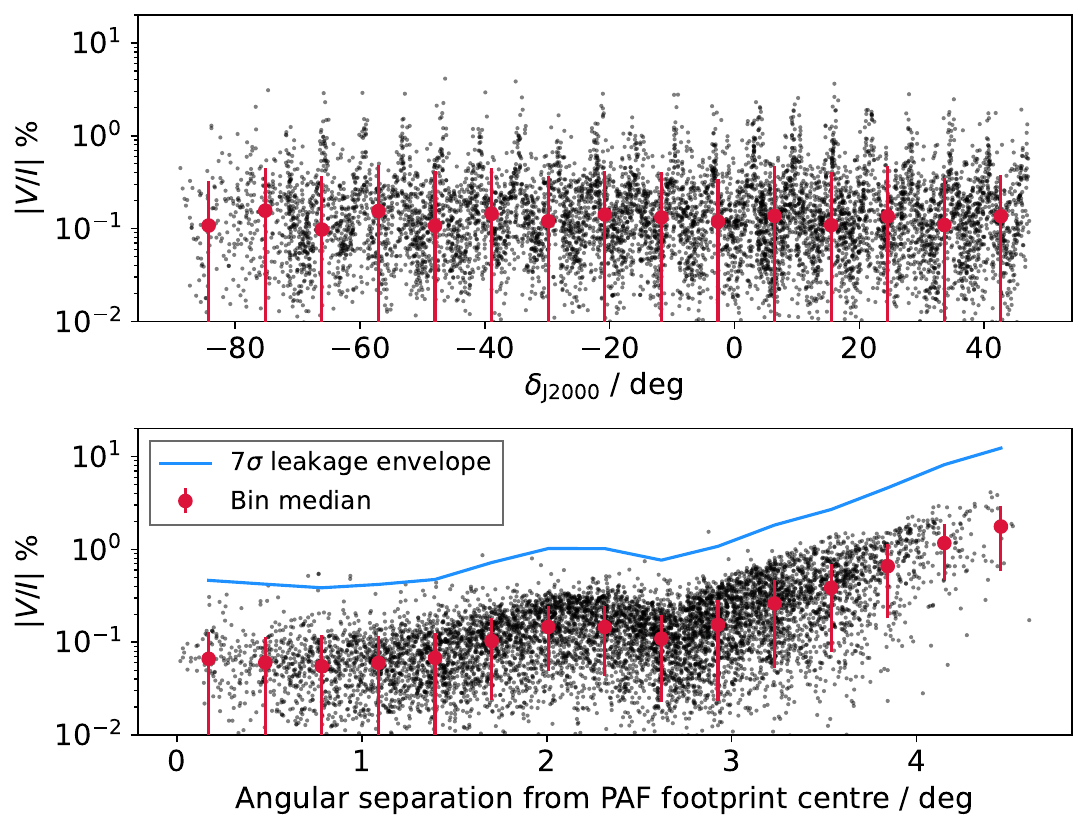}
    \caption{\label{fig:circ:coords} Residual absolute fractional circular polarization ($|V/I$) for unresolved Stokes I sources above $1000\sigma_\text{rms}$ as an estimate of residual Stokes I leakage. \emph{Top panel.} The fractional circular polarization as a function of declination. \emph{Bottom panel.} The same as a function of angular separation from the centre of the PAF footprint. The red circles are the median absolute $|V/I|$ in coordinate bins. The blue line in the bottom panel shows the leakage threshold we use to define search for real circularly polarized sources.}
\end{figure}

Before counting Stokes V sources, we investigate the residual widefield leakage of Stokes I into Stokes V. We can expect some residual widefield leakage as we have seen this in previous RACS epochs \citepalias{racs-mid} and data from the VAST survey using the same PAF footprint and frequency \citep[][]{Pritchard2024}. {Due to the third roll axis of the ASKAP dishes, the parallactic angle is fixed in time during each observation, and sources within the field maintain fixed positions with respect to the primary beam. We therefore do not expect the recovered widefield leakage patterns shown in Figure~\ref{fig:circ:tile} to vary over a single observation \citep[see ][for further detail on RACS and ASKAP polarimetry]{Thomson2023}. As we only use an unpolarized source for calibration and removal of leakage (PKS~B1934$-$638) we are not able to remove on-axis leakage of Stokes Q or U into V with the same process used for Stokes I into V. A maximum of $\approx 10\%$ of the Stokes Q/U brightness has been observed to leak into V, though has only been observed for very bright (> Jy) highly linearly-polarized sources (e.g.\ the Vela pulsar). For modelling the overall leakage, we have only considered Stokes I leakage into V, as we do not produce Stokes U or Q images as part of the standard RACS processing.} 

{To assess the leakage of Stokes I into V}, we initially consider Stokes I sources above $500\sigma_\text{rms}$ and look at the fractional circular polarization reported for these sources in the concatenated catalogue. We also only consider unresolved sources, based on the definition in {Section~\ref{sec:resolved}}. First, we show the residual leakage $V/I$ as a function of position on the tile (as in Section~\ref{sec:tile_brightness} for the brightness scale), split by the first three holography groups in Figure~\ref{fig:circ:tile}. Each grouping shows the same general widefield leakage pattern, similar to the fairly consistent residual brightness scale error. 

We then consider only Stokes I sources above $1000\sigma_\text{rms}$ and in Figure~\ref{fig:circ:coords} show the absolute fractional circular polarization $|V/I|$ as a function of declination and as a function of angular separation from the centre of PAF footprint. As a function of declination, we see regular spikes in the leakage which correspond to the higher leakage seen at the edge of each image (i.e.\ far from the PAF footprint centre). Despite this and the lack of widefield leakage correction, the image centres feature low residual leakage $< 1$\%. While the residual leakage does vary in sign (see Figure~\ref{fig:circ:tile}), in the absence of knowing where exactly the sign changes occur we opt to consider only an absolute leakage threshold for determining real circularly polarized sources. To obtain the residual leakage, we bin $|V/I|$ as a function of angular separation from the footprint centre and take the median $|V/I|$ for each bin. The medians for each bin are shown as red markers in Figure~\ref{fig:circ:coords}. We then set $7 \times \text{median}\left|V/I\right|$ as a threshold above which to consider real emission, linearly interpolating between the binned points. The resultant threshold is shown on the bottom panel of Figure~\ref{fig:circ:coords}. We note that leakage is only considered in the absolute sense, and to properly consider a signed leakage would require position dependence in both $l$ and $m$ \citep[e.g.][]{Pritchard2024}. The final threshold ranges from $\approx 0.4$--$12$\%

This residual leakage is overall lower than what was reported for RACS-high (cf.~$\approx 2$\% for the image centre, and $\approx 6$\% towards the image edges), despite RACS-high also lacking widefield leakage corrections, though probing brighter Stokes I sources allows for a lower estimate. The results we see here are more in line with RACS-mid, which after widefield leakage correction results in a residual leakage of $\approx 0.9$\% for most of the survey. We note, however, that the residual leakage at the image edges for RACS-low2 is more significant than for RACS-mid, which is where the widefield leakage correction becomes important.

\subsection{Circularly polarized sources}\label{sec:circ:sources}

\begin{figure}[t]
    \centering
    \includegraphics[width=1\linewidth]{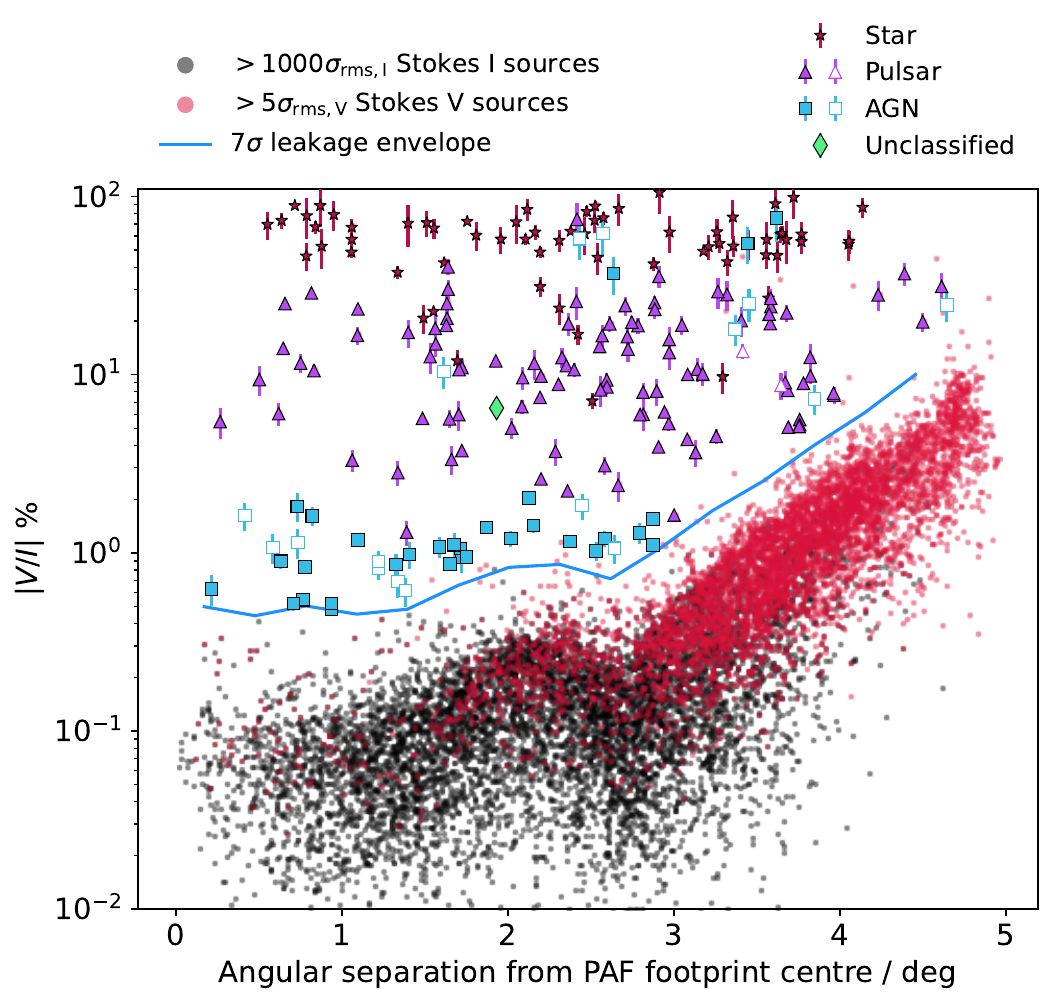}
    \caption{\label{fig:circ:distance} Fractional circular polarization of sources as a function of angular separation from the PAF footprint centre, coloured by source type. Open markers indicate probable associations. We also show the distribution of presumably unpolarized bright sources (black circles, as in Figure~\ref{fig:circ:coords}) and sources with Stokes V flux density above $5\sigma_\text{rms,V}$ (red circles). The leakage threshold is shown as a blue line, as in Figure~\ref{fig:circ:coords}.}
\end{figure}

{As most radio sources in the sky do not have significant circularly polarized emission at MHz--GHz frequencies \citep[e.g.][]{Lenc2018,Callingham2023} we do not expect many real circularly polarized sources in the RACS-low2 Stokes V images. Most sources with significant Stokes V flux density are likely to be leakage from Stokes I, with the remaining likely to be either stellar systems, pulsars, or (less likely) AGN. We do not expect to see large-scale circularly polarized emission due to the lack of sensitivity on very large angular scales and the low fractional polarization expected from such emission \citep{Ensslin2017}.} 

{Stellar emission generated by incoherent processes at radio wavelengths is generally too faint to detect in stars more than a few parsec away \citep[e.g.][]{Gary1985,Gudel2002}. Coherent mechanisms, like the electron cyclotron maser or plasma emission, can generate emission with much higher brightness temperatures \citep[e.g.][]{Dulk1985}. The emission from these coherent processes can be highly circularly polarized \citep[up to 100\%, e.g.][]{Hallinan2007} and provide insights into the stellar magnetosphere, as well as constraints on the magnetoionic properties of the local plasma environment. Pulsars have also been observed with both linear and circularly polarized emission, with reasonably high ($\gtrsim 10\%$) fractional polarization \citep[e.g.][]{Sobey2021,Anumarlapudi2023}, though pulsar emission mechanisms remain uncertain \citep[e.g.][]{Melrose2021}.}

{In comparison, circular polarization from AGN is either related to synchrotron processes directly or through conversion of linear polarization to circular polarization through Faraday rotation. AGN are expected to show far less than $\approx 1$\% fractional circular polarization \citep[e.g.][]{Rayner2000,Macquart2002,Taylor2024,Kramer2025}, which has historically fallen below the leakage threshold in previous untargeted circular polarisation surveys \citep[e.g.][]{Pritchard2021}.}

\subsubsection{Selection criteria}

We refine the catalogue of Stokes V measurements described in Section~\ref{sec:catalogue:stokesv} to remove measurements that are likely to be noise, artefacts, or sources of leakage. We begin by selecting sources with significant Stokes V signal, i.e.\begin{equation}\label{eq:crit1}
|S_{\text{peak},V}|/\sigma_{\text{rms},V}>5 \, ,
\end{equation}
where $S_\text{peak,V}$ is the peak Stokes V flux density, and sources that have Stokes V emission significantly above the relevant leakage threshold satisfying \begin{equation}\label{eq:crit2}
\left| \frac{S_{\text{peak,}V}}{S_{\text{peak,}I}} \right| - \sigma_{\text{lower}}  > 7 \times \text{median}\left(|V/I|\right) \, ,
\end{equation}
where $\sigma_{\text{lower}}$ is the lower bound of the measurement uncertainty in $|S_{\text{peak,}V} / S_{\text{peak,}I}|$. {Finally, as much circularly polarized emission is expected to be from spatial regions smaller than the RACS-low2 angular resolution, and to lower the likelihood of cataloguing artefacts near bright extended sources, we only consider unresolved Stokes I sources (using the definition from Section~\ref{sec:resolved}).} This yields \NsourcesV\ Stokes V measurements, and we show the distribution of these across the sky on Figure~\ref{fig:circ:rms}{, and in Figure~\ref{fig:circ:distance} as a function of angular separation from the centre of the PAF footprint with the leakage threshold. On Figure~\ref{fig:circ:distance},} we also show the distribution of bright Stokes I sources and those with significant ($>5\sigma_\text{rms,V}$) Stokes V emission to illustrate how the sample of circularly polarized sources is constructed. This sample includes duplicate measurements of the same source, but is unlikely to be a complete list of circularly polarized sources detected in RACS-low2 data as we only consider unresolved Stokes I sources and sources that are detected in the Stokes I images above $5\sigma_{\text{rms},I}$.

\begin{figure}[t]
    \centering
    \includegraphics[width=1\linewidth]{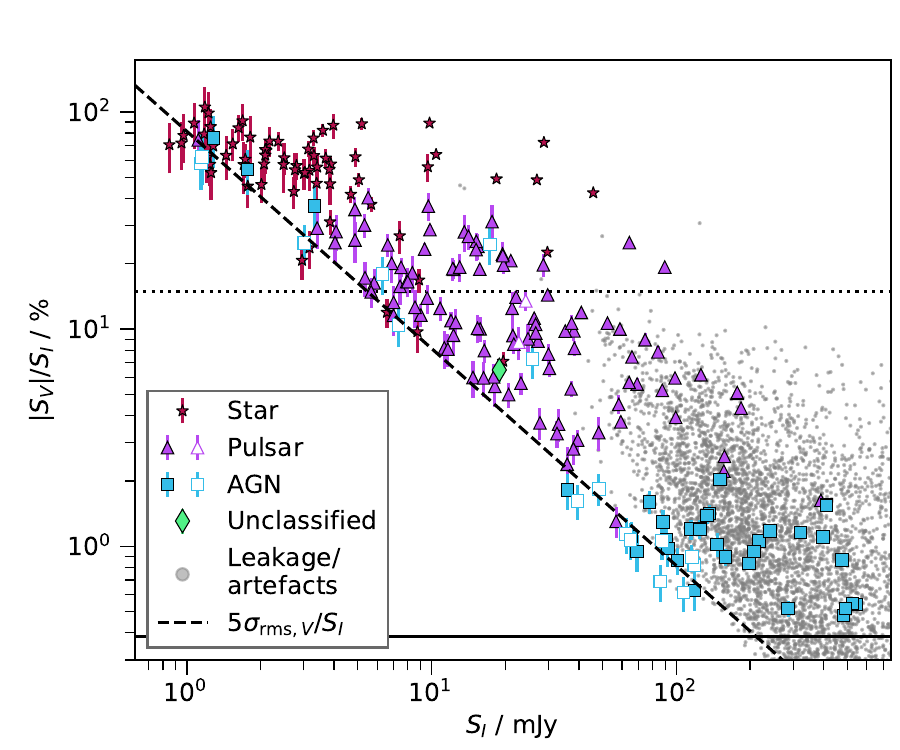}
    \caption{\label{fig:stokesv_sources} Fractional circular polarization of sources as a function of Stokes I flux density, coloured by source type. Open markers reflect probable associations (mostly AGN---see Section~\ref{sec:circ:sources:agn}). The solid and dotted horizontal lines indicate the minimum and maximum leakage thresholds, respectively. Only unresolved sources with $S_V > 5\sigma_{\text{rms},V}$ are shown. The dashed line drawn at $5\sigma_{\text{rms},V}/S_I$ assumes a median Stokes V rms noise from the catalogue of $\sigma_{\text{rms},V} = 163$\,\textmu Jy\,PSF$^{-1}$.}
\end{figure}

\begin{figure*}[p]
    \centering
    \begin{subfigure}[b]{1\linewidth}
    \includegraphics[width=1\linewidth]{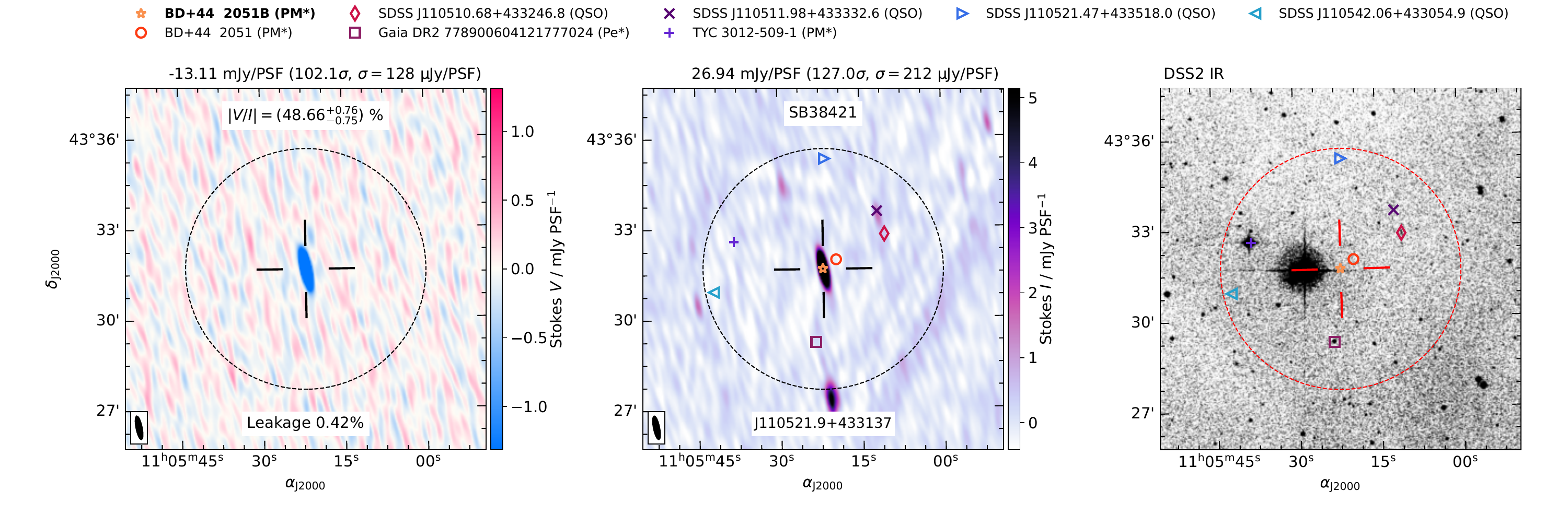}
    \caption{\label{fig:circ:example:BD+44}}
    \end{subfigure}\\%
    \begin{subfigure}[b]{1\linewidth}
    \includegraphics[width=1\linewidth]{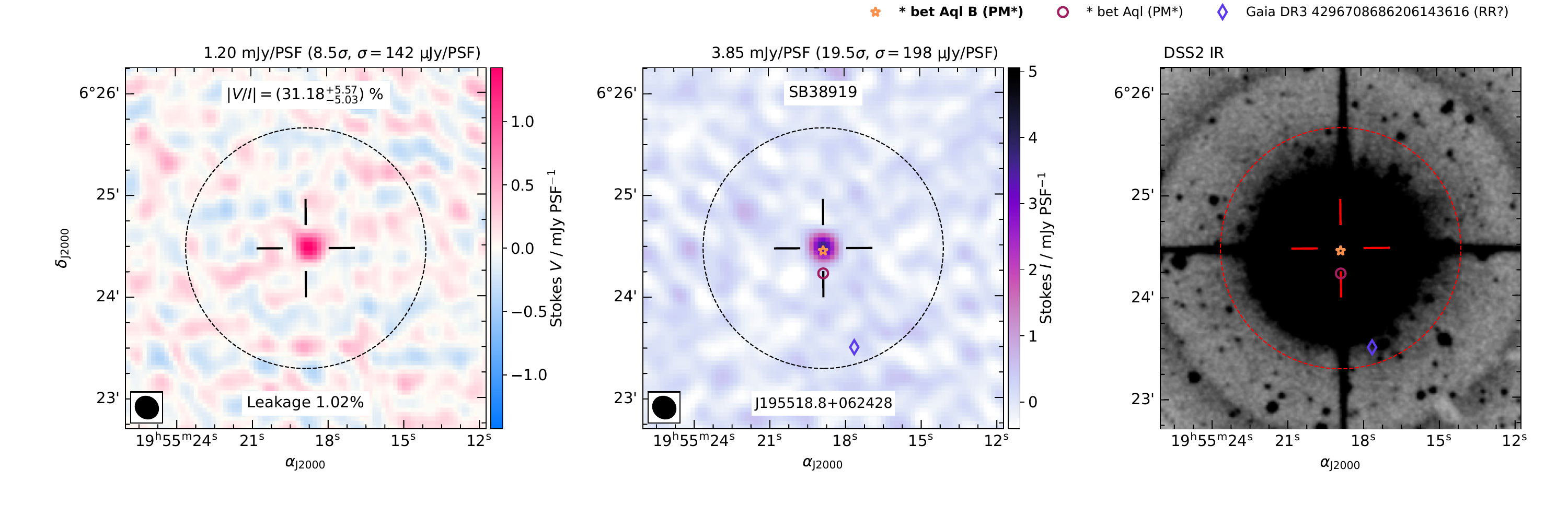}
    \caption{\label{fig:circ:example:freespace2ref}}
    \end{subfigure}\\%
    \begin{subfigure}[b]{1\linewidth}
    \includegraphics[width=1\linewidth]{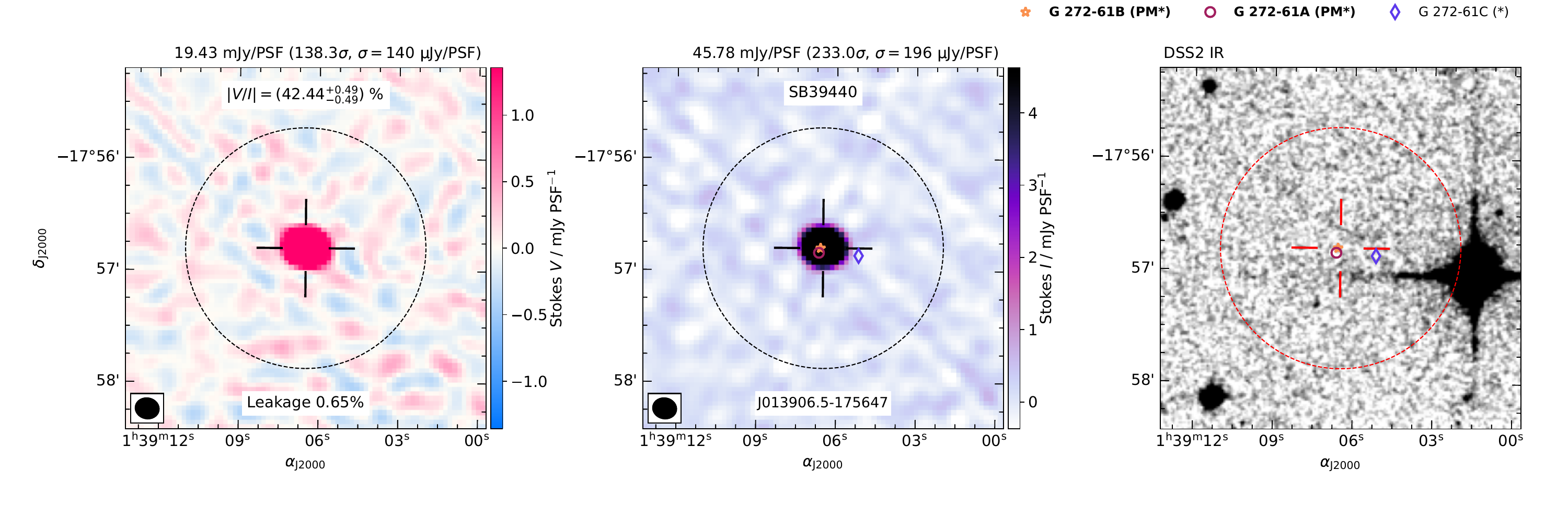}
    \caption{\label{fig:circ:example:uvcet}}
    \end{subfigure}\\%
    \caption{\label{fig:circ:examples1} Example cutouts used for visual inspection of the circularly polarized sources, showcasing three example radio stars: BD$+$44~2051 (\ref{fig:circ:example:BD+44}), *~bet~Aql (\ref{fig:circ:example:freespace2ref}), and UV~Ceti (\ref{fig:circ:example:uvcet}). \textit{Left panels}: Stokes V image; \textit{centre panels}: Stokes I image; \textit{right panels}: DSS2 infrared image. On each panel, the dashed circle is drawn at $5\times\theta_\text{M}$, and the crosshairs point to the RACS-low2 Stokes I source position. Markers shown on the centre and right panels are sources within the dashed circle from SIMBAD (after proper motion correction, where relevant) with the sources ordered in the legend by angular separation from the source---the source names in the legend also include the object type as reported in SIMBAD, and the yellow star symbol is generally the closest source. Source names in bold font are within $0.5\times\sqrt{\theta_\text{M}\theta_\text{m}}$ of the radio position. The ellipses in the bottom left corner of the Stokes V and Stokes I images show the PSF size.}
\end{figure*}

\begin{figure*}[p]
    \centering
    \begin{subfigure}[b]{1\linewidth}
    \includegraphics[width=1\linewidth]{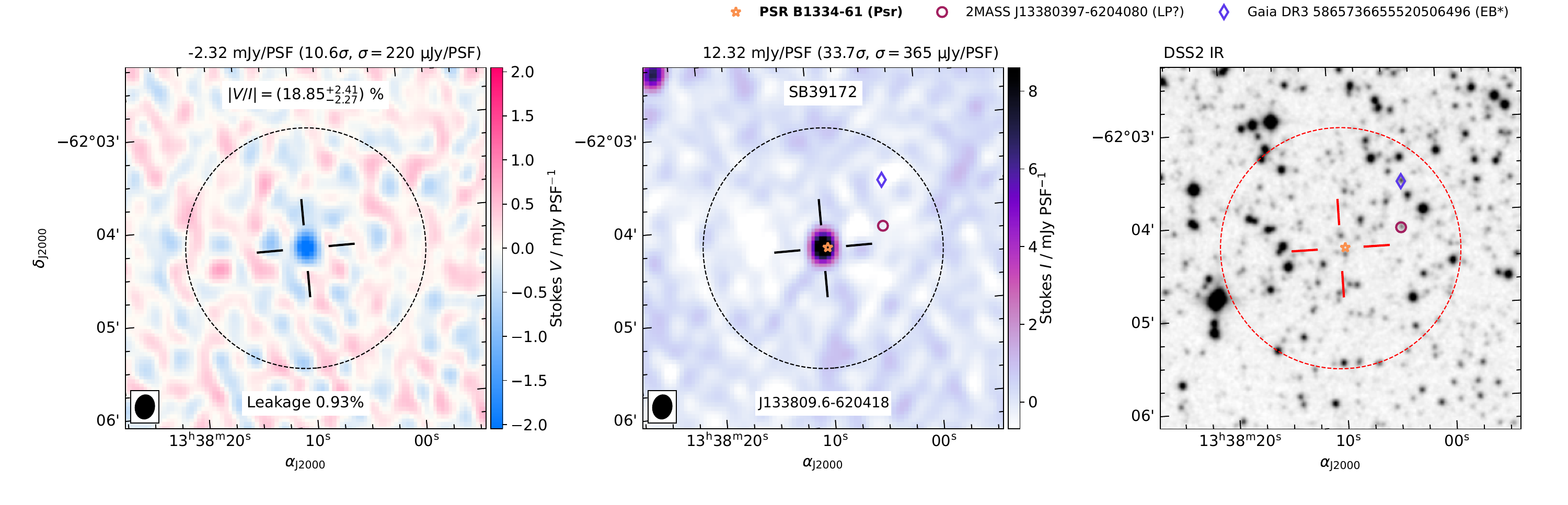}
    \caption{\label{fig:circ:example:psr}}
    \end{subfigure}\\%
    \begin{subfigure}[b]{1\linewidth}
    \includegraphics[width=1\linewidth]{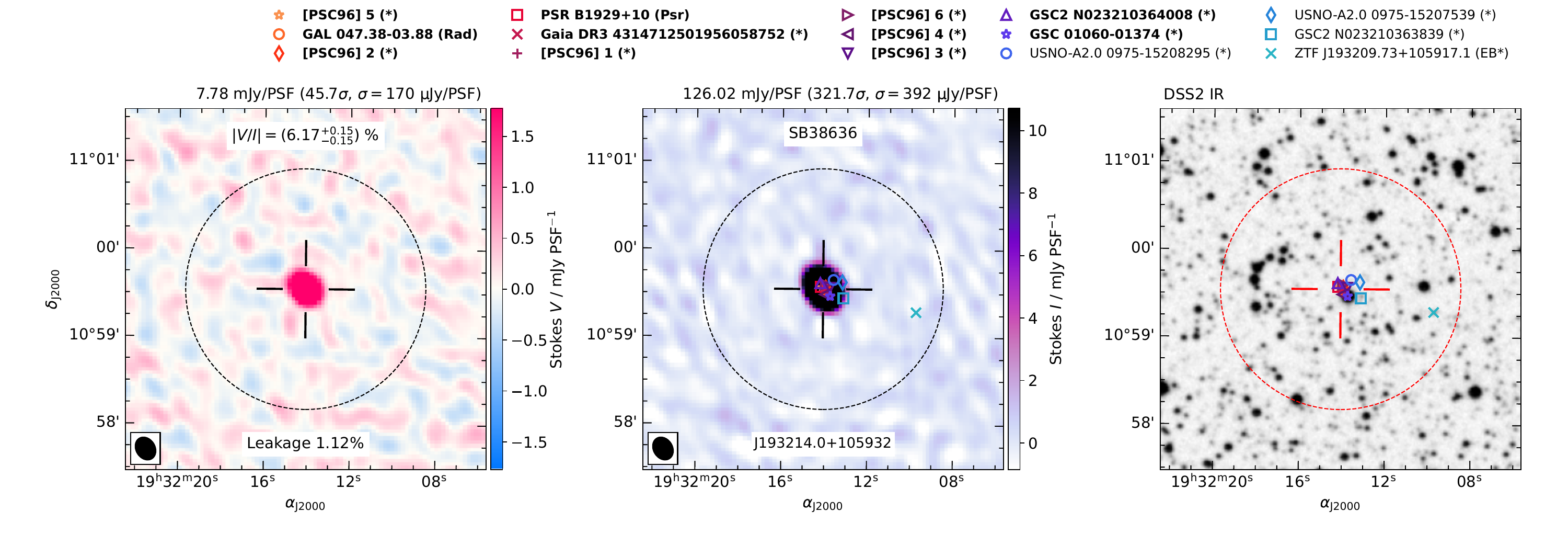}
    \caption{\label{fig:circ:example:psr2}}
    \end{subfigure}\\%
    \begin{subfigure}[b]{1\linewidth}
    \includegraphics[width=1\linewidth]{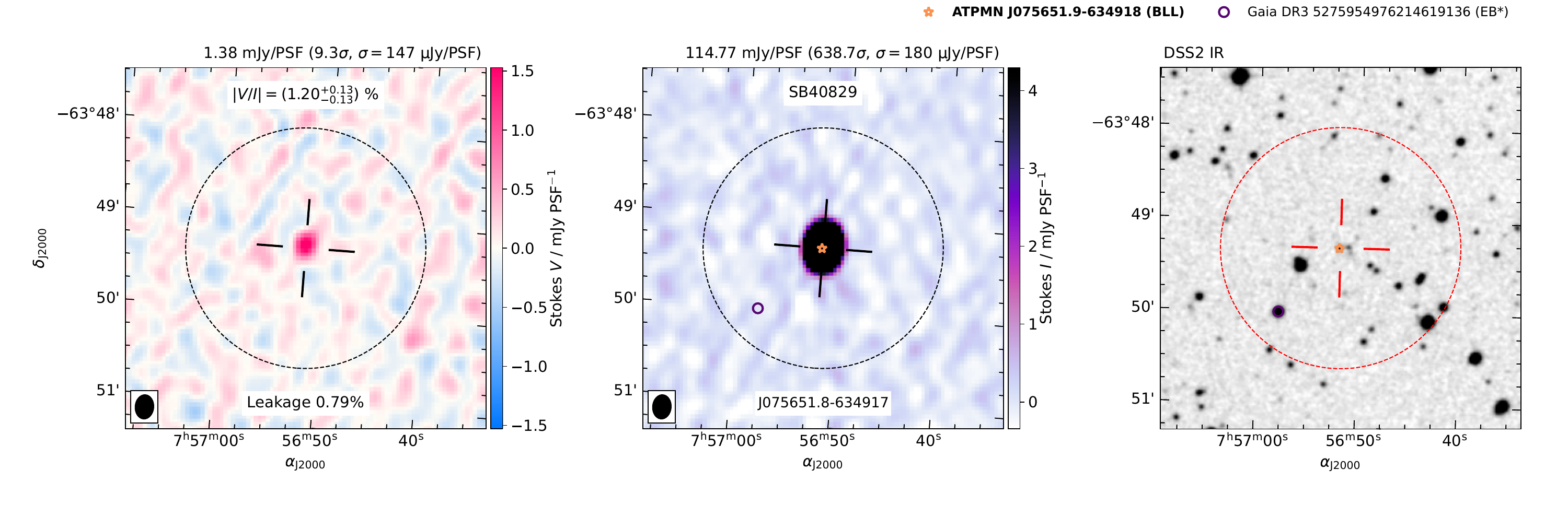}
    \caption{\label{fig:circ:example:agn}}
    \end{subfigure}\\%
    \caption{\label{fig:circ:examples2} Example cutouts used for visual inspection of the circularly polarized sources, showing the pulsars PSR~B1929$+$10 (\ref{fig:circ:example:psr}) and PSR~B1929$+$10 (\ref{fig:circ:example:psr2}), and the AGN ATPMN~J075651.9$-$634918 (\ref{fig:circ:example:agn}). The other details are the same as in Figure~\ref{fig:circ:examples1}.}
\end{figure*}

For the purpose of assessing the individual circularly polarized sources selected above, we produce small cutouts from the Stokes I and V images, alongside a cutout from the Digitized Sky Survey 2 (DSS2) infrared band, and query the SIMBAD astronomical database \footnote{Via the \texttt{Simbad} module in \texttt{astroquery} \citep{astroquery}.} \citep{simbad} for all sources within 8\arcmin \footnote{A large search radius here is to ensure stars with high proper motions are able to be matched.}. We then correct the positions of sources with proper motion (PM) measurements to the epoch of our observations, and plot all sources within $5 \times \theta_\text{M}$, where $\theta_\text{M}$ is the FWHM of the major axis of the PSF for the given radio image. We consider the RACS-low2 source associated with the astronomical source if it lies within $0.5\times\sqrt{\theta_\text{M}\theta_\text{m}}$ of the source positions after PM correction. While this varies by source, the average radius is $\approx 7$\arcsec\ (note the astrometric error is of order $1$\arcsec--$2$\arcsec\ and the image pixel size is $2.5\arcsec \times 2.5\arcsec$).

These cutouts with nearby sources overlaid are used for visual inspection to assess (1) the general quality of our selection criteria noted above to avoid leakage sources and to check for residual imaging artefacts, and (2) to assess the automated cross-matching to sources in the SIMBAD database. We manually inspect all cross-matches for validity, and with extra considerations for each source type discussed in the following sections. We show examples of these sources in Figure~\ref{fig:circ:examples1} (specifically radio stars) and Figure~\ref{fig:circ:examples2}. In total, we identify \StarMeasurements\ Stokes V measurements to be associated with stars, \PulsarMeasurments\ to (likely) be associated with pulsars, and \AGNMeasurements\ to (likely) be associated with AGN. One final source is not associated with any known object and not likely to be an AGN. This unassociated source was recently found with VASTER, the short-timescale imaging and transient detection pipeline using ASKAP data (Wang et al., submitted), and a detailed follow-up will be presented in future work (Wang et al., in prep.).   

The details of the columns included in the reduced Stokes V catalogue of \NsourcesV\ are shown in Table~\ref{tab:columns:stokesv} in Appendix~\ref{app:columns}, including host identification and Stokes V measurement information.  

\subsubsection{Artefacts}\label{sec:circ:sources:artefacts}

While Equation~\ref{eq:crit1} and the Stokes I and V $5\sigma_\text{rms}$ detection thresholds should remove artefacts in general, a small number of residual artefacts or spurious sources are found. In particular, we find two example spurious sidelobe artefacts near bright sources, both close to the detection thresholds in Stokes V. These sidelobe artefacts are removed from the final catalogue. We also remove one source that is partially cut-off at an image boundary, resulting in an incorrect measurement in Stokes I. We also see an example where a $\approx 1$\,Jy Stokes I source (4C~34.04) was catalogued as two separate sources where the second, minor source had a flux density of $\approx 5.6$\,mJy. The Stokes V measurement for the minor source was $\approx -2.5$\,mJy, though should be associated with the brighter part of 4C~34.04. The major component of 4C~34.04 is considered resolved so is not included in our Stokes V source list. These artefacts are not included in the Stokes V catalogue and are not included in the \NsourcesV\ Stokes V measurements. 

\subsubsection{Stars}\label{sec:circ:sources:stars}

Previous circular polarization searches of RACS data have been effective in finding radio stars, with 47, 63, and 59 radio stars found in RACS-low1, RACS-mid, and RACS-high, respectively \citep{Driessen2024}. In total we report \StarUnique\ unique stars (or stellar systems) with \StarMeasurements\ total measurements coincident with the RACS-low2 circularly polarized sources. {Of these, 36 have been previously detected in ASKAP data, and 13 also have a previous Stokes V detection \citep{Driessen2024}. Stars in the SRSC that were detected in ASKAP observations other than RACS-low1, RACS-mid, RACS-high, and the VAST pilot survey do not have recorded, associated Stokes V detections in the current version of the catalogue. Therefore, there are likely more stars with Stokes V detections in other ASKAP observations.}

The radio stars largely occupy the high-fractional polarization and low-flux density quadrant on Figure~\ref{fig:stokesv_sources}, as in other Stokes V source searches \citep[e.g.][]{Pritchard2021}. We find that 40 of the \StarMeasurements\ measurements have a positive sign. In Figure~\ref{fig:circ:examples1} we show examples of known radio stars detected in the RACS-low2 data, showing the plots used in the visual inspection process with the Stokes I, Stokes V, and DSS2 IR images and sources detected within $5\theta_M$ of the RACS-low2 position.

We also cross-match the Stokes V sources to the Sydney Radio Star Catalogue \citep[SRSC;][]{Driessen2024}, applying PM corrections where appropriate. We find \SRSCMatches\ unique matches with SRSC stars. We note that \SRSCNoMatch\ radio stars detected here are not present in the SRSC. While variability in the radio emission may result in non-detections in other ASKAP datasets, we note that the minimum cross-match radii are also slightly larger in this work than those used for the SRSC by a factor of two or more. For our radio star subsample, the range of search radii are 5.9\arcsec\ to 12.3\arcsec\ (median 7.0\arcsec), with matches ranging from to $\approx 0.1$--$7.7$\arcsec, whereas the maximum separation for radio-to-optical cross-matches in the SRSC is 4\arcsec. 

We note that G~190$-$27 (SRSC 00841) is separated by $\approx 7.7$\arcsec\ which is within our previously-defined match radius and also detected across RACS-mid, RACS-high, and RACS-low3 {in total intensity} with similar properties \citep[][{and also detected with circular polarization data from RACS-high}]{Driessen2024}. \citet{Yiu2024} also identified G~190$-$27 as radio-emitting from a cross-match between VLASS and the \emph{Gaia} Catalogue of Nearby Stars, so we consider the RACS-low2 detection likely to correspond to G~190$-$27 as well. Given the high declination of the source, $\approx +41\degr$, this does highlight the care that needs to be taken in cross-matching where the RACS astrometric precision becomes affected by an undiagnosed elevation-dependent offset despite best-effort corrections in catalogue space (Section~\ref{sec:astrometry}), a large PSF, and the beam-to-beam variation common to all current RACS data.

An additional cross-match is also performed against \emph{Gaia} Data Release 3 \citep[\emph{Gaia} DR3;][]{Gaia2021a} finding Gaia DR3 4167934571550235264 a likely host for RACS-LOW2 J172350.6$-$075711, which matches within $\approx 0.8$\arcsec\ after PM correction though is not present in the SIMBAD search. Gaia DR3 4167934571550235264 is in the \emph{Gaia} catalogue of Nearby Stars \citep[GCNS;][]{2021A&A...649A...6G} and has a geometric distance of $109.9_{-0.6}^{+0.7}$\,pc \citep{2021AJ....161..147B}.

Included in the previously-detected stars is the archetypal flare star UV~Ceti \citep[as G~272$-$61B; e.g.][]{Lynch2017,Plant2024}, which is also recorded in the SRSC. We show the RACS-low2 radio detections in the bottom panel of Figure~\ref{fig:circ:example:uvcet}. We note that our search does not distinguish between the stars and very close companions (e.g. companion stars or exoplanets) and would reasonably require higher astrometric precision to distinguish in some cases. In addition to UV~Ceti, we also detect circularly polarization emission from the similarly variable M-dwarf BD$+$44~2051B, originally detected at radio wavelengths by \citet[][and included in the SRSC]{White1989}, highlighting the utility of the high-declination RACS-low2 Stokes V data despite the previously-noted larger astrometric uncertainties.

\subsubsection{Pulsars}\label{sec:circ:sources:pulsars}

We identify \PulsarUnique\ unique pulsars detected with \PulsarMeasurments\ total measurements. {Of these known pulsars, 44 have also been detected in Stokes V data from RACS-low1 \citep{Anumarlapudi2023}, and 23 in Stokes V data from RACS-mid (K.\ Rose, private communication), with 15 detected in both RACS-low1 and RACS-mid.} We find that 53 of the \PulsarMeasurments\ fractional polarization measurements have a positive sign. In addition to searches with the SIMBAD astronomical database, we also used the Pulsar Survey Scraper \footnote{\url{https://pulsar.cgca-hub.org/}} \citep{Kaplan2022} to search from known pulsars within $0.5\sqrt{\theta_M\theta_m}$\arcsec\ of the RACS-low2 positions. PM measurements are typically not available for the pulsar surveys, and due to distance any proper motion would likely be insignificant, so the separate search through the pulsar surveys does not include any PM corrections. This extra search only results in matches to the ATNF  \footnote{Australia Telescope National Facility.} pulsar catalogue \citep{Manchester2005} \footnote{\url{https://www.atnf.csiro.au/research/pulsar/psrcat/}.}. 

After the SIMBAD match and ATNF pulsar catalogue matches, we manually refine the cross-matches of a small number of sources. We find that RACS-LOW2 J193214.0$+$105932 matches to a number of stars in the field near PSR~B1929$+$10 reported by \citet{Pavlov1996}. The closest match is separated by 1.0\arcsec, whereas the pulsar is separated by 1.2\arcsec. The difference here is within the astrometric uncertainties for RACS-low2, and given the fractional polarization of $\approx +6$\% we consider PSR B1929$+$10 a likely match. We show the cutouts used for visual inspection in Figure~\ref{fig:circ:example:psr2}, which highlights the crowded region and the general complexity in associating astronomical sources to radio components without precise astrometry.

{We also identify PSR~B1737$-$30 to match RACS-low2 sources across four separate SBIDs. This pulsar was not originally matched from the SIMBAD database search due to a $\approx 21\arcsec$ offset from the RACS-low2 position in each SBID image. The coordinates reported in SIMBAD are from \citet{Lower2020}, though other work provides different astrometry for the pulsar \citep[e.g.][]{Fomalont1997,Lower2021} consistent with the RACS-low2 positions within $\approx 1$\arcsec--$2\arcsec$. The ATNF pulsar catalogue also reports the position from \citet{Fomalont1997}, which we assume to be correct so include the pulsar in the final Stokes V catalogue.}

Additional manual pulsar cross-identifications include PSR B1706$-$44 as the likely host of RACS-LOW2 J170942.8$-$442908 instead of a closer nearby star \citep{Chakrabarty1998}. The ATNF pulsar catalogue also reports the pulsar to be closer in this case. We consider PSR J1901$+$0436 to be associated with RACS-LOW2 J190132.4$+$043509 (detected in two SBIDs with the same RACS-low2 source name). This pulsar is $\approx 14$\arcsec\ from the RACS-low2 position which falls outside of the automated cross-match radius of 6.5\arcsec\ for the two images. We also note \citet{Anumarlapudi2023} updated the position for PSR B1353-62, based on observations with Murriyang, CSIRO's Parkes radio telescope, which we now associate with RACS-LOW2 J135655.3$-$623006. Finally, \citet{Ahmad2025} reported a new pulsar detected in an untargeted image-based pulsar search with ASKAP, PSR J2223-0654 (also followed-up with Murriyang), which we associate with RACS-LOW2 J222341.1$-$065441. 

Pulsars tend to be detected with lower fractional circular polarization than radio stars \citep[e.g.][]{Gould1998,Han1998}, and we see the pulsar population largely fill an intermediate fractional polarization space on in Figure~\ref{fig:stokesv_sources} as in \citet{Pritchard2021}, continuing down to the detection and leakage limits. 

\subsubsection{AGN and galaxies}\label{sec:circ:sources:agn}

\begin{figure}[t]
    \centering
    \includegraphics[width=1\linewidth]{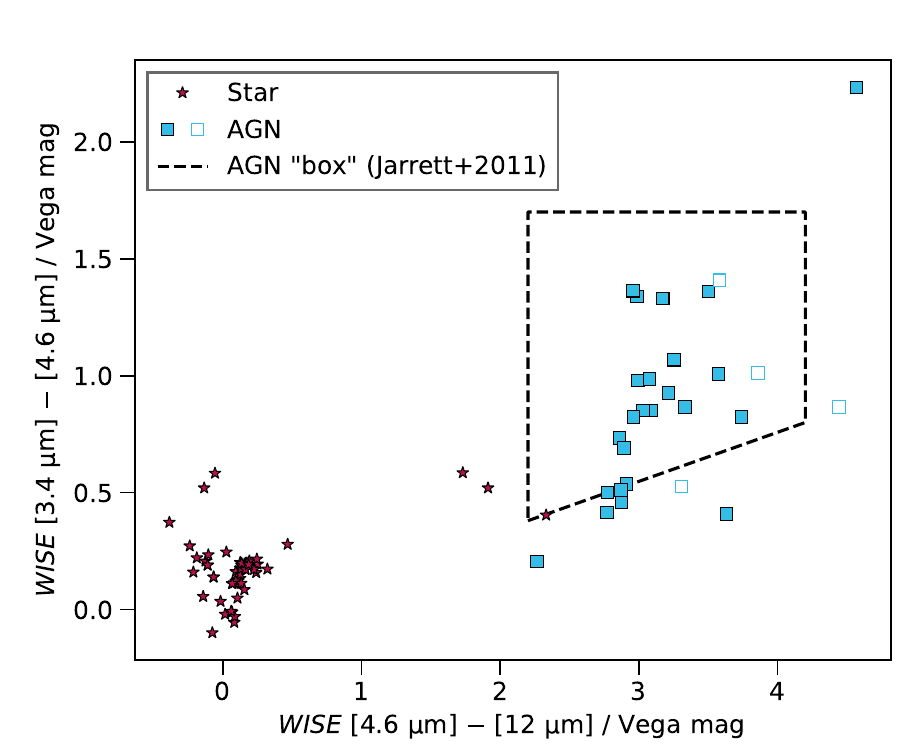}
    \caption{\label{fig:wise} \emph{WISE} colour-colour plot showing the the star and AGN associations in the RACS-low2 circular polarization catalogue. The open markers for some AGN associations are for `probable' AGN. {The dashed region is the empirically-derived location of AGN \citep{Jarrett2011}.}}
\end{figure}

In addition to the original SIMBAD database search, we also query the NASA/IPAC Extragalactic Database \footnote{The NASA/IPAC Extragalactic Database (NED) is operated by the Jet Propulsion Laboratory, California Institute of Technology, under contract with the National Aeronautics and Space Administration.} using the same search radius, and consider any source with a match to an AGN, galaxy, absorptions line system, or quasar classification or if it has an extragalactic redshift to be an AGN. 

We also cross-matched the Stokes V sources to the All\emph{WISE} \footnote{Wide-field Infrared Survey Explorer; \citet{Wright2010}.} catalogue \citep{Cutri2014:AllWISE} using the same cross-matching radius used for other catalogues/databases. {After cross-matching, we take \emph{WISE} magnitudes in the 3.4, 4.6, and 12\,\textmu m bands above signal-to-noise ratios of 2 and plot them on the standard three-band \emph{WISE} colour-colour {plot} \citep[e.g.][]{Jarrett2011,Jarrett2017} in Figure~\ref{fig:wise}. We also show the empirically-derived AGN region from \citet{Jarrett2011}, highlighting the approximate expected location of AGN on the colour-colour plot.} We do not show the pulsar cross-matches as the \emph{WISE} source is unlikely to be associated with the pulsar. We note that some of the stars may not be correctly cross-matched to \emph{WISE} sources resulting in some spurious \emph{WISE} colours in Figure~\ref{fig:wise}. The sources not specifically classified as AGN or stars have similar \emph{WISE} colours to the AGN sample, {and lie within or near the empirical AGN region,} making them likely to be AGN. We consider any sources that are coincident with a previously-detected radio source (but not confirmed as an AGN or similar source in NED or the SIMBAD database) to likely be AGN as well.

Finally, some AGN are recorded with spuriously-high fractional polarization (i.e.\ much greater than the leakage threshold). In these cases, the Stokes I and V measurements are close to the detection thresholds and the Stokes V measurement is likely to be artificially increased due to noise. As the Stokes V measurement is a absolute peak within a box surrounding the Stokes I position, it is sensitive to noise peaks if there is no actual source present. These cases are classified as `AGN'/`AGN?' since the sources are still associated with AGN even if the Stokes V measurement is likely to be incorrect. These sources can be seen clearly in Figure~\ref{fig:circ:distance} and \ref{fig:stokesv_sources} occupying the same regions as stars and pulsars.

In total, we find \AGNUnique\ unique AGN, with \AGNMeasurements\ total measurements, with 27 of the measurements having a positive fractional polarization. It is likely most of the AGN detections that are not spurious noise detections are the result of leakage. We show an example BL Lac, ATPMN J075651.9-634918 (RACS-LOW2 J075651.8$-$634917), in Figure~\ref{fig:circ:example:agn}. We note that AGN with detected circular polarization typically have flat spectra \citep[e.g.][]{Rayner2000,Taylor2024,Callingham2023}, so to further investigate our sample we cross-matched the AGN to the RACS-mid catalogue, obtaining spectral indices between 887.5 and 1367.5\,MHz. The median spectral index for our sample is $\alpha_{887}^{1367} = -0.11 \pm 0.67$, showing a sample of flat-spectrum sources are present. While the large range is inconclusive, it is consistent with there being real circularly polarized AGN in our sample. The leakage threshold derived from bright Stokes I sources as a function of angular separation from the image centres may not adequately capture all aspects of the actual leakage in our data. While we have not taken signed leakage into account (see e.g.\ Figure~\ref{fig:circ:tile}), the absolute leakage model should capture an upper limit to the leakage of Stokes I into V. {As previously noted, we have not assessed leakage of Stokes Q/U into V.} While a specific set of SBIDs from RACS-mid were found to have higher leakage than other SBIDs \citepalias[see Section 2.4.3 in][]{racs-mid2},  we cannot identify other observation-, time-, or position-dependence to the measured $|V/I|$ for bright sources that allows us to further refine the leakage model for the full field images.

\subsubsection{Matches with V-LoTSS}

We also cross-matched the Stokes V sources to V-LoTSS \citep{Callingham2023} after PM correction (if available). Three sources match within $0.5\sqrt{\theta_M\theta_m}$: ILT~J082651.49$+$263720.4 (PSR~B0823$+$26), ILT~J111151.47$+$333213.4 (V* CW UMa; SRSC 00367 in the SRSC), and ILT~J133447.94$+$371056.4 (V* BH CVn; SRSC 00533). We note that a radio star reported in both V-LoTSS and in the SRSC, BD+44 2051B (Figure~\ref{fig:circ:example:BD+44}), has an angular separation of $\approx 28.7\arcsec$ between the RACS and V-LoTSS coordinates after PM correction (cf.\ search radius $\approx 12.1\arcsec$ for this source) so is not considered a match. Given that the RACS positions match to SIMBAD and the SRSC  within $<1\arcsec$ it is likely that it is the same source.

\section{Ongoing work with RACS}

RACS is an ongoing project, and observing and processing of the second mid-band epoch, RACS-mid2, has recently completed. RACS-mid2 uses appropriate holography for each new set of beams formed throughout the observing epoch (a feature that was mostly missing during RACS-mid1 and brings it in line with observing improvements introduced since RACS-mid1). Additional features have also been introduced in the processing pipeline, notably offset field imaging to improve the handling of bright sources just outside of the beam area imaged, and initial phase self-calibration against a RACS-mid1 sky-model to allow basic phase referencing. These new enhancements allowed RACS-mid2 to be processed completely automated, and the resulting RACS-mid2 raw data products have already been deposited and released in CASDA. They are currently in the process of being curated for use in an all-sky data product.

The third low-band epoch, RACS-low3, had also seen initial processing and data deposited onto CASDA through the Observatory's automated processing pipeline. The main motivating factor of RACS-low3 was to acquire that shared the same on-sky beam positions as RACS-mid1/2 and RACS-high. By leveraging consistent beam positions it would be possible to perform multi-frequency synthesis simultaneously across all bands without needing to invoke an $A$-term gridder, significantly reducing the computational footprint. After the initial processing, the uncalibrated RACS-low3 data were reprocessed with an alternative calibration and imaging workflow named \texttt{Flint} \footnote{\url{https://github.com/flint-crew/flint}.} (Galvin et al., in prep.), as a form of stress-testing for the alternate pipeline. 

While carrying out the \texttt{Flint}-based processing a robust set of astrometry corrections were derived on a per-beam basis. The final mosaic outputs from that work will form an initial sky-model capable of aligning data across all three {ASKAP} bands, enabling improved multi-band image synthesis, with the astrometrically-corrected RACS-low3 catalogue now in use by the main ASKAP operational pipeline for phase referencing. 

While the RACS-low3 reprocessing has allowed us to revisit the astrometric accuracy of the RACS data, other work has also explored improvements to RACS astrometry. Namely, \citet{Jaini2025} showed methods to improve the astrometry of the RACS-low1 catalogue and RACS-low3 per-beam source lists, specifically making use of careful cross-matching to infrared sources (detected by \emph{WISE}) to calculate positional offsets. Upcoming work by Jaini et al. (in prep) is also aiming to improve the astrometry for RACS-mid1 and RACS-high per-beam source lists for use in Fast Radio Burst localisation, improving on the methods introduced for their work on the RACS-low3 data. 

Stokes V images and source catalogues for all RACS epochs across the low, mid, and high bands will be described in a future data release paper in this series. These data products will provide an opportunity to study the time-variable and spectral properties of circularly polarised sources across the entire accessible sky, for example allowing comparative studies of the radio star population between the three ASKAP bands.

\section{Data availability}
As with other RACS data releases, all data products are available via the CSIRO ASKAP Science Data Archive (CASDA) or through the CSIRO Data Access Portal (DAP). Individual SBID images and calibrated visibility datasets (along with metadata and auxiliary data files) are available through CASDA. Users can request data through the CASDA observation search for specific SBIDs, the RACS project in general (AS110), or search for ASKAP observations satisfying specific requirements (near coordinates or frequencies). The catalogues and full-sensitivity Stokes I mosaics are also available through CASDA (\url{https://doi.org/10.25919/m8t2-b486}), and the Stokes I catalogue and the Stokes V catalogue will also be made available through VizieR.

\section{Summary}

ASKAP has been used to conduct The Rapid ASKAP Continuum Survey (RACS), observing the whole sky up to $\approx +50\degr$ in 15-min snapshot observations across its frequency range of 700--1800\,MHz. This work presents a second epoch of the survey at 887.5\,MHz, RACS-low2. RACS-low2 makes use of a number of observatory and processing improvements since the first epoch to produce imaging and catalogue data products including both Stokes I and V polarizations. 

After mosaicking neighbouring observations, we present a consolidated Stokes I catalogue that covers the sky up to $\approx +48\degr$. The images have a median rms noise of $195_{-32}^{+48}$\,\textmu Jy\,PSF$^{-1}$, and an angular resolution that largely varies as a function of declination, with a median of $15.2\arcsec \times 13.0\arcsec$. We report 3\,922\,151 sources in the Stokes I catalogue, and find that the brightness scale is accurate to $\approx 7$\% after comparison to external surveys and known calibrator sources. As with other published RACS data, the astrometric uncertainty remains at $\approx 1$--$2\arcsec$ when comparing to radio surveys with high astrometric precision, and we find an increase in uncertainty at high declination even after some attempts to model the declination-dependent offset in catalogue space. 

We also produce a catalogue of Stokes V sources based on Stokes I positions, after source-finding on individual images prior to mosaicking neighbouring fields. We estimate the residual widefield Stokes I leakage into Stokes V using bright Stokes I sources, finding a position-dependent threshold ranging from $\approx 0.4$--$12$\% that is used to help define real circularly polarized sources. We obtain a list of \NsourcesV\ sources with significant Stokes V emission. After automated and manual cross-matching to external catalogues and databases, we find that \StarMeasurements\ sources match to stars, \PulsarMeasurments\ match to pulsars, and \AGNMeasurements\ match to AGN, with much of the AGN cross-matches likely to residual leakage, and with some sources detected multiple times in repeated/overlapping observations.

The data products, including calibrated visibilities, images, and catalogues, are made publicly available through CASDA. 

\begin{acknowledgement}
{The authors would like to thank the referee for their comments and suggestions that have improved the manuscript. SWD would also like to thank Natasha Hurley-Walker, Marcus Lower, and Akash Anumarlapudi for answering pulsar-related questions.}
This scientific work uses data obtained from Inyarrimanha Ilgari Bundara / the Murchison Radio-astronomy Observatory. We acknowledge the Wajarri Yamaji People as the Traditional Owners and native title holders of the Observatory site. CSIRO’s ASKAP radio telescope is part of the Australia Telescope National Facility (\url{https://ror.org/05qajvd42}). Operation of ASKAP is funded by the Australian Government with support from the National Collaborative Research Infrastructure Strategy. ASKAP uses the resources of the Pawsey Supercomputing Research Centre. Establishment of ASKAP, Inyarrimanha Ilgari Bundara, the CSIRO Murchison Radio-astronomy Observatory and the Pawsey Supercomputing Research Centre are initiatives of the Australian Government, with support from the Government of Western Australia and the Science and Industry Endowment Fund. This work was supported by resources provided by the Pawsey Supercomputing Research Centre with funding from the Australian Government and the Government of Western Australia. This project was supported by resources and expertise provided by CSIRO IMT Scientific Computing.

This research has made use of the NASA/IPAC Extragalactic Database (NED), which is operated by the Jet Propulsion Laboratory, California Institute of Technology, under contract with the National Aeronautics and Space Administration.

This publication makes use of data products from the Wide-field Infrared Survey Explorer, which is a joint project of the University of California, Los Angeles, and the Jet Propulsion Laboratory/California Institute of Technology, funded by the National Aeronautics and Space Administration. 

The Digitized Sky Surveys were produced at the Space Telescope Science Institute under U.S. Government grant NAG W-2166. The images of these surveys are based on photographic data obtained using the Oschin Schmidt Telescope on Palomar Mountain and the UK Schmidt Telescope. The plates were processed into the present compressed digital form with the permission of these institutions. The Second Palomar Observatory Sky Survey (POSS-II) was made by the California Institute of Technology with funds from the National Science Foundation, the National Geographic Society, the Sloan Foundation, the Samuel Oschin Foundation, and the Eastman Kodak Corporation. The Oschin Schmidt Telescope is operated by the California Institute of Technology and Palomar Observatory. The UK Schmidt Telescope was operated by the Royal Observatory Edinburgh, with funding from the UK Science and Engineering Research Council (later the UK Particle Physics and Astronomy Research Council), until 1988 June, and thereafter by the Anglo-Australian Observatory. The blue plates of the southern Sky Atlas and its Equatorial Extension (together known as the SERC-J), as well as the Equatorial Red (ER), and the Second Epoch [red] Survey (SES) were all taken with the UK Schmidt.

This work made use of a number of \texttt{python} packages, including \texttt{astropy} \citep{astropy:2018}, \texttt{mat\-plot\-lib} \citep{Hunter2007}, \texttt{numpy} \citep{numpy2020}, \texttt{scipy} \citep{scipy}, \texttt{astroquery} \citep{astroquery}, and \texttt{cmasher} \citep{cmasher}.  Some of the results in this paper have been derived using the \texttt{healpy} \citep{healpy} and HEALPix package. We make use of \texttt{ds9} \citep{ds9} and \texttt{topcat} \citep{topcat} for visualisation, as well as the ``Aladin sky atlas'' developed at CDS, Strasbourg Observatory, France \citep{aladin1,aladin2} for obtaining catalogue data. This research has made use of the SIMBAD database, operated at CDS, Strasbourg, France.

\end{acknowledgement}


\bibliography{bib}

\appendix

\section{Subtracted sources}\label{app:subtracted}

\begin{figure}[t]
    \centering
    \includegraphics[width=1\linewidth]{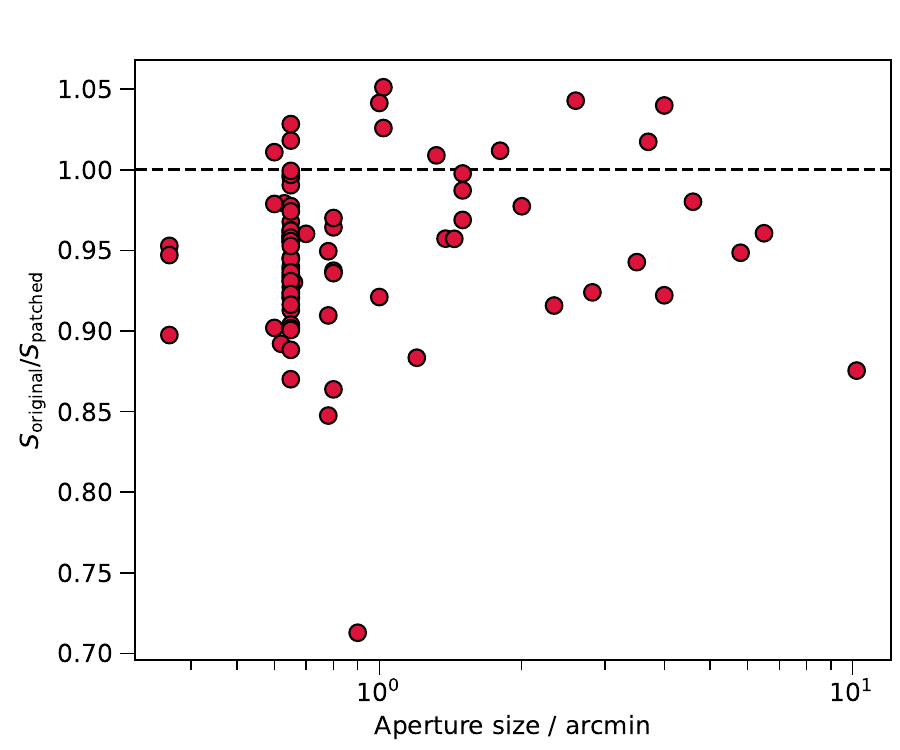}
    \caption{\label{fig:patched_comparison} The flux density ratio of sources listed in Table~\ref{tab:subtracted} in the original images (on CASDA) compared to the consolidated catalogue (after patching), as a function of the aperture size used for subtraction. Note the clustering at 0.65\,arcmin is due to that being the default aperture size for a usual point source.}
\end{figure}

As noted in Section~\ref{sec:patching}, bright sources that were subtracted or peeled were sometimes over-subtracted from PAF beam datasets when within the field of view of the image. This resulted in a marginal reduction in brightness after mosaicking of adjacent PAF beam images. Table~\ref{tab:subtracted} lists all sources that were included in the subtraction process, though not all sources were necessarily reduced in brightness. In Figure~\ref{fig:patched_comparison} we show the flux density difference between the original and patched sources after running \texttt{PyBDSF} on each image, noting that the original values are what would be obtained when using the images on CASDA, and the patched values are those present in the consolidated Stokes I catalogue.

\begin{table}[p]
\caption{The list of subtracted sources, with the specific coordinates and radii used to define the apertures within which the sources were subtracted. \label{tab:subtracted}}
\begin{tabular}{llll}\toprule
Source    & $\alpha_\text{J2000}$      &  $\delta_\text{J2000}$    & Radius \\
                &(hh:mm:ss.ss)& (dd:mm:ss.s) & (arcmin)       \\\midrule
PKS B0003$-$003 & 00:06:22.59 & -00:04:24.7 & 0.65           \\
PKS B0023$-$263 & 00:25:49.16 & -26:02:12.6 & 0.60           \\
3C 17           & 00:38:20.37 & -02:07:42.7 & 1.50           \\
PKS B0039$-$445 & 00:42:08.98 & -44:14:00.5 & 0.65           \\
PKS B0043$-$424 & 00:46:17.60 & -42:07:55.8 & 2.60           \\
3C 29           & 00:57:35.44 & -01:24:03.3 & 4.00           \\
3C 33           & 01:08:52.99 & +13:20:08.7 & 4.60           \\
3C 43           & 01:29:59.75 & +23:38:20.3 & 1.00           \\
3C 48           & 01:37:41.00 & +33:09:34.0 & 0.60           \\
PKS B0237$-$233 & 02:40:08.17 & -23:09:15.7 & 0.65           \\
M 77            & 02:42:40.68 & -00:00:47.9 & 1.50           \\
3C 78           & 03:08:26.48 & +04:06:38.6 & 4.00           \\
PKS B0316$+$162 & 03:18:57.80 & +16:28:32.7 & 0.80           \\
3C 84           & 03:19:48.00 & +41:30:42.0 & 1.44           \\
3C 98           & 03:58:53.76 & +10:25:47.4 & 6.50           \\
3C 105          & 04:07:19.56 & +03:42:05.6 & 5.80           \\
3C 103          & 04:08:03.34 & +43:00:24.4 & 0.65           \\
PKS B0407$-$658 & 04:08:20.00 & -65:45:09.0 & 0.36           \\
PKS B0409$-$752 & 04:08:48.00 & -75:07:19.0 & 0.78           \\
3C 111          & 04:18:21.00 & +38:01:52.0 & 2.82           \\
PKS B0420$-$625 & 04:20:56.06 & -62:23:39.7 & 0.65           \\
3C 119          & 04:32:36.00 & +41:38:26.0 & 1.02           \\
3C 123          & 04:37:04.00 & +29:40:15.0 & 0.78           \\
PKS B0458$-$020 & 05:01:12.81 & -01:59:14.3 & 0.65           \\
3C 133          & 05:02:58.48 & +25:16:25.1 & 0.65           \\
Pictor A        & 05:19:48.00 & -45:45:53.0 & 6.30           \\
3C 138          & 05:21:09.89 & +16:38:22.1 & 0.65           \\
Taurus A        & 05:34:31.00 & +22:00:59.0 & 4.20           \\
PKS B0531$+$194 & 05:34:44.51 & +19:27:21.5 & 0.65           \\
M 42            & 05:35:18.00 & -05:23:11.0 & 4.50           \\
Orion B         & 05:41:41.00 & -01:53:58.0 & 4.95           \\
PKS B0605$-$063 & 06:07:45.96 & -06:23:00.3 & 0.65           \\
3C 154          & 06:13:50.43 & +26:04:36.1 & 1.50           \\
3C 161          & 06:27:09.86 & -05:53:08.4 & 0.80           \\
4C $+$10.20     & 06:32:15.30 & +10:22:02.2 & 0.65           \\
PKS B0637$-$752 & 06:35:46.51 & -75:16:16.8 & 0.65           \\
PKS B0723$-$008 & 07:25:50.00 & -00:54:55.0 & 0.36           \\
PKS B0741$-$063 & 07:44:21.66 & -06:29:35.9 & 0.65           \\
3C 196          & 08:13:36.03 & +48:13:02.6 & 1.00           \\
4C $+$37.24     & 08:31:10.02 & +37:42:09.6 & 0.65           \\
PKS B0842$-$754 & 08:41:27.03 & -75:40:27.9 & 0.65           \\
3C 216          & 09:09:33.50 & +42:53:46.5 & 0.65           \\
Hydra A         & 09:18:05.00 & -12:05:42.0 & 1.20           \\
3C 219          & 09:21:08.62 & +45:38:57.4 & 3.70           \\
3C 237          & 10:08:00.03 & +07:30:16.3 & 0.65           \\
PKS B1151$-$348 & 11:54:21.79 & -35:05:29.1 & 0.65           \\\bottomrule
\end{tabular}
\end{table}

\setcounter{table}{3}
\begin{table}[p]
\caption{\emph{Continued}.}
\begin{tabular}{llll}\toprule
Source    & $\alpha_\text{J2000}$      &  $\delta_\text{J2000}$    & Radius \\
                &(hh:mm:ss.ss)& (dd:mm:ss.s) & (arcmin)       \\\midrule
3C 273          & 12:29:06.29 & +02:03:01.6 & 1.00           \\
Virgo A         & 12:30:49.00 & +12:23:28.0 & 10.20          \\
3C 275          & 12:42:19.66 & -04:46:20.7 & 0.65           \\
3C 279          & 12:56:11.17 & -05:47:21.5 & 0.65           \\
3C 280          & 12:56:57.00 & +47:20:19.9 & 0.80           \\
3C 283          & 13:11:40.10 & -22:17:04.4 & 0.65           \\
Centaurus A     & 13:25:28.00 & -43:00:40.0 & 6.60           \\
3C 287          & 13:30:37.69 & +25:09:10.9 & 0.80           \\
3C 286          & 13:31:08.51 & +30:30:38.5 & 0.65           \\
PKS B1345$+$125 & 13:47:33.36 & +12:17:24.2 & 0.65           \\
3C 293          & 13:52:18.29 & +31:26:40.2 & 3.50           \\
3C 298          & 14:19:08.18 & +06:28:34.8 & 0.66           \\
PKS B1453$-$109 & 14:55:54.87 & -11:08:42.8 & 0.65           \\
PKS B1508$-$055 & 15:10:53.59 & -05:43:07.4 & 0.65           \\
PKS B1510$-$089 & 15:12:50.53 & -09:05:59.8 & 0.65           \\
PKS B1524$-$136 & 15:26:59.44 & -13:51:00.2 & 0.65           \\
B2 1611$+$34    & 16:13:41.06 & +34:12:47.9 & 0.80           \\
3C 345          & 16:42:58.00 & +39:48:32.0 & 1.02           \\
Hercules A      & 16:51:08.00 & +04:59:33.0 & 2.34           \\
3C 349          & 16:59:28.62 & +47:02:48.7 & 2.80           \\
PKS B1730$-$130 & 17:33:02.71 & -13:04:49.5 & 0.65           \\
PKS B1814$-$637 & 18:19:34.90 & -63:45:49.0 & 0.80           \\
4C $+$39.56     & 18:21:20.86 & +39:42:44.7 & 0.65           \\
3C 380          & 18:29:31.00 & +48:44:46.0 & 1.02           \\
PKS B1830$-$210 & 18:33:39.89 & -21:03:39.8 & 0.65           \\
3C 388          & 18:44:02.72 & +45:33:27.3 & 2.00           \\
PKS B1859$-$235 & 19:02:49.28 & -23:29:51.3 & 0.65           \\
PKS B1921$-$293 & 19:24:51.06 & -29:14:30.1 & 0.65           \\
PKS B1932$-$464 & 19:35:56.00 & -46:20:39.0 & 0.63           \\
PKS B1934$-$638 & 19:39:25.00 & -63:42:45.0 & 0.78           \\
Cygnus A        & 19:59:28.00 & +40:44:02.0 & 1.80           \\
3C 409          & 20:14:27.00 & +23:34:53.0 & 0.62           \\
3C 410          & 20:20:06.56 & +29:42:14.2 & 0.65           \\
PKS B2032$-$350 & 20:35:47.86 & -34:54:09.2 & 0.60           \\
B2 2050$+$36    & 20:52:52.05 & +36:35:35.3 & 0.65           \\
3C 433          & 21:23:44.00 & +25:04:14.0 & 1.32           \\
3C 438          & 21:55:52.29 & +38:00:29.6 & 0.65           \\
PKS B2152$-$699 & 21:57:07.00 & -69:41:15.0 & 1.38           \\
PKS B2155$-$152 & 21:58:06.15 & -15:01:08.1 & 0.65           \\
PKS B2203$-$188 & 22:06:10.42 & -18:35:38.7 & 0.70           \\
PKS B2216$-$038 & 22:18:52.04 & -03:35:36.9 & 0.65           \\
3C 446          & 22:25:47.00 & -04:57:01.0 & 0.36           \\
PKS B2230$+$114 & 22:32:36.41 & +11:43:50.9 & 0.65           \\
3C 454          & 22:53:57.00 & +16:08:53.0 & 0.90           \\
PKS B2323$-$407 & 23:26:34.13 & -40:27:15.0 & 0.65           \\
PKS B2326$-$477 & 23:29:17.70 & -47:30:19.1 & 0.65           \\
PKS B2331$-$417 & 23:34:26.19 & -41:25:24.3 & 0.65           \\
PKS B2356$-$611 & 23:59:02.00 & -60:54:52.0 & 4.50           \\\bottomrule
\end{tabular}
\end{table}

\section{Catalogue columns}\label{app:columns}
The RACS-low2 Stokes I source and component catalogues follow the same structure as the equivalent RACS-high catalogues. Table~\ref{tab:columns:source} details the columns of the Stokes I source catalogue and Table~\ref{tab:columns:component} shows the same for the Stokes I component catalogue. Table~\ref{tab:columns:stokesv} details the columns in the Stokes V source catalogue, which mirrors the Stokes I source catalogue with additional columns pertaining to individual observations (SBID and observation date) along with Stokes V properties and source associations.  Note that `beam$^{-1}$' units are equivalent to `PSF$^{-1}$' used elsewhere in this paper to be consistent with the representation of the units in the FITS tables.

\begin{table*}
    \centering
    \caption{\label{tab:columns:source} Columns in the RACS-low2 Stokes I source catalogue.}
    \begin{adjustbox}{max width=\textwidth}
    \begin{tabular}{l l c l}\toprule
        Index & Name & Unit & Description \\\midrule
        0 & \texttt{Name} &  & Source name with the convention: \texttt{RACS-LOW2 JHHMMSS.S$\pm$DDMMSS} \\
        1 & \texttt{Source\_ID} &  & \makecell[l]{Unique identifier for the source based on the \texttt{Field\_ID} (if drawn from a full-sensitivity mosaic) \\ or the SBID for sources that were patched from individual PAF beam images (Section~\ref{sec:patching}).} \\
        2 & \texttt{Field\_ID} & & Field name: \texttt{RACS\_HHMM$\pm$DD}. \\
        3 & \texttt{RA} & {deg} & J2000 right ascension of the source. \\
        4 & \texttt{Dec} & {deg} & J2000 declination of the source. \\
        5 & \texttt{Dec\_corr} & {deg} & J2000 declination, with declination-dependent offset correction (Section~\ref{sec:astrometry}). \\
        6 & \texttt{E\_RA} & {deg} & Uncertainty on \texttt{RA} from fitting the source position. \\
        7 & \texttt{E\_Dec} & {deg} & Uncertainty on \texttt{Dec} from fitting the source position. \\
        8 & \texttt{E\_Dec\_corr} & {deg} & Uncertainty on \texttt{Dec\_corr} from fitting the source position and declination offset model. \\
        9 & \texttt{Total\_flux} & mJy & Total flux density of the source. \\
        10 & \texttt{E\_Total\_flux\_PyBDSF} & mJy & Uncertainty in the total flux density from PyBDSF fitting. \\
        11 & \texttt{E\_Total\_flux} & mJy & Quadrature sum of the brightness scale and PyBDSF uncertainties for total flux density. \\
        12 & \texttt{Peak\_flux} & mJy\,beam$^{-1}$ & Peak flux density of the source. \\
        13 & \texttt{E\_Peak\_flux\_PyBDSF} & mJy\,beam$^{-1}$ & Uncertainty in the peak flux density from PyBDSF fitting. \\
        14 & \texttt{E\_Peak\_flux} & mJy\,beam$^{-1}$ &  Quadrature sum of the brightness scale and PyBDSF uncertainties for peak flux density. \\
        15 & \texttt{Maj\_axis} & arcsec & Size of the major axis of the source. \\
        16 & \texttt{Min\_axis} & arcsec & Size of the minor axis of the source. \\
        17 & \texttt{PA} & deg & Position angle of the source, east of north. \\
        18 & \texttt{E\_Maj\_axis} & arcsec & Uncertainty in the source major axis. \\
        19 & \texttt{E\_Min\_axis} & arcsec & Uncertainty in the source minor axis. \\
        20 & \texttt{E\_PA} & deg & Uncertainty in the source PA. \\
        21 & \texttt{DC\_Maj\_axis} & arcsec & Deconvolved size of the major axis of the source. \\
        22 & \texttt{DC\_Min\_axis} & arcsec & Deconvolved size of the minor axis of the source. \\
        23 & \texttt{DC\_PA} & deg & Deconvolved position angle of the source, east of north. \\
        24 & \texttt{E\_DC\_Maj\_axis} & arcsec & Uncertainty in the deconvolved source major axis. \\
        25 & \texttt{E\_DC\_Min\_axis} & arcsec & Uncertainty in the deconvolved source minor axis. \\
        26 & \texttt{E\_DC\_PA} & deg & Uncertainty in the deconvolved source PA. \\
        27 & \texttt{Noise} & mJy\,beam$^{-1}$ & Local estimate of the rms noise. \\
        28 & \texttt{Tile\_l} & deg & Direction cosine $l$ of the source with respect to the field centre. \\
        29 & \texttt{Tile\_m} & deg & Direction cosine $m$ of the source with respect to the field centre. \\
        30 & \texttt{Tile\_sep} & deg & Angular distance from the field centre. \\
        31 & \texttt{PSF\_Maj} & arcsec & FWHM of the PSF major axis of the field. \\
        32 & \texttt{PSF\_Min} & arcsec & FWHM of the PSF minor axis of the field. \\
        33 & \texttt{PSF\_PA} & deg & PA of the PSF of the field. \\
        34 & \texttt{S\_Code} & & \makecell[l]{Source structure classification provided by PyBDSF. Note this is not modified when regrouping \\ source components so represents the original code assigned during source finding.} \\
        35 & \texttt{N\_Gaussians} & & Number of Gaussian components comprising the source. \\
        36 & \texttt{Flag} & &  Source type flag. See Section~\ref{sec:resolved}. \\
         \bottomrule
    \end{tabular}
    \end{adjustbox}
\end{table*}

\begin{table*}
    \centering
    \begin{adjustbox}{max width=\textwidth}
    \caption{\label{tab:columns:component} Columns in the RACS-low2 Stokes I component catalogue.}
    \begin{tabular}{l l c l}\toprule
        Index & Name & Unit & Description \\\midrule
        0 & \texttt{Gaussian\_ID} &  & \makecell[l]{Unique identifier for the component based on the \texttt{Field\_ID} (if drawn from a full-sensitivity mosaic) \\ or the SBID for sources that were patched from individual PAF beam images (Section~\ref{sec:patching}).} \\
        1 & \texttt{Source\_ID} &  & Unique identifier for the source the component belongs to (see Table~\ref{tab:columns:source}). \\
        2 & \texttt{Field\_ID} & & Field name: \texttt{RACS\_HHMM$\pm$DD}. \\
        3 & \texttt{RA} & {deg} & J2000 right ascension of the component. \\
        4 & \texttt{Dec} & {deg} & J2000 declination of the component. \\
        5 & \texttt{Dec\_corr} & {deg} & J2000 declination, with declination-dependent offset correction (Section~\ref{sec:astrometry}). \\
        6 & \texttt{E\_RA} & {deg} & Uncertainty on \texttt{RA} from fitting the component position. \\
        7 & \texttt{E\_Dec} & {deg} & Uncertainty on \texttt{DEC} from fitting the component position. \\
        8 & \texttt{E\_Dec\_corr} & {deg} & Uncertainty on \texttt{Dec\_corr} from fitting the component position and declination offset model. \\
        9 & \texttt{Total\_flux} & mJy & Total flux density of the component. \\
        10 & \texttt{E\_Total\_flux\_PyBDSF} & mJy & Uncertainty in the total flux density from PyBDSF fitting. \\
        11 & \texttt{E\_Total\_flux} & mJy & Quadrature sum of the brightness scale and PyBDSF uncertainties for total flux density. \\
        12 & \texttt{Peak\_flux} & mJy\,beam$^{-1}$ & Peak flux density of the component. \\
        13 & \texttt{E\_Peak\_flux\_PyBDSF} & mJy\,beam$^{-1}$ & Uncertainty in the peak flux density from PyBDSF fitting. \\
        14 & \texttt{E\_Peak\_flux} & mJy\,beam$^{-1}$ &  Quadrature sum of the brightness scale and PyBDSF uncertainties for peak flux density. \\
        15 & \texttt{Maj\_axis} & arcsec & Size of the major axis of the component. \\
        16 & \texttt{Min\_axis} & arcsec & Size of the minor axis of the component. \\
        17 & \texttt{PA} & deg & Position angle of the component, east of north. \\
        18 & \texttt{E\_Maj\_axis} & arcsec & Uncertainty in the component major axis. \\
        19 & \texttt{E\_Min\_axis} & arcsec & Uncertainty in the component minor axis. \\
        20 & \texttt{E\_PA} & deg & Uncertainty in the component PA. \\
        21 & \texttt{DC\_Maj\_axis} & arcsec & Deconvolved size of the major axis of the component. \\
        22 & \texttt{DC\_Min\_axis} & arcsec & Deconvolved size of the minor axis of the component. \\
        23 & \texttt{DC\_PA} & deg & Deconvolved position angle of the component, east of north. \\
        24 & \texttt{E\_DC\_Maj\_axis} & arcsec & Uncertainty in the deconvolved component major axis. \\
        25 & \texttt{E\_DC\_Min\_axis} & arcsec & Uncertainty in the deconvolved component minor axis. \\
        26 & \texttt{E\_DC\_PA} & deg & Uncertainty in the deconvolved component PA. \\
        27 & \texttt{Noise} & mJy\,beam$^{-1}$ & Local estimate of the rms noise. \\
        28 & \texttt{Tile\_l} & deg & Direction cosine $l$ of the component with respect to the field centre. \\
        29 & \texttt{Tile\_m} & deg & Direction cosine $m$ of the component with respect to the field centre. \\
        30 & \texttt{Tile\_sep} & deg & Angular distance from the field centre. \\
        31 & \texttt{PSF\_Maj} & arcsec & FWHM of the PSF major axis of the field. \\
        32 & \texttt{PSF\_Min} & arcsec & FWHM of the PSF minor axis of the field. \\
        33 & \texttt{PSF\_PA} & deg & PA of the PSF of the field. \\
        34 & \texttt{S\_Code} & & \makecell[l]{Host source structure classification provided by PyBDSF. Note this is not modified when \\ regrouping source components so represents the original code assigned during source finding.} \\
        35 & \texttt{Flag} & &  Host source type flag. See Section~\ref{sec:resolved}. \\
         \bottomrule
    \end{tabular}
    \end{adjustbox}
\end{table*}

\begin{table*}
    \centering
    \caption{\label{tab:columns:stokesv} Columns in the RACS-low2 Stokes V source catalogue.}
    \begin{adjustbox}{max width=\textwidth}
    \begin{tabular}{l l c l}\toprule
        Index & Name & Unit & Description \\\midrule
        0 & \texttt{Name} &  & Source name with the convention: \texttt{RACS-LOW2 JHHMMSS.S$\pm$DDMMSS} \\
        1 & \texttt{Source\_ID} &  & Unique identifier for the source based on the \texttt{Field\_ID}. \\
        2 & \texttt{Field\_ID} & & Field name: \texttt{RACS\_HHMM$\pm$DD}. \\
        3 & \texttt{RA} & {deg} & J2000 right ascension of the source. \\
        4 & \texttt{Dec} & {deg} & J2000 declination of the source. \\
        5 & \texttt{Dec\_corr} & {deg} & J2000 declination, with declination-dependent offset correction (Section~\ref{sec:astrometry}). \\
        6 & \texttt{E\_RA} & {deg} & Uncertainty on \texttt{RA} from fitting the source position. \\
        7 & \texttt{E\_Dec} & {deg} & Uncertainty on \texttt{Dec} from fitting the source position. \\
        8 & \texttt{E\_Dec\_corr} & {deg} & Uncertainty on \texttt{Dec\_corr} from fitting the source position and declination offset model. \\
        9 & \texttt{Total\_flux} & mJy & Total Stokes I flux density of the source. \\
        10 & \texttt{E\_Total\_flux\_PyBDSF} & mJy & Uncertainty in the total Stokes I flux density from PyBDSF fitting. \\
        11 & \texttt{E\_Total\_flux} & mJy & Quadrature sum of the brightness scale and PyBDSF uncertainties for the total Stokes I flux density. \\
        12 & \texttt{Peak\_flux} & mJy\,beam$^{-1}$ & Peak Stokes I flux density of the source. \\
        13 & \texttt{E\_Peak\_flux\_PyBDSF} & mJy\,beam$^{-1}$ & Uncertainty in the peak Stokes I flux density from PyBDSF fitting. \\
        14 & \texttt{E\_Peak\_flux} & mJy\,beam$^{-1}$ &  Quadrature sum of the brightness scale and PyBDSF uncertainties for the peak Stokes I flux density. \\
        15 & \texttt{Maj\_axis} & arcsec & Size of the major axis of the source. \\
        16 & \texttt{Min\_axis} & arcsec & Size of the minor axis of the source. \\
        17 & \texttt{PA} & deg & Position angle of the source, east of north. \\
        18 & \texttt{E\_Maj\_axis} & arcsec & Uncertainty in the source major axis. \\
        19 & \texttt{E\_Min\_axis} & arcsec & Uncertainty in the source minor axis. \\
        20 & \texttt{E\_PA} & deg & Uncertainty in the source PA. \\
        21 & \texttt{DC\_Maj\_axis} & arcsec & Deconvolved size of the major axis of the source. \\
        22 & \texttt{DC\_Min\_axis} & arcsec & Deconvolved size of the minor axis of the source. \\
        23 & \texttt{DC\_PA} & deg & Deconvolved position angle of the source, east of north. \\
        24 & \texttt{E\_DC\_Maj\_axis} & arcsec & Uncertainty in the deconvolved source major axis. \\
        25 & \texttt{E\_DC\_Min\_axis} & arcsec & Uncertainty in the deconvolved source minor axis. \\
        26 & \texttt{E\_DC\_PA} & deg & Uncertainty in the deconvolved source PA. \\
        27 & \texttt{Noise} & mJy\,beam$^{-1}$ & Local estimate of the rms noise. \\
        28 & \texttt{Tile\_l} & deg & Direction cosine $l$ of the source with respect to the field centre. \\
        29 & \texttt{Tile\_m} & deg & Direction cosine $m$ of the source with respect to the field centre. \\
        30 & \texttt{Tile\_sep} & deg & Angular distance from the field centre. \\
        31 & \texttt{Gal\_lon} & deg & Galactic longitude of the source. \\
        32 & \texttt{Gal\_lat} & deg & Galactic latitude of the source. \\
        33 & \texttt{PSF\_Maj} & arcsec & FWHM of the PSF major axis of the field. \\
        34 & \texttt{PSF\_Min} & arcsec & FWHM of the PSF minor axis of the field. \\
        35 & \texttt{PSF\_PA} & deg & PA of the PSF of the field. \\
        36 & \texttt{S\_Code} & & Source structure classification provided by PyBDSF. \\
        37 & \texttt{N\_Gaussians} & & Number of Gaussian components comprising the source. \\
        38 & \texttt{Flag} & &  Source type flag. See Section~\ref{sec:resolved}. \\
        39 & \texttt{Scan\_start\_MJD} & yr & MJD of the observation start time for the SBID. \\
        40 & \texttt{Scan\_length} & s & Observation length. \\
        41 & \texttt{SBID} & & Scheduling block ID. \\
        42 & \texttt{Peak\_flux\_V} &  mJy\,beam$^{-1}$ & Stokes V peak flux density of the source. \\
        43 & \texttt{Noise\_V} & mJy\,beam$^{-1}$ & Stokes V local rms noise estimate, also used as an uncertainty in the Stokes V peak flux density measurement. \\
        44 & \texttt{Peak\_VI} & \% & Fractional circular polarization based on peak flux densities as a percentage. \\
        45 & \texttt{E\_Peak\_VI} & \% & Uncertainty in the fractional circular polarization as a percentage. \\
        46 & \texttt{E\_Peak\_VI\_upper} & \% & Upper uncertainty in the absolute fractional circular polarization as a percentage. \\
        47 & \texttt{E\_Peak\_VI\_lower} & \% & Lower uncertainty in the absolute fractional circular polarization as a percentage.\\
        48 & \texttt{Leakage} & \% & Estimate of the absolute Stokes I leakage into V as a percentage. \\
        49 & \texttt{Source\_Association} & & \makecell[l]{Astrophysical source associated with the Stokes V source from automated and \\ manual cross-matching and identification (Section~\ref{sec:circ:sources}).} \\
        50 & \texttt{Source\_Separation} & arcsec & Angular separation from astrophysical source after correcting for proper motion where relevant. \\
        51 & \texttt{Source\_Type} & & \makecell[l]{Basic source type. One of `star' (Section~\ref{sec:circ:sources:stars}), `pulsar' (Section~\ref{sec:circ:sources:pulsars}), `AGN' (Section~\ref{sec:circ:sources:agn}), and `unclassified' \\ where we cannot assign a likely source type. '?' indicates a probable type.}\\
        52 & \texttt{Search\_Radius} & arcsec & Initial database search radius for automated searches.\\
         \bottomrule
    \end{tabular}
    \end{adjustbox}
\end{table*}

\section{{Source counts data}}

{Table~\ref{tab:counts} reports the RACS-low2 source counts scaled to 1.4\,GHz assuming $\alpha=-0.83$, in raw form ($N$), with Euclidean normalisation, and with with corrections applied for incompleteness as a function of declination and flux density bin. Figure~\ref{fig:counts} shows both the uncorrected and corrected normalised counts.}

\begin{table*}[t!p]
    \centering
    \caption{Tabulated source counts for the RACS-low2 catalogue.\label{tab:counts}}
    \begin{adjustbox}{max width=\textwidth}
    \begin{tabular}{l c c c c}\toprule
    $S_\text{1.4\,GHz}$ & $S_\text{1.4\,GHz,centre}$ & $N$ & Normalised $N$ & Normalised, corrected $N$ \\
    (mJy) & (mJy) & & (Jy$^{1.5}$\,sr$^{-1}$) & (Jy$^{1.5}$\,sr$^{-1}$) \\\midrule
$0.82$--$1.1$ & $0.96$ & $237\,730 \pm 490$ &$2.3824 \pm 0.0049$ &$9.744 \pm 0.02$ \\
$1.1$--$1.5$ & $1.3$ & $432\,140 \pm 660$ &$6.823 \pm 0.01$ &$12.978 \pm 0.02$ \\
$1.5$--$2.0$ & $1.8$ & $495\,100 \pm 700$ &$12.314 \pm 0.018$ &$15.112 \pm 0.021$ \\
$2.0$--$2.7$ & $2.4$ & $458\,750 \pm 680$ &$17.976 \pm 0.027$ &$18.636 \pm 0.028$ \\
$2.7$--$3.7$ & $3.2$ & $378\,810 \pm 620$ &$23.385 \pm 0.038$ &$22.987 \pm 0.037$ \\
$3.7$--$5.0$ & $4.4$ & $300\,420 \pm 550$ &$29.217 \pm 0.053$ &$32.299 \pm 0.059$ \\
$5.0$--$6.8$ & $5.9$ & $242\,200 \pm 490$ &$37.108 \pm 0.075$ &$38.49 \pm 0.078$ \\
$6.8$--$9.2$ & $8.0$ & $197\,180 \pm 440$ &$47.6 \pm 0.11$ &$47.56 \pm 0.11$ \\
$9.2$--$12$ & $11$ & $163\,660 \pm 400$ &$62.23 \pm 0.15$ &$65.84 \pm 0.16$ \\
$12$--$17$ & $15$ & $135\,410 \pm 370$ &$81.12 \pm 0.22$ &$81.88 \pm 0.22$ \\
$17$--$23$ & $20.$ & $110\,600 \pm 330$ &$104.38 \pm 0.31$ &$107.71 \pm 0.32$ \\
$23$--$31$ & $27$ & $89\,490 \pm 300$ &$133.06 \pm 0.44$ &$133.15 \pm 0.45$ \\
$31$--$42$ & $36$ & $72\,280 \pm 270$ &$169.32 \pm 0.63$ &$173.93 \pm 0.65$ \\
$42$--$57$ & $49$ & $57\,430 \pm 240$ &$211.93 \pm 0.88$ &$219.61 \pm 0.92$ \\
$57$--$77$ & $67$ & $44\,590 \pm 210$ &$259.2 \pm 1.2$ &$267.4 \pm 1.3$ \\
$77$--$100$ & $90.$ & $34\,570 \pm 190$ &$316.6 \pm 1.7$ &$320.6 \pm 1.7$ \\
$100$--$140$ & $120$ & $26\,340 \pm 160$ &$380.0 \pm 2.3$ &$375.3 \pm 2.3$ \\
$140$--$190$ & $170$ & $19\,440 \pm 140$ &$441.8 \pm 3.2$ &$438.8 \pm 3.1$ \\
$190$--$260$ & $220$ & $13\,990 \pm 120$ &$501.2 \pm 4.2$ &$505.8 \pm 4.3$ \\
$260$--$350$ & $300$ & $9\,876 \pm 99$ &$557.2 \pm 5.6$ &$536.6 \pm 5.4$ \\
$350$--$470$ & $410$ & $6\,758 \pm 82$ &$600.7 \pm 7.3$ &$574.6 \pm 7.0$ \\
$470$--$640$ & $560$ & $4\,427 \pm 67$ &$619.9 \pm 9.3$ &$611.5 \pm 9.2$ \\
$640$--$870$ & $750$ & $2\,959 \pm 54$ &$653 \pm 12$ &$655 \pm 12$ \\
$870$--$1\,200$ & $1\,000$ & $1\,752 \pm 42$ &$609 \pm 15$ &$610 \pm 15$ \\
$1\,200$--$1\,600$ & $1\,400$ & $1\,096 \pm 33$ &$600 \pm 18$ &$600 \pm 18$ \\
$1\,600$--$2\,200$ & $1\,900$ & $651 \pm 26$ &$562 \pm 22$ &$562 \pm 22$ \\
$2\,200$--$2\,900$ & $2\,500$ & $386 \pm 20$ &$525 \pm 27$ &$525 \pm 27$ \\
$2\,900$--$3\,900$ & $3\,400$ & $229 \pm 15$ &$490 \pm 32$ &$490 \pm 32$ \\
$3\,900$--$5\,300$ & $4\,600$ & $144 \pm 12$ &$486 \pm 40$ &$486 \pm 40$ \\
$5\,300$--$7\,200$ & $6\,300$ & $76 \pm 8$ &$404 \pm 46$ &$404 \pm 46$ \\
$7\,200$--$9\,800$ & $8\,500$ & $42 \pm 6$ &$352 \pm 54$ &$352 \pm 54$ \\
$9\,800$--$13\,000$ & $12\,000$ & $21 \pm 4$ &$277 \pm 60$ &$277 \pm 60$ \\
$13\,000$--$18\,000$ & $16\,000$ & $8 \pm 2$ &$166 \pm 59$ &$166 \pm 59$ \\
$18\,000$--$24\,000$ & $21\,000$ & $10 \pm 3$ &$330 \pm 100$ &$330 \pm 100$ \\
$24\,000$--$33\,000$ & $29\,000$ & $4 \pm 2$ &$210 \pm 100$ &$210 \pm 100$ \\
$33\,000$--$45\,000$ & $39\,000$ & $2 \pm 1$ &$160 \pm 110$ &$160 \pm 110$ \\
$45\,000$--$60\,000$ & $52\,000$ & $1 \pm 1$ &$130 \pm 130$ &$130 \pm 130$ \\
$60\,000$--$82\,000$ & $71\,000$ & $5 \pm 2$ &$1\,010 \pm 450$ &$1\,010 \pm 450$ \\
$82\,000$--$110\,000$ & $96\,000$ & $3 \pm 1$ &$950 \pm 550$ &$950 \pm 550$ \\
$110\,000$--$150\,000$ & $130\,000$ &  - &  - &  - \\
$150\,000$--$200\,000$ & $180\,000$ & $1 \pm 1$ &$790 \pm 790$ &$790 \pm 790$ \\
$200\,000$--$270\,000$ & $240\,000$ &  - &  - &  - \\
$270\,000$--$370\,000$ & $320\,000$ &  - &  - &  - \\
$370\,000$--$500\,000$ & $440\,000$ & $1 \pm 1$ &$3\,100 \pm 3\,100$ &$3\,100 \pm 3\,100$ \\
$500\,000$--$680\,000$ & $590\,000$ &  - &  - &  - \\
$680\,000$--$920\,000$ & $800\,000$ & $1 \pm 1$ &$7\,600 \pm 7\,600$ &$7\,600 \pm 7\,600$ \\
$920\,000$--$1\,200\,000$ & $1\,100\,000$ &  - &  - &  - \\
$1\,200\,000$--$1\,700\,000$ & $1\,500\,000$ &  - &  - &  - \\
$1\,700\,000$--$2\,300\,000$ & $2\,000\,000$ &  - &  - &  - \\
$2\,300\,000$--$3\,100\,000$ & $2\,700\,000$ & $1 \pm 1$ &$47\,000 \pm 47\,000$ &$47\,000 \pm 47\,000$ \\\bottomrule
    
    \end{tabular}
    \end{adjustbox}
\end{table*}

\end{document}